\documentclass[11pt]{article}
\usepackage[a4paper,hmargin=1.0in,vmargin=1.0in]{geometry}
\usepackage{amsmath,amsthm,amssymb,setspace,sectsty,mathtools,xspace}
\usepackage[round]{natbib}
\usepackage[usenames,dvipsnames]{xcolor}
\usepackage[linktocpage=true,pagebackref=true,colorlinks,allcolors=blue,bookmarks=true, bookmarksopen,bookmarksnumbered]{hyperref}
\usepackage{tikz}
\usetikzlibrary{trees,decorations.pathreplacing,calc,arrows.meta, positioning, shapes}

\newtheoremstyle{DStheorem}
  {\topsep}
  {\topsep}
  {\itshape}
  {0pt}
  {\scshape}
  {.}
  { }
  {\thmname{#1}\thmnumber{ #2}\thmnote{ (#3)}}
\theoremstyle{DStheorem}
\newtheorem{theorem}{Theorem}[section]
\newtheorem{lemma}[theorem]{Lemma}

\newtheorem{corollary}[theorem]{Corollary}
\newtheorem{observation}[theorem]{Observation}

\let\oldproofname=\proofname
\renewcommand{\proofname}{\rm\sc{\oldproofname}}

\newcommand{\bs}[1]{\boldsymbol{#1}}
\newcommand{\eps}{\epsilon}

\newcommand{\prpar}[1]{{\rm Pr} [ #1 ]}
\newcommand{\prsub}[2]{{\rm Pr}_{#1} \left[ #2 \right]}

\newcommand{\ex}[1]{{\mathbb E} \left[ #1 \right]}
\newcommand{\expar}[1]{{\mathbb E} [ #1 ]}
\newcommand{\exsub}[2]{{\mathbb E}_{#1} \left[ #2 \right]}

\newcommand{\opt}{\mathrm{OPT}}
\newcommand{\myapp}{\mathrm{app}}
\newcommand{\mydepth}{\mathrm{depth}}
\newcommand{\myroot}{\mathrm{root}}
\newcommand{\myvalue}{\mathrm{value}}
\newcommand{\mytree}{\mathrm{tree}}
\newcommand{\mystate}{\mathrm{state}}
\newcommand{\lrproperty}{L$\geq$R\xspace}
\newcommand{\vrlproperty}{V$+$R$\geq$L\xspace}
\newcommand{\myblock}{\mathrm{block}}
\newcommand{\mynew}{\mathrm{new}}

\sectionfont{\large} \subsectionfont{\normalsize}
\allowdisplaybreaks
\makeindex
\newcommand{\changelocaltocdepth}[1]{%
  \addtocontents{toc}{\protect\setcounter{tocdepth}{#1}}%
  \setcounter{tocdepth}{#1}%
}

\begin{document}

\begin{titlepage}

\title{Approximation Schemes for Sequential Hiring Problems}
\author{%
Danny Segev\thanks{School of Mathematical Sciences and Coller School of Management, Tel Aviv University, Tel Aviv 69978, Israel. Email: {\tt segevdanny@tauex.tau.ac.il}. Supported by Israel Science Foundation grant 1407/20.}%
\and%
Uri Stein\thanks{School of Mathematical Sciences, Tel Aviv University, Tel Aviv 69978, Israel. Email: {\tt uriystein@gmail.com}.}}
\date{}
\maketitle

\pagenumbering{Roman}

\begin{abstract}
The main contribution of this paper resides in providing novel algorithmic advances and analytical insights for the sequential hiring problem, a recently introduced dynamic optimization model where a firm adaptively fills a limited number of positions from a pool of applicants with known values and acceptance probabilities. While earlier research established a strong foundation --- notably an LP-based $(1 - \frac{e^{-k}k^k}{k!})$-approximation by Epstein and Ma (Operations Research, 2024) --- the attainability of superior approximation guarantees has remained a central open question.

Our work addresses this challenge by establishing the first polynomial-time approximation scheme for sequential hiring, proposing an $O(n^{O(1)} \cdot T^{2^{\tilde{O}(1/\epsilon^{2})}})$-time construction of semi-adaptive policies whose expected reward is within factor $1 - \eps$ of optimal. To overcome the constant-factor optimality loss inherent to earlier literature, and to circumvent intrinsic representational barriers of adaptive policies, our approach is driven by the following innovations:
\begin{itemize}
    \item \emph{The block-responsive paradigm}: We introduce block-responsive policies, a new class of decision-making strategies, selecting ordered sets (blocks) of applicants rather than single individuals, while still allowing for internal reactivity.

    \item \emph{Adaptivity and efficiency}: We prove that these policies can nearly match the performance of general adaptive policies while utilizing polynomially-sized decision trees.

    \item \emph{Efficient construction}: By developing a recursive enumeration-based framework, we resolve the problematic ``few-positions'' regime, bypassing a fundamental hurdle that hindered previous approaches.
\end{itemize}
\end{abstract}


\end{titlepage}

\newpage
\setcounter{page}{2}
\tableofcontents

\newpage
\pagestyle{plain}
\pagenumbering{arabic}
\setcounter{page}{1}

\section{Introduction} \label{sec:intro}

In today's highly competitive business environment, efficacious hiring pipelines stand as critical determinants of organizational success. As highlighted by surveys on global recruiting trends such as those published by \cite{linkedin_survey} and \cite{glassdoor_survey}, developing and implementing state-of-the-art mechanisms to attract, assess, and onboard top talent is paramount for companies striving to maintain a competitive edge. On the academic front, multiple disciplines have been continuously concerned with providing comprehensive insights into recruitment strategies and best practices. In this context, one prominent research direction focuses on the mathematical optimization of various components within hiring pipelines, aiming to enhance the efficiency and effectiveness of talent acquisition processes.

By leveraging a multitude of modern-day optimization techniques, organizations can refine various stages along these pipelines, such as applicant screening, interview scheduling, and selection strategies, aiming to circumvent hiring bottlenecks, improve candidate quality, and lower recruitment costs. 
Optimization has also been extremely useful in its predictive capacity, allowing companies to anticipate candidate success and retention rates, further aligning recruitment outcomes with organizational goals.
Faced with recurring talent shortages and fluctuating job markets, organizations are progressively adopting data-driven strategies to enhance their hiring processes. This trend is underscored in academic discourse, as evidenced by the work of \cite{Dakin89}, \cite{Chapman05}, and \cite{konig22}, for example. 
By harnessing data-driven methodologies, firms can accurately forecast and quantify applicants' compatibility with a particular organization, as well as the likelihood that they will accept a given job offer.
Consequently, one of the most fundamental questions along these lines is: How can organizations leverage such insights to optimize hiring pipelines for maximum efficacy and agility?

Operating within this framework, the current paper is dedicated to improving on the best-known performance guarantees for the sequential hiring problem, initially introduced by \cite*{PurohitGR19} and subsequently studied by \cite{EpsteinM22}.
At a high level, this dynamic optimization model considers a single firm that wishes to fill a limited number of identical positions via a given pool of available applicants,
each associated with a pre-estimated value as well as with an acceptance probability.
Equipped with these input ingredients, our firm of interest adaptively extends offers, one after the other, across a limited time frame. In this regard, the fundamental challenge is to determine on the fly which applicants will receive an offer as well as to judiciously sequence these events in real time, taking into account the decisions made on the applicants' side up to that point. Our objective is to identify an optimal hiring policy in the sense of yielding the best-possible expected reward.

In the remainder of this exposition, Section~\ref{subsec:model_description} provides a formal description of our model formulation. In Section~\ref{subsec:results_questions}, we elaborate on earlier work within this framework and delineate our motivating research questions. Section~\ref{subsec:contributions_highlights} presents the principal contributions of this paper and highlights selected technical ideas. Lastly, we provide a succinct overview of related literature in Section~\ref{subsec:related_work}.

\subsection{Model description} \label{subsec:model_description}

Broadly speaking, in the sequential hiring problem, a firm aims to fill up to $k$ identical positions via a dynamic process that proceeds along a sequence of $T$ stages. At each stage, the firm strategically extends an offer to a selected applicant and instantly receives a definitive response: Either acceptance or rejection.
When an applicant accepts an offer, she is allocated one of the available positions, and the firm gains an applicant-specific reward; this decision is binding and irrevocable, meaning that it cannot be reversed at a later point in time. Conversely, when an applicant declines, we cannot extend further offers to this applicant in subsequent stages.

To capture the finer details of this model, we assume that the pool of available applicants is known in advance, consisting of $n$ applicants who will be referred to as $1, \ldots, n$. 
Each applicant $i \in [n]$ is associated with a deterministic value $v_i \geq 0$, representing our firm's projected utility due to hiring this applicant. 
In addition, applicant $i$ has a known probability of accepting a given offer, designated by $p_i$. 
It is worth emphasizing that these two parameters are stage-invariant and position-invariant, meaning that they remain unchanged throughout the entire hiring process. 
For ease of notation, each applicant's decision is represented by a binary random variable $X_i$, indicating whether applicant $i$ accepts a given offer, with $\prpar{ X_i = 1 } = p_i$. We assume that applicants' decisions are unrelated, in the sense that $X_1, \ldots, X_n$ are mutually independent.

\paragraph{Hiring policies.} Given these input parameters, the random process we consider evolves along a sequence of $T$ discrete stages, indexed by $1, \ldots, T$, according to system dynamics guided by so-called adaptive hiring policies. To formalize this notion, for any stage $t \in [T]$, let us make use of $\mathcal{A}_t$ to denote the set of applicants available at the beginning of stage $t$, referring to those who have not been extended an offer up until now. In addition, $k_t$ will stand for the number of open positions at the beginning of this stage. 
As described below, our system state at any point in time will be captured by triples of the form $(t, k_t, \mathcal{A}_t)$. With this representation, an adaptive hiring policy is simply a function $\mathcal{H} : [T] \times [k] \times 2^{[n]} \to [n]$ that, given any state $(t, k_t, \mathcal{A}_t)$ with $t \leq T$, $k_t \geq 1$, and $\mathcal{A}_t \neq \emptyset$, decides on the available applicant $\mathcal{H}(t, k_t, \mathcal{A}_t) \in \mathcal{A}_t$ who will be given an offer in the current stage.

\paragraph{System dynamics.} At the beginning of stage $1$,  all $k$ positions are still open and all applicants are available to choose from, meaning that our starting state $(1, k_1, \mathcal{A}_1)$ has $k_1 = k$ and $\mathcal{A}_1 = [n]$. Then, at each stage $t \in [T]$, assuming that $k_t \geq 1$ and $\mathcal{A}_t \neq \emptyset$, the following sequence of steps occurs:
\begin{enumerate}
    \item {\em Offering step}: We first let our hiring policy $\mathcal{H}$ pick one of the available applicants, $\mathcal{H}(t, k_t, \mathcal{A}_t)$, who is instantaneously offered one of the $k_t$ still-open positions.
    
    \item {\em Applicant decision step}: Then, applicant $\mathcal{H}(t, k_t, \mathcal{A}_t)$ decides to either accept or reject this offer,  corresponding to $X_{\mathcal{H}(t, k_t, \mathcal{A}_t)} = 1$ and $X_{\mathcal{H}(t, k_t, \mathcal{A}_t)} = 0$. In the former scenario, we collect the value $v_{\mathcal{H}(t, k_t, \mathcal{A}_t)}$ of the accepting applicant and increment our cumulative reward by this quantity. Conversely, when an offer is rejected, we do not collect any value. In either case, our cumulative reward is incremented by $X_{\mathcal{H}(t, k_t, \mathcal{A}_t)} \cdot v_{\mathcal{H}(t, k_t, \mathcal{A}_t)}$.
    
    \item {\em State update}: Finally, we proceed to stage $t + 1$ with the remaining set of applicants $\mathcal{A}_{t+1} = \mathcal{A}_{t}\setminus \{\mathcal{H}(t, k_t, \mathcal{A}_t)\}$ and with the updated number of open positions, $k_{t+1} = k_t - X_{\mathcal{H}(t, k_t, \mathcal{A}_t)}$.
\end{enumerate}
This process terminates as soon as we arrive at a terminal state $(t, k_t, \mathcal{A}_t)$ that corresponds to completing our planning horizon ($t = T+1$), or filling all positions ($k_t = 0$), or exhausting the pool of available applicants ($\mathcal{A}_t = \emptyset$).

\paragraph{Reward function and objective.} For any hiring policy $\mathcal{H} : [T] \times [k] \times 2^{[n]} \to [n]$ and for any state $(t, k_t, \mathcal{A}_t)$, we make use of $R_{\mathcal{H}}(t, k_t, \mathcal{A}_t)$ to denote the cumulative reward collected by the policy $\mathcal{H}$ along the sequence of stages $t, \ldots, T$, starting at state $(t, k_t, \mathcal{A}_t)$. Based on the preceding discussion, this random variable can be written as:
\[ R_{\mathcal{H}}(t, k_t, \mathcal{A}_t) ~~=~~ \underbrace{ X_{\mathcal{H}(t, k_t, \mathcal{A}_t)} \cdot v_{\mathcal{H}(t, k_t, \mathcal{A}_t)} }_{ \text{immediate reward} } + \underbrace{ R_{\mathcal{H}}(t+1, k_t - X_{\mathcal{H}(t, k_t, \mathcal{A}_t)}, \mathcal{A}_{t}\setminus \{\mathcal{H}(t, k_t, \mathcal{A}_t)\}) }_{ \text{cumulative future reward} } \ . \]
Consequently, letting $\mathcal{R}_{\mathcal{H}}(t, k_t, \mathcal{A}_t) = \expar{ R_{\mathcal{H}}(t, k_t, \mathcal{A}_t) }$ be the corresponding expected cumulative reward, it follows that the latter function can be recursively expressed as
\begin{eqnarray}
\mathcal{R}_{\mathcal{H}}(t, k_t, \mathcal{A}_t) & = &  p_{\mathcal{H}(t, k_t, \mathcal{A}_t)} \cdot v_{\mathcal{H}(t, k_t, \mathcal{A}_t)} + \ex{R_{\mathcal{H}}(t+1, k_t - X_{\mathcal{H}(t, k_t, \mathcal{A}_t)}, \mathcal{A}_{t}\setminus \{\mathcal{H}(t, k_t, \mathcal{A}_t)\})} \nonumber \\ 
& = & p_{\mathcal{H}(t, k_t, \mathcal{A}_t)} \cdot (v_{\mathcal{H}(t, k_t, \mathcal{A}_t)} + \mathcal{R}_{\mathcal{H}}(t+1, k_t - 1, \mathcal{A}_{t}\setminus \{\mathcal{H}(t, k_t, \mathcal{A}_t)\})) \nonumber\\
&& \mbox{} + (1 - p_{\mathcal{H}(t, k_t, \mathcal{A}_t)}) \cdot \mathcal{R}_{\mathcal{H}}(t+1, k_t, \mathcal{A}_{t}\setminus \{\mathcal{H}(t, k_t, \mathcal{A}_t)\}) \ . \label{eq:reward_2}
\end{eqnarray}
It is worth noting that all expectations above are taken over the randomness in the accept/reject choice of applicant $\mathcal{H}(t, k_t, \mathcal{A}_t)$ as well as over the additional applicants' choices across future stages. As previously mentioned, terminal states of this recursion correspond to having $t = T+1$ or $k_t = 0$ or $\mathcal{A}_t = \emptyset$, in which case $\mathcal{R}_{\mathcal{H}}(t, k_t, \mathcal{A}_t) = 0$. For convenience, we designate the total expected reward along the entire process by $\mathcal{R}(\mathcal{H}) = \mathcal{R}_{\mathcal{H}}(1, k, [n])$. Our goal is to compute a hiring policy $\mathcal{H}$ whose total expected reward $\mathcal{R}(\mathcal{H})$ is maximized.

\subsection{Existing results and open questions} \label{subsec:results_questions} 
\paragraph{Introductory work.}
The sequential hiring problem was originally introduced by \cite{PurohitGR19}, who aimed to rigorously address one of the most fundamental questions in hiring pipelines: How should a firm make offering decisions under uncertainty on the applicants' side, ensuring that positions are filled with the best set of applicants?
Their main algorithmic contribution consists of designing an efficiently representable hiring policy whose expected reward is within factor $\frac{1}{2}$ of optimal.
Technically speaking, \cite{PurohitGR19} utilized dynamic programming ideas to identify an optimal policy within the family of greedy policies, where applicants are extended offers in weakly-decreasing order of their values;
they further showed that this policy outperforms any non-adaptive policy.
Their approximation guarantee is established by proving the existence of a non-adaptive policy within this family whose expected reward is within factor $\frac{1}{2}$ of optimal, following the stochastic probing approach of \cite{Gupta17}.
In addition, the authors proposed a $\frac{1}{8}$-approximation for the so-called parallel hiring problem, as well as a $\frac{1}{10}$-approximation for the stochastic knapsack problem, both falling outside the scope of our paper.

\paragraph{Stronger guarantees.}
Subsequently, \cite{EpsteinM22} designed a non-adaptive policy that collects a fraction of at least $1 - \frac{e^{-k}k^k}{k!}$ of the optimal expected reward.
This performance guarantee significantly improves on that of \cite{PurohitGR19}, since the function $k \mapsto 1 - \frac{e^{-k}k^k}{k!}$ is monotone increasing, starting at $1-\frac{1}{e} \approx 0.632$ for $k=1$ and converging to $1$ as $k$ tends to infinity.
To derive this result, \cite{EpsteinM22} initially focused on obtaining improved guarantees for the ProbeTop$K$ problem, which is a fundamental setting in stochastic optimization; see, e.g., \cite{Gupta13}, \cite{Gupta16}, and \cite{Gallego22}. 
In particular, with respect to a well-known linear relaxation, they fused the dependent rounding framework of \cite{GandhiKPS06} with additional insights, showing that their resulting policy can be migrated to the sequential hiring problem while preserving its performance guarantee.
On a different front, \cite{EpsteinM22} have also proposed a $(1-\frac{1}{e})$-approximation for the parallel hiring problem.

\paragraph{Open questions.}
In light of these state-of-the-art results, quite a few intriguing questions remain wide open.
These inquiries about improved approximation guarantees and fine-grained structural understanding, which are the main driving forces behind our work, can be concisely summarized as follows:
\begin{enumerate}    
    \item {\em Devising an approximation scheme?} Given that the sequential hiring problem can be efficiently approximated within a constant factor, can we  devise an approximation scheme?
    Somewhat unsurprisingly, such results are extremely rare in stochastic combinatorial optimization.
        
    \item {\em Structural characterization of near-optimal policies?} Unraveling useful structural properties of near-optimal hiring policies stands at the heart of attaining improved approximation guarantees. Do optimal or near-optimal policies possess unique features that set them apart? Can we exploit these properties for algorithmic purposes? 

    \item {\em Representation of near-optimal policies in polynomial space?} 
    Taking into account that conventional representations of optimal dynamic policies are exponentially-sized, is it possible to describe carefully-constructed near-optimal policies in polynomial space?
\end{enumerate}

\subsection{Main contributions and technical highlights} \label{subsec:contributions_highlights}

In what follows, we take  readers through a guided tour of our cornerstone results, touching upon selected notions that will appear further down the road.
In particular, we elaborate on a number of well-hidden structural properties as well as on their algorithmic usefulness, which are essential to our overall framework. It is worth pointing out that this discussion is by no means an in-depth examination of all moving parts; rather, it serves as a high-level overview of the upcoming contents.

\paragraph{Approximation scheme.}
The primary algorithmic contribution of this paper resides in developing a polynomial-time approximation scheme (PTAS) for the sequential hiring problem. 
Specifically, for any fixed $\epsilon > 0$, our algorithmic approach determines in polynomial time an efficiently-representable hiring policy whose expected reward is within factor $1-\epsilon$ of the best-possible expected reward of any adaptive policy. The specifics of this result can be briefly summarized as follows. 

\begin{theorem} \label{thm:ptas}
For any $\epsilon > 0$, the sequential hiring problem can be approximated within factor $1 - \epsilon$ of optimal. Our algorithm admits an $O(n^{ O(1) } \cdot T^{2^{\tilde{O}(1/\epsilon^2)}})$-time implementation. 
\end{theorem}


\paragraph{Unveiling structural properties of optimal hiring policies.} 
In Section~\ref{sec:background}, we start off with a succinct discourse on how hiring policies can be represented through binary decision trees.
From this perspective, we introduce two pivotal characteristics of optimal policies, the \lrproperty-property and the \vrlproperty-property, identifying trees that satisfy both as being ``canonical''.
At a high level, the \lrproperty-property asserts that when an applicant decides to reject an offer, the future expected reward of any reasonable policy from this point on should be at least as good as the analogous expected reward had the offer in question been accepted.
This reasoning is very intuitive, as any rejection favorably impacts subsequent stages by leaving an additional open position to be filled.
In contrast, the \vrlproperty-property informally states that for any offer extended by a reasonable policy, the expected immediate and future reward gained by accepting this offer must be at least as good as the future reward gained by rejecting it.
At least intuitively, this property makes sense; otherwise, why are we even making such an offer?
Along these lines, we prove that any decision tree can be converted into a canonical form without compromising its expected reward.

\paragraph{Preliminary approximation scheme.} 
In Section~\ref{sec:quasi_ptas}, we integrate the above-mentioned structural properties along with suitable perturbations to several input parameters, ending up with a quasi-polynomial time approximation scheme (QPTAS). 
This approach to addressing sequential hiring involves two fundamental alterations to the underlying set of applicants.
First, we round up the acceptance probability of each applicant who is very unlikely to accept an offer, so that all such applicants are accepting with ``non-tiny'' probabilities; their associated values are adjusted as well.
Second, we round down the remaining acceptance probabilities so that the number of distinct probabilities over all applicants becomes $O_{\epsilon}(\log T)$.
Such alterations allow us to dramatically trim down the exponentially-sized state-space of natural dynamic programs, thereby deriving an optimal hiring policy with respect to these modified parameters in $O ( n^{O(1)} \cdot T^{O_{\eps}(\log T)} )$ time. Leveraging several insights gleaned from canonical hiring policies, we prove that this policy can be efficiently migrated back to our original instance, losing only an $\epsilon$-related fraction of its expected reward.

\paragraph{Block-responsive policies.} 
Toward obtaining a truly polynomial-time approach, Section~\ref{sec:block_responsive} introduces the concept of block-responsive policies.
Here, rather than selecting a single applicant at each stage, such policies select ordered subsets of applicants, called blocks, thereby enabling a potential reduction in the overall size of their decision trees. To ensure that such policies are implementable in terms of our original system dynamics, offers within each block are sequentially extended in a predetermined order; once an applicant accepts an offer, all remaining applicants are immediately skipped over.
Quite surprisingly, we prove the existence of a block-responsive policy such that, on the one hand, its decision tree consists of only $(\frac{1}{\eps})^{O(k)}$ blocks, and on the other hand, its expected reward is within factor $1- O(\eps^3 k )$ of optimal, across all adaptive policies.
This finding indicates that, when the number of available positions is sufficiently small, there exists a polynomially-sized block-responsive policy whose expected reward is near-optimal. Once again, we emphasize that this result is not constructive; rather, it is merely an existence proof.

\paragraph{Devising a polynomial-time implementation.}
To explore the algorithmic consequences of block-responsive policies, in Section~\ref{section:PTAS}, we devise a true polynomial-time approximation scheme (PTAS) for the sequential hiring problem.
To this end, we begin by classifying any given instance into distinct parametric regimes, based on its number of open positions, $k$.
Specifically, for sufficiently large values of $k$, we show that the existing approach of \cite{EpsteinM22} already provides an approximation scheme.
In contrast, the core difficulty of sequential hiring resides in the yet-uncharted regime where $k$ is small. Here, we devise a recursive enumeration-based approach to efficiently construct a polynomially-sized family of candidate block-responsive policies.
Based on the analysis conducted in Section~\ref{sec:block_responsive}, we prove that this family contains at least one near-optimal hiring policy.    

\subsection{Related work} \label{subsec:related_work}
From a conceptual standpoint, as well as from a technical one, our work resides at the intersection of several interrelated research domains, including optimal stopping, block-adaptive policies, and stochastic probing.
Noting that each of these domains has accumulated an extensive body of literature by now, we cannot do justice and present an exhaustive overview. Therefore, the remainder of this section mainly focuses on conveying the general spirit of these topics.

\paragraph{Secretary problems and prophet inequalities.}
Secretary problems are among the most renowned research directions in optimal stopping theory \citep{Lindley61, Dyn63, Chow64}. Broadly speaking, a decision-maker is presented with a finite set of applicants who arrive in random order, with their values revealed upon arrival. The decision-maker faces constraints on the number of applicants she can observe and must decide which applicants to accept, adhering to a predefined set of rules that govern the composition of the accepted set. Our objective is to maximize the combined value of accepted applicants, benchmarked against the optimal offline decision.
For additional background on this topic, we refer readers to a number of excellent surveys \citep{freeman83, ferguson89, hill09}.

Yet another pivotal concept in the realm of online decision-making is that of ``prophet inequalities'', first introduced by \cite{krengel77}.
In such scenarios, we are typically presented with a finite set of independent distributions, from which random variables are sequentially drawn according to a predetermined order. Upon observing each realization, we should decide whether to accept or reject it, subject to an underlying set of constraints on the collection of accepted variables.
Our objective is to maximize the expected combined value of the accepted variables, benchmarked against a `prophet' who possesses foreknowledge of all realizations and can thus make the optimal offline decision.
The extensive research on this topic can be appreciated by consulting surveys such as those of \cite{ hill83}, \cite{Hill1992ASO}, \cite{lucier17}, and \cite{correa19}, as well as the work of 
\cite{ Cahn84}, \cite{Hajiaghayi2007AutomatedOM}, \cite{michal20}, \cite{michal22}, and \cite{arnosti23}, for instance.

Despite sharing some common features, our problem of interest fundamentally differs from classic secretary problems and prophet inequalities.
First, in sequential hiring, decisions must be made prior to observing an applicant’s outcome. 
Specifically, offers are extended irrevocably, without knowing whether the applicant will accept or reject them. 
Additionally, sequential hiring policies are evaluated against an optimal policy that achieves the maximum expected reward, whereas secretary problems and prophet inequalities are benchmarked against a hypothetical prophet with full knowledge of all outcomes in advance.

\paragraph{Block-adaptive policies.}
In Sections~\ref{sec:block_responsive} and \ref{section:PTAS}, our algorithmic approach revolves around transforming near-optimal hiring policies into being ``block-responsive'', with negligible loss in their expected reward.
As explained in Section~\ref{subsec:results_questions}, rather than selecting a single applicant at each stage, such policies select ordered subsets of applicants, referred to as blocks.
Within each block, offers are sequentially extended in a predetermined order, such that once an applicant accepts an offer, all remaining applicants are immediately skipped over.

What inspired us to consider policies of this nature is the known concept of ``block-adaptive'' policies, originally introduced by \cite*{BhalgatGK11}.
These reduced-adaptivity policies aim to address representational issues by utilizing a small number of decision points, mostly by selecting a set of elements at each such point and operating within this set in a non-adaptive way. 
Block-adaptive policies have turned out to be very useful in stochastic combinatorial optimization, as demonstrated by the work of \cite{LiY13}, \cite{FuLX18}, and \cite{Segev021}.

As explained in Section~\ref{sec:block_responsive}, the key distinction between block-responsive and block-adaptive policies lies in their level of adaptivity. 
Block-adaptive policies limit flexibility by selecting a fixed set of actions in advance, meaning that every element within a chosen block must be probed. In contrast, block-responsive policies allow adjustments within a block, potentially affecting our overall sequence of offering decisions in a dramatic way, since once an applicant accepts an offer, all remaining applicants in that block are skipped without receiving offers.

\paragraph{Stochastic probing.}
Stochastic probing problems have gained significant traction in recent years, finding applications across various domains such as routing, online dating, and healthcare management, among others.
Generally speaking, we are given a universe of elements $E$, where each element $e \in E$ is associated with a random variable $X_e$ whose distribution is assumed to be known;
the value of $X_e$ can be determined by probing $e$.
Our objective is to devise a policy that dynamically decides which elements to probe and accept, aiming to maximize the expected total value of  accepted elements. Here, any solution is subject to ``outer'' constraints, limiting the set of elements we can probe, as well as to ``inner'' constraints, limiting those that can be accepted.
The performance of our policy is evaluated against an optimal policy that achieves the maximum possible expected total value of its accepted elements.
To date, numerous approaches have been developed to rigorously address stochastic probing problems, including LP-based methods \citep{Guha07, guha10, Gupta13}, submodularity-based techniques \citep{AsadpourN16, chen16, Gallego22}, and direct decision-tree arguments \citep{Gupta17, bradac19}.


\section{Technical Background} \label{sec:background}

In what follows, we introduce a number of fundamental concepts related to hiring policies, laying the groundwork for technical discussions in upcoming sections.
Specifically, in Section~\ref{subsec:trees}, we will elaborate on how the inner workings of any hiring policy can be represented via an appropriate decision tree. 
Subsequently, Section~\ref{subsec:dec_tree_example} will provide a small-scale example to illustrate this concept and some of its related measures.
In Section~\ref{subsec:structural_properties}, we will establish two structural properties intrinsic to optimal decision trees; these will play a crucial role within our algorithmic approach.

\subsection{Decision tree representation} \label{subsec:trees}

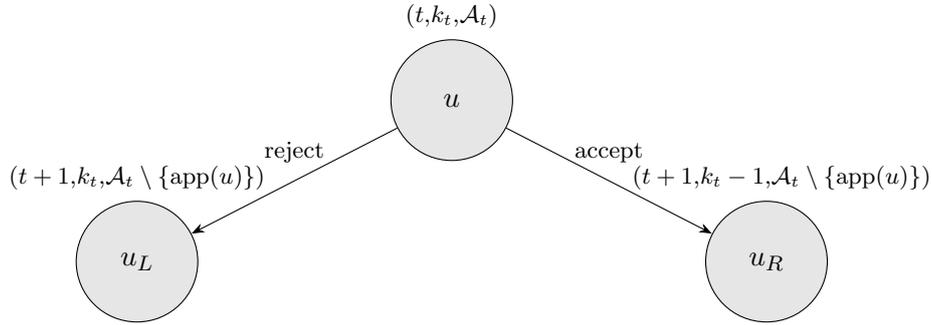
\begin{figure}[htbp!]
    \centering
    \begin{tikzpicture}[
  >=Stealth,                      
  every node/.style={
    circle, draw=black,           
    fill=gray!20,                 
    minimum size=16mm,            
    font=\bfseries,               
    align=center
  },
  node distance=1cm and 3cm     
]

\node (root) {$u$}
    node(rootlabel) [above=0.1mm of root, font=\footnotesize, style={rectangle, draw=none, fill=none, minimum size=4mm}]
    {($t$,$k_t$,$\mathcal{A}_t$)};

\node (n2) [below left=of root] {$u_L$}
    node(n2label) [above=0.1mm of n2, font=\footnotesize, style={rectangle, draw=none, fill=none, minimum size=4mm}]
    {($t+1$,$k_t$,$\mathcal{A}_t \setminus \{\myapp(u)\}$)};

\node (n3) [below right=of root] {$u_R$}
  node[above=0.1mm of n3, xshift=2mm, font=\footnotesize, style={rectangle, draw=none, fill=none, minimum size=4mm}]
    {($t+1$,$k_t-1$,$\mathcal{A}_t \setminus \{\myapp(u)\}$)};

\draw[->] (root) -- node[midway, above=3pt, font=\footnotesize, style={rectangle, draw=none, fill=none, minimum size=4mm}]{reject} (n2);
\draw[->] (root) -- node[midway, above=3pt, font=\footnotesize, style={rectangle, draw=none, fill=none, minimum size=4mm}]{accept} (n3);

\end{tikzpicture}
  
    \caption{Internal nodes of the decision tree $\mathcal{T}_{\mathcal{H}}$.}
    \label{fig:internal_nodes}
\end{figure}

Following conventional constructs in dynamic optimization, we represent any hiring policy through an equivalent recursive relation, captured by an augmented binary decision tree. To better understand the upcoming discussion, we advise the reader to concurrently consult the small-scale example in Section~\ref{subsec:dec_tree_example}. In our context, for any hiring policy $\mathcal{H}$, its decision tree representation $\mathcal{T}_{\mathcal{H}}$ is recursively defined as follows:
\begin{itemize}
    \item {\em Functions:} Each node of ${\cal T}_{\cal H}$ will be associated with two functions, $\mystate(\cdot)$ and $\myapp(\cdot)$, whose intended roles are to encode its corresponding system state and chosen applicant, respectively. We mention in passing that, due to employing ${\cal H}$ as our hiring policy, one always has $\myapp(\cdot) = {\cal H}( \mystate(\cdot) )$, meaning that the former notation is being utilized mostly for convenience. 
   
    \item {\em Internal nodes:} The root of the tree ${\cal T}_{\cal H}$, denoted as $\myroot(\mathcal{T}_{\mathcal{H}})$, corresponds to the starting state of our hiring process, meaning that $\mystate(\myroot(\mathcal{T}_{\mathcal{H}})) = (1, k_1, \mathcal{A}_1)$. Consequently, the first applicant to whom $\mathcal{H}$ makes an offer is given by $\myapp(\myroot(\mathcal{T}_{\mathcal{H}})) = \mathcal{H}(1, k_1, \mathcal{A}_1)$. Beyond this root, each internal node $u$ has two downward arcs connecting $u$ to its left and right children, denoted as $u_L$ and $u_R$, respectively. As schematically shown in Figure~\ref{fig:internal_nodes}, the left downward arc captures the event where $\myapp(u)$ rejects our offer, while the right downward arc captures the opposite event, where this offer is accepted. As such, when $\mystate(u) = (t, k_t, \mathcal{A}_t)$, we have $\mystate(u_L) = (t + 1, k_t, \mathcal{A}_t \setminus \{ \myapp(u) \})$ and $\mystate(u_R) = (t + 1, k_t - 1, \mathcal{A}_t \setminus \{ \myapp(u) \})$. This construction proceeds by recursion until reaching a leaf node, which is separately discussed below. 
    For ease of notation, we make use of $\mathcal{T}_{\mathcal{H}}(u)$ to designate the subtree rooted at node $u$, noting that it can be viewed as the decision tree of the hiring policy $\mathcal{H}$, starting at $\mystate(u)$.

    \item {\em Leaves:} Within the decision tree $\mathcal{T}_{\mathcal{H}}$, each leaf $u$ serves as an indication that, upon reaching $\mystate(u)$, the hiring policy ${\cal H}$ cannot extend further offers. In other words, $\mystate(u) = (t, k_t, \mathcal{A}_t)$ happens to be a terminal system state, with  $t = T+1$ or $k_t = 0$ or $\mathcal{A}_t = \emptyset$. 
    For uniform treatment in subsequent analysis, each leaf will be associated with a ``virtual'' applicant, denoted as applicant $0$.
    The latter possesses a value of $v_0 = 0$ and an acceptance probability of $p_0 = 0$.
\end{itemize}

\paragraph{Salient attributes.} Given the construction above, let us briefly highlight several structural attributes related to root-to-leaf paths that will become useful later on. To this end, consider a downward path $P$ that connects $\myroot(\mathcal{T}_{\cal H})$ to some leaf node. We first observe that $P$ involves at most $k$ right turns (i.e., right downward arcs), as each such turn corresponds to accepting an offer, and the hiring policy $\cal H$ may fill up to $k$ positions. In addition, the applicants $\{ \myapp(u) \}_{ u \in P }$ are distinct, since every root-to-leaf path within $\mathcal{T}_{\mathcal{H}}$ stands for one particular sequence of offers made by the hiring policy $\mathcal{H}$, and since the latter extends at most one offer to each applicant. Finally, following the same reasoning, $P$ consists of at most $T$ arcs, implying that the depth of $\mathcal{T}_{\cal H}$ is at most $T$.

\paragraph{Valid trees.} Knowing in advance that subsequent sections will repeatedly alter the structure of various decision trees, it is worth highlighting several basic properties that must be maintained to ensure that a given tree ${\cal T}$ still represents a concrete hiring policy. Based on the preceding discussion, these properties can be briefly summarized as follows:
\begin{enumerate}
    \item {\em At most $T$ stages}: $\mydepth(\mathcal{T}) \leq T$. \label{tree_def_0}
    
    \item {\em Filling at most $k$ positions}: Each root-to-leaf path has at most $k$ right turns. \label{tree_def_1}
    
    \item {\em At most one offer per applicant}: On each root-to-leaf path, all applicants are distinct.  \label{tree_def_2}
    
    \item {\em Consistency}: For any pair of nodes $u \neq v$ with $\mystate(u) = \mystate(v)$, we have $\myapp(u) = \myapp(v)$. \label{tree_def_3}
\end{enumerate}
Any decision tree that satisfies properties~\ref{tree_def_0}-\ref{tree_def_3} will be called valid. 
    
\begin{figure}[htbp!]
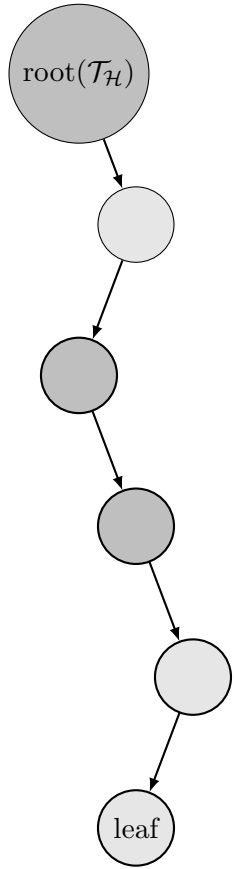

    \centering
 \tikz [level distance=2cm, edge from parent/.style={draw,->,>=latex, thick}]
 \node [font=\bfseries, style={circle, circle, draw=black, fill=gray!50, minimum size=10mm}] {$\myroot(\mathcal{T}_{\mathcal{H}})$}  
    child [missing]
    child { node [font=\bfseries, style={circle, circle, draw=black, fill=gray!20, minimum size=10mm}] { }
        child { node [font=\bfseries, style={circle, circle, draw=black, fill=gray!50, minimum size=10mm}] { }
            child [missing]
            child { node [font=\bfseries, style={circle, circle, draw=black, fill=gray!50, minimum size=10mm}] { }
                child [missing]
                child{ node [font=\bfseries, style={circle, circle, draw=black, fill=gray!20, minimum size=10mm}] { }
                    child { node [ style={circle, circle, draw=black, fill=gray!20, minimum size=10mm}] {leaf}}
                    child [missing]
                    }
                }
            }
        child [missing]
            };

\caption{Active nodes along the path $P \in \mathcal{P}(\mathcal{T}_{\mathcal{H}})$ are darker gray, while inactive nodes are marked in light gray.}
    \label{fig:path_active_nodes}
\end{figure}

\paragraph{The induced distribution $\bs{\pi_{\cal{T}_{\cal H}}}$ over root-to-leaf paths.} As previously mentioned, each root-to-leaf path in the decision tree $\mathcal{T}_{\cal H}$ represents one possible outcome of employing the hiring policy $\mathcal{H}$, consisting of a unique sequence of offers along with the specific decision made by each applicant involved. As a result, $\mathcal{T}_{\cal H}$ induces a distribution $\pi_{\mathcal{T}_{\cal H}}$ over the collection of such paths. To better understand this statement, let ${\cal P}(\mathcal{T}_{\cal H})$ denote the collection of all root-to-leaf paths in $\mathcal{T}_{\cal H}$.
For each path $P \in {\cal P}(\mathcal{T}_{\cal H})$, we say that node $u \in P$ is active when its outgoing arc along $P$ corresponds to a right turn, meaning that applicant $\myapp(u)$ has accepted our offer. Conversely, $u$ is inactive when its outgoing arc is a left turn, corresponding to a rejection.
Figure~\ref{fig:path_active_nodes} provides a simple example of this notion.
We define $\alpha(P)$ and $\bar{\alpha}(P)$ as the sets of active and inactive nodes along $P$, respectively. As such, when employing the hiring policy ${\cal H}$, the probability of traversing through the path $P$ is exactly
\[ \prsub{{\cal T_{\cal H}}}{P}~~=~~ \prod_{u \in \alpha(P)} p_{\myapp(u)} \cdot \prod_{u \in \bar \alpha(P) } (1 - p_{\myapp(u)}) \ . \]
With this observation, $\pi_{\mathcal{T}_{\cal H}}$ is defined as the distribution over $\cal P(\cal T_{\cal H})$, where the probability mass of each path $P \in \cal P(\cal T_{\cal H})$ is given by $\prsub{{\cal T_{\cal H}}}{P}$. 

\paragraph{Expected reward of decision trees.} We conclude this exposition by explaining how the expected reward of a decision tree $\mathcal{T}_{\mathcal{H}}$ is formally defined. For this purpose, given a root-to-leaf path $P \in \cal P(\cal T_{\cal H})$, we utilize $\myvalue(P)$ to designate the total reward collected along this path, namely, $\myvalue(P) = \sum_{u \in \alpha(P)} v_{\myapp(u)}$. 
Here, we only consider the cumulative value of active nodes, corresponding to applicants who accepted our offers. With this definition in place, let us introduce the expected cumulative reward of the decision tree $\mathcal{T}_{\cal H}$, given by 
\begin{equation} 
   \mathcal{R}_{\mytree}(\mathcal{T}_{\cal H}) ~~=~~ \exsub{P \sim \pi_{\mathcal{T}_{\mathcal{H}}}} {\myvalue(P)} \ . \label{eq:reward_def} 
\end{equation}
In this representation, we view the random reward attained by the policy ${\cal H}$ as $\myvalue(P)$, where $P$ is a random root-to-leaf path, sampled according to the distribution $\pi_{\mathcal{T}_{\cal H}}$.
This definition encapsulates the notion of computing the expected reward $\mathcal{R}(\mathcal{H})$ of $\cal H$ conditional on its possible outcomes, implying that $\mathcal{R}_{\mytree}(\mathcal{T}_{\mathcal{H}}) = \mathcal{R}(\mathcal{H})$. 

\paragraph{Expected reward of subtrees.}
For notational convenience, it is useful to extend representation~\eqref{eq:reward_def} from the entire tree ${\cal T}_{\cal H}$ to any of its subtrees. To this end, for every node $u$, we define $\mathcal{R}_{\mytree}(\mathcal{T}_{\mathcal{H}}(u))$ as the expected reward associated with the subtree $\mathcal{T}_{\mathcal{H}}(u)$, meaning that
\[ \mathcal{R}_{\mytree}(\mathcal{T}_{\mathcal{H}}(u)) ~~=~~ \exsub{P \sim \pi_{\mathcal{T}_{\mathcal{H}}(u)}} {\myvalue(P)} \ , \]
where $\pi_{\mathcal{T}_{\mathcal{H}}(u)}$ stands for the induced distribution over all root-to-leaf paths in $\mathcal{T}_{\mathcal{H}}(u)$.
Yet another way of writing the expected reward of $\mathcal{T}_{\mathcal{H}}(u)$ is by conditioning on the decision of applicant $\myapp(u)$, standing at the root of this subtree, similarly to equation~\eqref{eq:reward_2}. Specifically, recalling that $u_L$ and $u_R$ are the left and right children of $u$, respectively, we have 
\begin{equation} 
\mathcal{R}_{\mytree}(\mathcal{T}_{\mathcal{H}}(u)) ~~=~~ p_{\myapp(u)} \cdot \left( v_{\myapp(u)} + \mathcal{R}_{\mytree}(\mathcal{T}_{\mathcal{H}}(u_R)) \right) + (1 - p_{\myapp(u)}) \cdot \mathcal{R}_{\mytree}(\mathcal{T}_{\mathcal{H}}(u_L)) \ .  \label{eq:reward_tree_recursive}
\end{equation}

\subsection{Small-scale example} \label{subsec:dec_tree_example}

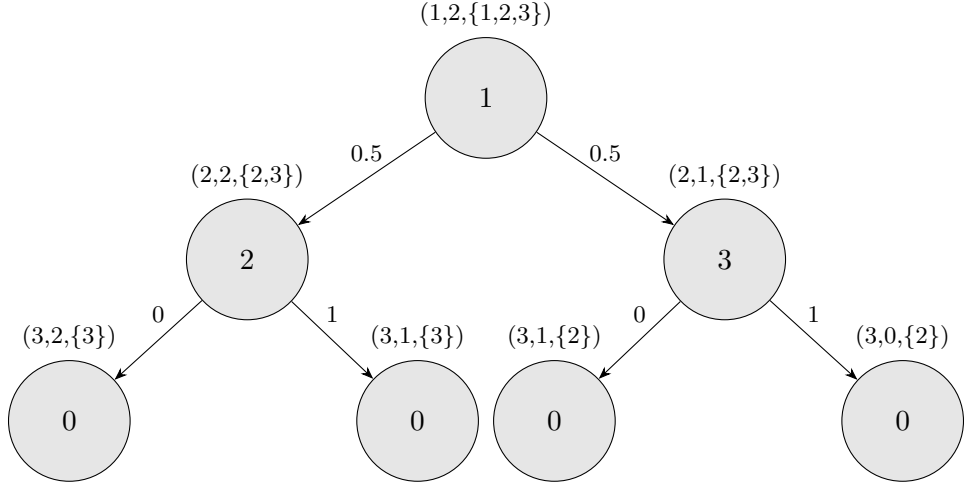
\begin{figure}[htbp!]
\centering
\begin{tikzpicture}[
  >=Stealth,                      
  every node/.style={
    circle, draw=black,           
    fill=gray!20,                 
    minimum size=16mm,            
    align=center
  },
  node distance=1cm and 2cm     
]

\node (root) {1}
    node(rootlabel) [above=0.1mm of root, font=\footnotesize, style={rectangle, draw=none, fill=none, minimum size=4mm}]
    {(1,2,\{1,2,3\})};

\node (n2) [below left=of root] {2}
    node(n2label) [above=0.1mm of n2, font=\footnotesize, style={rectangle, draw=none, fill=none, minimum size=4mm}]
    {(2,2,\{2,3\})};

\node (n3) [below right=of root] {3}
  node[above=0.1mm of n3, font=\footnotesize, style={rectangle, draw=none, fill=none, minimum size=4mm}]
    {(2,1,\{2,3\})};

\node (l1) [below left=1cm and 1.2cm of n2] {0}
  node[above=0.1mm of l1, font=\footnotesize, style={rectangle, draw=none, fill=none, minimum size=4mm}]
    {(3,2,\{3\})};

\node (l2) [below right=1cm and 1.2cm of n2, xshift=-1mm] {0}
  node[above=0.1mm of l2, font=\footnotesize, style={rectangle, draw=none, fill=none, minimum size=4mm}]
    {(3,1,\{3\})};

\node (l3) [below left=1cm and 1.2cm of n3, xshift=1mm] {0}
  node[above=0.1mm of l3, font=\footnotesize, style={rectangle, draw=none, fill=none, minimum size=4mm}]
    {(3,1,\{2\})};

\node (l4) [below right=1cm and 1.2cm of n3] {0}
  node[above=0.1mm of l4, font=\footnotesize, style={rectangle, draw=none, fill=none, minimum size=4mm}]
    {(3,0,\{2\})};

\draw[->] (root) -- node[midway, above=3pt, font=\footnotesize, style={rectangle, draw=none, fill=none, minimum size=4mm}]{0.5} (n2);
\draw[->] (root) -- node[midway, above=3pt, font=\footnotesize, style={rectangle, draw=none, fill=none, minimum size=4mm}]{0.5} (n3);

\draw[->] (n2) -- node[midway, above=3pt, font=\footnotesize, style={rectangle, draw=none, fill=none, minimum size=4mm}]{0} (l1);
\draw[->] (n2) -- node[midway, above=3pt, font=\footnotesize, style={rectangle, draw=none, fill=none, minimum size=4mm}]{1} (l2);

\draw[->] (n3) -- node[midway, above=3pt, font=\footnotesize, style={rectangle, draw=none, fill=none, minimum size=4mm}]{0} (l3);
\draw[->] (n3) -- node[midway, above=3pt, font=\footnotesize, style={rectangle, draw=none, fill=none, minimum size=4mm}]{1} (l4);

\end{tikzpicture}
\caption{The decision tree $\mathcal{T}_{\mathcal{H}}$.} \label{fig:small_scale_example}
\end{figure}

To provide an illustrative example of the core definitions introduced in Section~\ref{subsec:trees}, let us consider the following instance.
Suppose we have a pool of three applicants, numbered 1--3, aiming to fill up to two open positions across two stages.
The applicants' acceptance probabilities are $p_1 = 0.5$, $p_2 = 1$, and $p_3 = 1$, whereas their values are $v_1 = 1$, $v_2 = 2$, and $v_3 = 3$.
In addition, suppose that the hiring policy $\mathcal{H}$ operates as follows:
\begin{itemize}
    \item In stage 1, a position is offered to applicant 1.
    \item When applicant 1 rejects this offer, $\mathcal{H}$ proceeds by offering a position to applicant 2 in stage 2.
    \item When applicant 1 accepts this offer, $\mathcal{H}$ offers a position to applicant 3 in stage 2.
\end{itemize}
The decision tree $\mathcal{T}_{\mathcal{H}}$ of this policy is presented in  Figure~\ref{fig:small_scale_example}, where the number appearing within each node $u$ represents $\myapp(u)$.
Next to each arc, we write its probability, being $p_{\myapp(u)}$ for a right turn (acceptance) and $1 - p_{\myapp(u)}$ for a left turn (rejection).
Finally, the triple above each node represents its state. In this case, the expected reward of $\mathcal{T}_{\mathcal{H}}$ is given by
\[ \mathcal{R}_{\mytree}(\mathcal{T}_{\mathcal{H}}) ~=~ \underbrace{0.5 \cdot 0 \cdot (0+0)}_{\mathrm{LL}} + \underbrace{0.5 \cdot 1 \cdot (0+2)}_{\mathrm{LR}} + \underbrace{0.5 \cdot 0 \cdot (1+0)}_{\mathrm{RL}} + \underbrace{0.5 \cdot 1 \cdot (1+3)}_{\mathrm{RR}} ~=~ 3 \ .  \]
Here, $\mathrm{LL}$ corresponds to the leftmost path, $(1,2,\{1,2,3\}) \to (2,2,\{2,3\}) \to (3,2,\{3\})$, $\mathrm{LR}$ corresponds to $(1,2,\{1,2,3\}) \to (2,2,\{2,3\}) \to (3,1,\{3\})$, and so on.

\subsection{Important structural properties} \label{subsec:structural_properties}

In what follows, we introduce and study two structural properties that may or may not be exhibited by an arbitrary decision tree, the \lrproperty-property and the \vrlproperty-property, which will play a crucial role in upcoming sections.
Along the way, we will prove that any decision tree can be converted into one satisfying both properties without compromising its expected reward. Moving forward, decision trees that simultaneously satisfy both properties will be referred to as being ``canonical''. Similarly, hiring policies whose decision trees are canonical will be called ``canonical'' as well.

\paragraph{The \lrproperty-property.} 
Let $\mathcal{H}$ be an arbitrary hiring policy, with $\mathcal{T}_{\mathcal{H}}$ standing for its decision tree. We say that $\mathcal{T}_{\mathcal{H}}$  satisfies the \lrproperty-property when, for every internal node $u$, the expected reward of its left subtree $\mathcal{T}_{\mathcal{H}}(u_L)$ dominates the expected reward of its right subtree $\mathcal{T}_{\mathcal{H}}(u_R)$, i.e., 
\begin{equation} \label{eqn:def_LR}
\mathcal{R}_{\mytree}(\mathcal{T}_{\mathcal{H}}(u_L)) ~~\geq~~ \mathcal{R}_{\mytree}(\mathcal{T}_{\mathcal{H}}(u_R)) \ .   
\end{equation}
The \lrproperty-property essentially asserts that for any such node, the expected future reward collected over its left subtree (corresponding to a rejection by $\myapp(u)$) is at least as good as the expected future reward of its right subtree (corresponding to an acceptance by $\myapp(u)$).
Intuitively, this property should be satisfied by ``reasonable'' policies, since both subtrees have exactly the same number of stages remaining and the same set of available applicants. 
However, these subtrees differ in their number of open positions, with the left subtree having an extra vacant position.

\paragraph{The \vrlproperty-property.} Our second structural property of interest connects the expected rewards $\mathcal{R}_{\mytree}(\mathcal{T}_{\mathcal{H}}(u_L))$ and $\mathcal{R}_{\mytree}(\mathcal{T}_{\mathcal{H}}(u_R))$ in the opposite direction.
Specifically, we say that $\cal T_{\mathcal{H}}$ satisfies the \vrlproperty-property when, for every internal node $u$, the expected reward of its left subtree $\mathcal{T}_{\mathcal{H}}(u_L)$ is upper-bounded by the value of $\myapp(u)$ plus the expected reward of its right subtree $\mathcal{T}_{\mathcal{H}}(u_R)$. In other words,
\begin{equation} \label{eqn:def_VRL}
\mathcal{R}_{\mytree}(\mathcal{T}_{\mathcal{H}}(u_L)) ~~\leq~~ v_{\myapp(u)} + \mathcal{R}_{\mytree}(\mathcal{T}_{\mathcal{H}}(u_R)) \ . 
\end{equation}
At least intuitively, when $\mathcal{T}_{\mathcal{H}}$ does not adhere to this property, i.e., when $\mathcal{R}_{\mytree}(\mathcal{T}_{\mathcal{H}}(u_L)) > v_{\myapp(u)} + \mathcal{R}_{\mytree}(\mathcal{T}_{\mathcal{H}}(u_R))$, the policy $\mathcal{H}$ arrives at $\mystate(u)$ and extends an offer where rejection is strictly preferable to acceptance, which seems like a suboptimal decision.

\paragraph{Replacing subtrees.} To examine these properties in greater depth, we will consider various ``replacements'' of subtrees within a given decision tree. 
In particular, suppose that $u \neq v$ are two nodes of ${\cal T}_{\cal H}$, such that $v$ is not an ancestor of $u$.  Then, replacing the subtree $\mathcal{T}_{\mathcal{H}}(u)$ rooted at $u$ by the subtree $\mathcal{T}_{\mathcal{H}}(v)$ rooted at $v$ corresponds to the following operation: 
\begin{itemize}
    \item First, we disconnect $u$ from its parent node and delete the subtree $\mathcal{T}_{\mathcal{H}}(u)$.

    \item Second, we connect this parent to an identical copy of the subtree $\mathcal{T}_{\mathcal{H}}(v)$.
\end{itemize}
Clearly, for this operation to create a valid tree, letting  
$\mystate(u) = (t_u, k_u, \mathcal{A}_u)$ and $\mystate(v) = (t_v, k_v, \mathcal{A}_v)$, we should ensure that $t_v \geq t_u$, $k_v \leq k_u$, and $\mathcal{A}_v \subseteq \mathcal{A}_u$.

\paragraph{Imposing both properties.} 
With these definitions in place, we argue that any decision tree can be converted to a canonical one, i.e., simultaneously satisfying the \lrproperty-property and the \vrlproperty-property, without any loss in its expected reward. The proof of this result is provided in Appendix~\ref{proof:lr_and_vrl_modification}. 

\begin{lemma} \label{lemma:lr_and_vrl_modification}
For any valid decision tree $\cal T$, there exists a canonical tree $\tilde{\cal T}$ with an expected reward of $\mathcal{R}_{\mytree}(\tilde{\cal T}) \geq \mathcal{R}_{\mytree}(\cal T)$. 
\end{lemma}
As an immediate corollary, we infer that any optimal decision tree can be altered into a canonical form while maintaining its expected reward.
\begin{corollary} \label{cor:rl_vrl_optimal}
There exists a canonical optimal decision tree.    
\end{corollary}

\section{Approximation Scheme in Quasi-Polynomial Time} \label{sec:quasi_ptas}

In this section, we propose a quasi-polynomial time approximation scheme for the sequential hiring problem. Further down the road, we will revamp the main ideas developed in this context to derive an improved approximation scheme, admitting a true polynomial-time implementation. The specific guarantees we are about to attain are formally stated in the next theorem, where to suppress cumbersome $\eps$-related dependencies, we make use of $O_{\eps}( f(n, T) ) = O(  (\frac{ 1 }{ \eps } )^{O(1)} \cdot f(n, T) )$.

\begin{theorem} \label{thm:quasi_ptas}
For any $\epsilon > 0$, the sequential hiring problem can be approximated within factor $1 - \eps$ of optimal. The running time of our algorithm is $O ( n^{O(1)} \cdot T^{O_{\eps}(\log T)} )$.
\end{theorem}

\paragraph{Outline.} In Section~\ref{subsec:partitioning_applicants}, we begin by partitioning applicants into disjoint classes according to their acceptance probabilities. Based on this classification, Section~\ref{subsec:structural_modifications} will be dedicated to introducing several structural alterations, enabling us to reduce the number of distinct acceptance probabilities to only $O_{\eps}(\log T)$ while losing an $O(\epsilon)$-fraction of the optimal reward along the way. In Section~\ref{subsec:dynamic_programming}, we devise a dynamic programming approach for computing an optimal policy in terms of this modified instance, crucially relying on having logarithmically-many acceptance probabilities. For our original instance, this policy will be shown to provide an approximation scheme.

\subsection{Partitioning applicants} \label{subsec:partitioning_applicants}

In what follows, we explain how to partition the overall set of applicants into $O(\frac{1}{\eps}\log\frac{T}{\eps})$ classes based on their acceptance probabilities. Later on, we will utilize this partition to restrict the number of distinct acceptance probabilities, a structural feature that will be important in developing a quasi-polynomial dynamic programming approach.

\paragraph{Defining applicant classes.} For simplicity of notation, let us introduce the threshold parameter $\gamma = \frac{\epsilon}{T}$, assuming without loss of generality that $\gamma < 1$. Given this parameter, we partition the collection of applicants into the acceptance-probability classes ${\cal C}_0, \ldots, {\cal C}_M$ as follows:
\begin{itemize}
    \item First, the class ${\cal C}_0$ consists of applicants whose acceptance probability is at most $\gamma$.
    \item Then, $\mathcal{C}_1, \dots, \mathcal{C}_M$ are created via a geometric partition of the remaining interval $(\gamma,1]$, by powers of $1+\eps$. In other words, the class ${\cal C}_1$ consists of applicants with $p_i \in (\gamma, (1 + \epsilon)\cdot \gamma]$. Then, ${\cal C}_2$ consists of those with $p_i \in ((1 + \epsilon)\cdot \gamma, (1 + \epsilon)^2\cdot \gamma]$, so on and so forth, up until reaching the final class ${\cal C}_M$. One can easily verify that $M = O(\frac{1}{\epsilon}\log\frac{1}{\gamma}) = O(\frac{1}{\eps}\log\frac{T}{\eps})$.
\end{itemize}

\subsection{Structural modifications} \label{subsec:structural_modifications}

\paragraph{Handling small acceptance probabilities.}
To limit the number of distinct acceptance probabilities, we begin by altering some of the features describing applicants who belong to class ${\cal C}_0$. 
For convenience of notation, any instance $I$ of the sequential hiring problem will be compactly characterized by the $5$-tuple $I = (n, V, P, k, T)$, where $V = (v_1, \ldots, v_n)$ and $P = (p_1, \ldots, p_n)$ respectively stand for the vectors of values and acceptance probabilities across all applicants. Given such an instance, we define its ``rounded-up'' counterpart $I^{\uparrow} = (n, V^{\uparrow}, P^{\uparrow}, k, T)$, where the latter two vectors will be modified to  $V^{\uparrow} = (v^{\uparrow}_1, \ldots, v^{\uparrow}_n)$ and $P^{\uparrow} = (p^{\uparrow}_1, \ldots, p^{\uparrow}_n)$.
Specifically, these alterations involve rounding up the acceptance probabilities of all $\mathcal{C}_0$-applicants, followed by rounding their values in the opposite direction, to keep the product of these two parameters unchanged:
\begin{itemize}
    \item  {\em Probabilities}: For every applicant $i \in {\cal C}_0$, we set her modified acceptance probability as $p^{\uparrow}_i = \gamma$. For any other applicant $i \in [n] \setminus {\cal C}_0$, we maintain $p^{\uparrow}_i = p_i$.
    
    \item  {\em Values}: For every applicant $i \in [n]$, letting $r_i = v_i p_i$, we set her modified value as $v^{\uparrow}_i = \frac{r_i}{p^{\uparrow}_i}$. \label{def:ri}
\end{itemize}
It is worth noting that since we have not made any changes to the set of applicants itself, to the number of stages $T$, and to the number of available positions $k$, every hiring policy for the original instance $I$ is valid for the rounded-up instance $I^{\uparrow}$ and vice versa. 
That said, due to altering the underlying acceptance probabilities and values, the expected reward of any given policy generally differs between these two instances.

\paragraph{Handling large acceptance probabilities.}

Next, we proceed by modifying the rounded-up instance $I^{\uparrow}$, ending up with the ``mixed-rounded'' instance $I^{\updownarrow}$.
The current alteration will round down the acceptance probabilities of the applicants within ${\cal C}_1, \ldots, {\cal C}_M$, ensuring that any given class has a uniform acceptance probability. Specifically, given the rounded-up instance $I^{\uparrow} = (n, V^{\uparrow}, P^{\uparrow}, k, T)$, we create $I^{\updownarrow} = (n, V^{\updownarrow}, P^{\updownarrow}, k, T)$ as follows.
For every $m \in [M]$, recalling that $\mathcal{C}_m = \{ i \in [n] \colon p_i \in ((1 + \epsilon)^{m-1} \cdot \gamma, (1 + \epsilon)^m \cdot \gamma] \}$ and that $p^{\uparrow}_i = p_i$ for every applicant $i \in \mathcal{C}_m$, we set her acceptance probability as $p^{\updownarrow}_i = (1+\eps)^{m-1} \cdot \gamma$.
For every applicant $i \in \mathcal{C}_0$, we keep $p^{\updownarrow}_i = p^{\uparrow}_i$.
That said, the values vector remains unchanged, i.e., $V^{\updownarrow} = V^{\uparrow}$.

\paragraph{Expected reward effects.}
As previously mentioned, any given policy $\mathcal{H}$ may have different expected rewards with respect to our original instance $I$ and its modified counterpart $I^{\updownarrow}$; we denote these expected rewards by $\mathcal{R}^{I}(\mathcal{H})$ and $\mathcal{R}^{I^{\updownarrow}}(\mathcal{H})$.
Moving forward, our objective is to prove that, under the right circumstances, these measures are nearly identical.
For this purpose, Lemma~\ref{lemma:rounding_up_down} considers the first direction, showing that the expected reward of any canonical policy for $I$ is nearly preserved with respect to $I^{\updownarrow}$.
In the opposite direction, Lemma~\ref{lemma:idu_smaller_than_i} argues that the expected reward of any canonical policy for $I^{\updownarrow}$ is always dominated by its expected reward in terms of $I$.
We provide the proofs of these claims in Appendices~\ref{app:proof_lemma_rounding_up_down} and \ref{proof:idu_smaller_than_i}.

\begin{lemma} \label{lemma:rounding_up_down}
Let $\mathcal{H}$ be a canonical hiring policy for $I$. Then, $\mathcal{R}^{I^{\updownarrow}}(\mathcal{H}) \geq (1 - 2\epsilon) \cdot \mathcal{R}^{I}(\mathcal{H})$.
\end{lemma}

\begin{lemma} \label{lemma:idu_smaller_than_i}
    Let $\mathcal{H}$ be a canonical hiring policy for $I^{\updownarrow}$.
    Then, $\mathcal{R}^{I}(\mathcal{H}) \geq \mathcal{R}^{I^{\updownarrow}}(\mathcal{H})$.
\end{lemma}

\subsection{Computing an optimal hiring policy for $\bs{I^{\updownarrow}}$} \label{subsec:dynamic_programming}

We conclude this section by explaining how to compute an optimal hiring policy for the mixed-rounded instance $I^{\updownarrow}$ via a dynamic programming approach.
To this end, we first argue about the existence of an optimal hiring policy that follows a very specific offering order within each class of applicants.

\paragraph{The order-by-value rule.}
We say that a hiring policy $\mathcal{H}$ follows the order-by-value rule when at any state $(t, k_t, \mathcal{A}_t)$, the chosen applicant $\mathcal{H}(t, k_t, \mathcal{A}_t)$ has the highest value out of the currently available applicants in her class, with ties consistently broken, say by lexicographical order. 
Namely, if $\mathcal{H}(t, k_t, \mathcal{A}_t) \in {\cal C}_m$, then $v_{\mathcal{H}(t, k_t, \mathcal{A}_t)} = \max \{v_i \colon i \in {\cal A}_t \cap {\cal C}_m \}$.
The next claim, whose proof is presented in Appendix~\ref{proof:optimal_order_by_value}, shows that there is an optimal hiring policy that adheres to this rule.

\begin{lemma} \label{lemma:optimal_Order_by_value}
For the mixed-rounded instance $I^{\updownarrow}$, there exists an order-by-value optimal hiring policy.
\end{lemma}

\paragraph{State space.}
Based on this characterization, we proceed by devising a rather straightforward dynamic program for computing an optimal hiring policy in $O( n^{O(1)} \cdot T^{O_{\eps}(\log T)} )$ time.
For this purpose, each state $(t, k_t, L_t)$ of our dynamic program consists of three parameters, with the first two being identical to our system states, $(t, k_t, \mathcal{A}_t)$.
In other words, $t$ stands for the current stage, whereas $k_t$ designates the number of open positions at the beginning of this stage.
That said, rather than making use of the subset $\mathcal{A}_t$, potentially taking $2^n$ values, we will be considering an $(M+1)$-dimensional vector $L_t = (\ell_{t, 0}, \dots, \ell_{t, M})$, representing the number of applicants selected from each class in earlier stages.
Formally, each $\ell_{t, m}$ captures the number of $\mathcal{C}_m$-applicants which were selected by our policy across stages $1, \dots, t-1$.

It is important to point out that in an order-by-value hiring policy, each state can be fully described via $(t, k_t, L_t)$, since the current subset of available applicants can be uniquely inferred by the number of applicants selected from each class in previous stages.
By Lemma~\ref{lemma:optimal_Order_by_value}, we know that there exists an order-by-value optimal hiring policy for $I^{\updownarrow}$, implying that the optimal policy determined by our dynamic program is indeed an optimal policy for $I^{\updownarrow}$.
With this representation, it is easy to verify that the underlying number of states is $O(k  T^{O(M)}) = O(k  T^{O(\frac{1}{\epsilon}\log \frac{T}{\epsilon})})$.

\paragraph{Value function and recursive equations.}
The value function of our dynamic program will be denoted by $F(t, k_t, L_t)$, standing for the maximum possible expected reward of an order-by-value hiring policy starting at state $(t, k_t, L_t)$.
In this scenario, our process begins at stage $t$, with $k_t$ open positions left, and with $\ell_{t, m}$ applicants already picked from each class $\mathcal{C}_m$ in earlier stages.
To express this function in recursive form, for any class $\mathcal{C}_m$, let $\mathcal{C}_m[1], \mathcal{C}_m[2], \dots$ be a permutation of its applicants, ordered by weakly-decreasing values, i.e., $v_{\mathcal{C}_m[1]} \geq v_{\mathcal{C}_m[2]} \geq \cdots$, with ties broken by lexicographical order.
Due to focusing on order-by-value policies, the only question we are facing is: 
Which class should be selected in the current stage to yield the best-possible expected reward? As a result,
\begin{align*}
    F(t, k_t, L_t) ~~=~~ \max_{0 \leq m \leq M} \left\{ p^{\updownarrow}_{\mathcal{C}_{m}[{\ell}_{t, m}+1]} \cdot (v^{\updownarrow}_{\mathcal{C}_{m}[{\ell}_{t, m}+1]} + F(t+1, k_t - 1, L_t +\mathrm{e}_m)) \right.\\
    + \left. (1 - p^{\updownarrow}_{\mathcal{C}_{m}[{\ell}_{t, m}+1]}) \cdot F(t+1, k_t, L_t +\mathrm{e}_m)  \right\}  \text{ }.
\end{align*}
Terminal states of this recursion correspond to $t = T+1$ or $k_t = 0$ or $\|L_t\|_1 = n$, in which case $F(t, k_t, L_t) = 0$.

\paragraph{Performance guarantee for $\bs{I}$.}
To complete the proof of Theorem~\ref{thm:quasi_ptas}, we show that the hiring policy $\mathcal{H}^{\updownarrow}$, computed by our dynamic program with respect to $I^{\updownarrow}$, is near-optimal in terms of the original instance $I$.
Our first claim, whose proof appears in Appendix~\ref{proof:dynamic_canonical}, establishes that $\mathcal{H}^{\updownarrow}$ is a canonical policy for $I^{\updownarrow}$. 

\begin{lemma} \label{lemma:dynamic_canonical}
    $\mathcal{H}^{\updownarrow}$ is a canonical hiring policy for $I^{\updownarrow}$.
\end{lemma}
Now, let $\mathcal{H}^*$ be a canonical optimal hiring policy for the original instance $I$, which is known to exist by Corollary~\ref{cor:rl_vrl_optimal}. Then, the expected reward of $\mathcal{H}^{\updownarrow}$ can be related to that of $\mathcal{H}^*$ by observing that
\begin{eqnarray}
    \mathcal{R}^{I}(\mathcal{H^{\updownarrow}}) ~~&\geq&~~ \mathcal{R}^{I^{\updownarrow}}(\mathcal{H^{\updownarrow}}) \label{eq:quasi_perf_1}\\
    &\geq&~~ \mathcal{R}^{I^{\updownarrow}}(\mathcal{H}^*) \label{eq:quasi_perf_2}\\
    &\geq&~~ (1-2\epsilon) \cdot \mathcal{R}^{I}(\mathcal{H}^*) \ .  \label{eq:quasi_perf_3}
\end{eqnarray}
Here, inequality~\eqref{eq:quasi_perf_1} follows from Lemma~\ref{lemma:idu_smaller_than_i}, since $\mathcal{H}^{\updownarrow}$ is a canonical policy for $I^{\updownarrow}$.
Inequality~\eqref{eq:quasi_perf_2} holds since $\mathcal{H}^{\updownarrow}$ is an optimal policy for $I^{\updownarrow}$.
Finally, inequality~\eqref{eq:quasi_perf_3} is a direct consequence of Lemma~\ref{lemma:rounding_up_down}.

\section{Block-Responsive  Policies} \label{sec:block_responsive}

As previously mentioned, a fundamental obstacle that stands between us and a polynomial-time approximation scheme for the sequential hiring problem is that optimal decision trees could be exponentially large.
To circumvent this issue, we introduce a novel class of ``block-responsive'' policies,  choosing a subset (or block) of applicants at each state rather than a single applicant. In turn, offers are sequentially extended to the applicants within each block, based on a predefined priority rule. 
Ultimately, this approach will allow us to compete against the best-possible expected reward achievable by general hiring policies via considerably smaller decision trees.

\paragraph{Outline.}
In what follows, we present the concept of block-responsive policies, proving the existence of a policy within this class that closely approximates the optimal unrestricted policy while significantly reducing its space complexity.
To this end, in Section~\ref{subsec:block_responsive}, we formally define block-responsive policies and introduce their decision tree representation. 
In Section~\ref{subsec:block_responsive_structural_properties}, we establish important structural properties of such policies.
Section~\ref{subsec:inter_summary_block} provides an intermediate summary, outlining an overall framework for designing approximation schemes via block‐responsive policies.
In Section~\ref{subsec:thm_block_policy}, we present our main structural result in this context, stating that there exists a near-optimal block-responsive policy with a small-sized decision tree. Due to its technical complexity, the proof of this result is provided in Sections~\ref{subsec:policy_transformation} and \ref{subsec:block_responsive_remaining_proof}.
Moving forward, Section~\ref{section:PTAS} will utilize these findings to develop a polynomial-time approximation scheme.

\subsection{Definitions and notation} \label{subsec:block_responsive}

In this section, we introduce the inner workings of block-responsive policies, allowing us to select an ordered set of applicants in each state, rather than a single applicant. 
These applicants are extended offers one after the other, stopping as soon as an offer is accepted. 
Formally, a block-responsive policy can be viewed as a function $\mathcal{H}^B$ that, given any state $(t, k_t, \mathcal{A}_t)$ with $k_t \geq 1$ and $\mathcal{A}_t \neq \emptyset$, selects a set of available applicants $\myblock_{\mathcal{H}^B}(t, k_t, \mathcal{A}_t) \subseteq \mathcal{A}_t$ who will be extended offers next, along with a complete order $\pi_{\mathcal{H}^B}(t, k_t, \mathcal{A}_t)$ on this set.
For ease of presentation, we proceed by elaborating on the dynamics of such policies in terms of our original instance $I$;
Section~\ref{subsec:inter_summary_block} will explain how any performance guarantee in relation to $I$ carries over to the mixed-rounded instance $I^{\updownarrow}$ and vice versa.

\paragraph{System dynamics.} 
At the beginning of stage $t=1$, all $k$ positions are still open, and all applicants are available to choose from, meaning that our starting state is $(1, k, [n])$. Then, at each stage $t \in [T]$, assuming that $k_t \geq 1$ and $\mathcal{A}_t \neq \emptyset$, the following sequence of steps occurs:
\begin{enumerate}
\item {\em Set selection}: First, the hiring policy $\mathcal{H}^B$ selects a subset of applicants $\myblock_{\mathcal{H}^B}(t, k_t, \mathcal{A}_t)$ from the pool of available applicants $\mathcal{A}_t$ as well as an order $\pi_{\mathcal{H}^B}(t, k_t, \mathcal{A}_t)$ on this subset.
For convenience, we refer to this ordered subset as the ``block'' $\mathcal{H}^B(t, k_t, \mathcal{A}_t)$, which includes both the selected applicants and their internal order.

\item {\em Intra-block offers}: 
Then, our hiring policy sequentially extends offers to applicants within its chosen block, according to the order $\pi_{\mathcal{H}^B}(t, k_t, \mathcal{A}_t)$.
That said, once an applicant accepts an offer, we will not be making further offers to the remaining applicants within this block. 
Here, the binary variable $X_{\mathcal{H}^B(t, k_t, \mathcal{A}_t)}$ is utilized to indicate whether one of the applicants in $\mathcal{H}^B(t, k_t, \mathcal{A}_t)$ accepts an offer, corresponding to $X_{\mathcal{H}^B(t, k_t, \mathcal{A}_t)} = 1$; in the opposite scenario, all applicants reject their offers, in which case $X_{\mathcal{H}^B(t, k_t, \mathcal{A}_t)} = 0$.

\item {\em State update}: Finally, regardless of whether all applicants were extended offers or not, we proceed to stage $t_\mynew = t + |\mathcal{H}^B(t, k_t, \mathcal{A}_t)|$, representing the point in time we land at due to processing the current block.
Our set of available applicants is updated by removing the current block, i.e., $\mathcal{A}_{t_\mynew} = \mathcal{A}_{t} \setminus \mathcal{H}^B(t, k_t, \mathcal{A}_t)$, whereas the remaining number of open positions becomes $k_{t_\mynew} = k_t - X_{\mathcal{H}^B(t, k_t, \mathcal{A}_t)}$.
\end{enumerate}
This process terminates as soon as we arrive at a state $(t, k_t, \mathcal{A}_t)$ that corresponds to completing our planning horizon ($t = T+1$), or filling all positions ($k_t = 0$), or exhausting the pool of available applicants ($\mathcal{A}_t = \emptyset$).

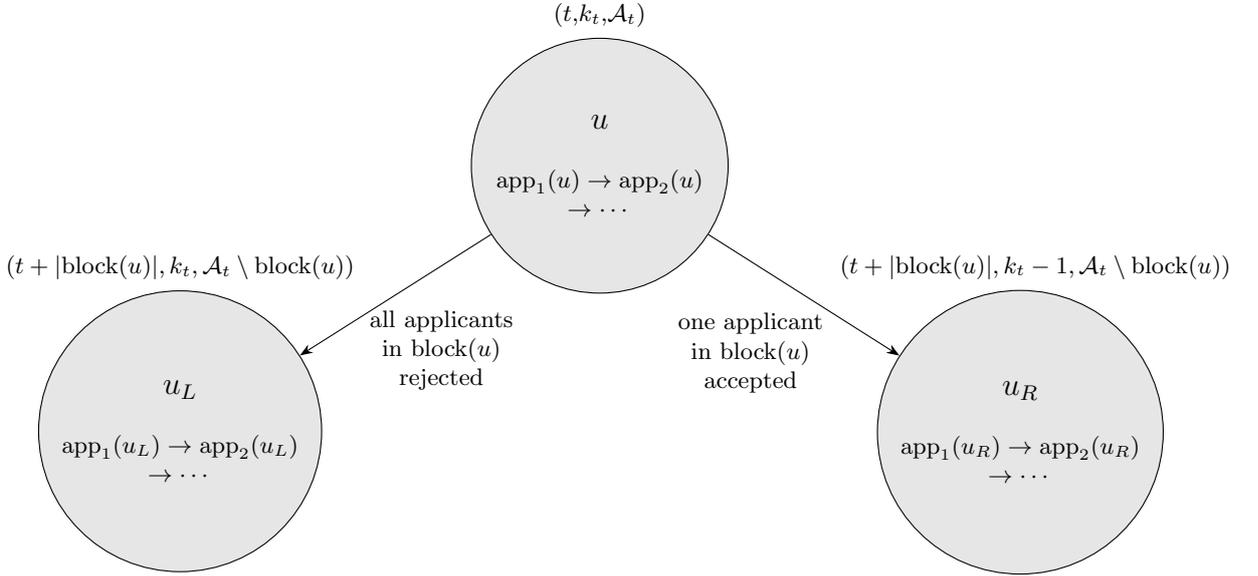
\begin{figure}[htbp!]
    \centering
    \begin{tikzpicture}[
  >=Stealth,                      
  every node/.style={
    circle, draw=black,           
    fill=gray!20,                 
    minimum size=16mm,            
    font=\bfseries,               
    align=center,
    font=\footnotesize
  },
  node distance=1cm and 3cm     
]

\node (root) {\large $u$\\ \\ \footnotesize$\myapp_1(u)\rightarrow \myapp_2(u)$ \\ $\rightarrow \cdots $}
    node(rootlabel) [above=0.1mm of root, font=\footnotesize, style={rectangle, draw=none, fill=none, minimum size=4mm}]
    {($t$,$k_t$,$\mathcal{A}_t$)};

\node (n2) [below left=of root] {\large $u_L$\\ \\ \footnotesize $\myapp_1(u_L)\rightarrow \myapp_2(u_L)$ \\ $\rightarrow \cdots $}
    node(n2label) [above=0.1mm of n2, font=\footnotesize, style={rectangle, draw=none, fill=none, minimum size=4mm}]
    {$(t + |\myblock(u)|, k_t, \mathcal{A}_t \setminus  \myblock(u) )$};

\node (n3) [below right=of root] {\large $u_R$\\ \\ \footnotesize $\myapp_1(u_R)\rightarrow \myapp_2(u_R)$ \\ $\rightarrow \cdots $}
  node[above=0.1mm of n3, xshift=2mm, font=\footnotesize, style={rectangle, draw=none, fill=none, minimum size=4mm}]
    {$(t + |\myblock(u)|, k_t - 1, \mathcal{A}_t \setminus \myblock(u) )$};

\draw[->] (root) -- node[midway, below right=0.1cm of root, xshift=-7mm, font=\footnotesize, style={rectangle, draw=none, fill=none, minimum size=4mm, text width=2.2cm}]{all applicants in $\myblock(u)$ rejected} (n2);
\draw[->] (root) -- node[midway, below=0.1cm of root, xshift=-7mm, font=\footnotesize, style={rectangle, draw=none, fill=none, minimum size=4mm, text width=2.2cm}]{one applicant in $\myblock(u)$ accepted} (n3);

\end{tikzpicture}
  
    \caption{Internal nodes of the block-responsive decision tree $\mathcal{T}_{\mathcal{H}^B}$.}
    \label{fig:block_internal_nodes}
\end{figure}

\paragraph{Decision tree representation.}
Similarly to standard single-selection hiring policies, one way of representing block-responsive policies is through a recursive relation that can be expressed as a binary decision tree. 
To better understand the upcoming discussion, we advise the reader to consult Figure~\ref{fig:block_internal_nodes}. In our setting, given a block-responsive policy $\mathcal{H}^B$, its decision tree representation $\mathcal{T}_{\mathcal{H}^B}$ is recursively defined as follows:
\begin{itemize}
    \item {\em Functions}: As explained below, each node of ${\cal T}_{\mathcal{H}^B}$ will be associated with three functions, $\mystate(\cdot)$, $\myblock(\cdot)$, and $\myapp_{\cdot}(\cdot)$, whose intended roles are to encode its corresponding system state, chosen block, and applicants' ranks within this block, respectively.
    In particular, the order by which applicants are processed is $\myapp_1(\cdot), \myapp_2(\cdot), \dots$.
    We mention in passing that, due to employing ${\mathcal{H}^B}$ as our block-responsive policy, one always has $\myblock(\cdot) = \mathcal{H}^B( \mystate(\cdot))$, meaning that the former notation is mainly for convenience. 
   
    \item {\em Internal nodes}: The root of the tree ${\cal T}_{\mathcal{H}^B}$, denoted as $\myroot(\mathcal{T}_{\mathcal{H}^B})$, corresponds to our starting state, meaning that $\mystate(\myroot(\mathcal{T}_{\mathcal{H}^B})) = (1, k_1, \mathcal{A}_1)$. Consequently, the first block $\mathcal{H}^B$ selects is given by $\myblock(\myroot(\mathcal{T}_{\mathcal{H}^B})) = \mathcal{H}^B(1, k_1, \mathcal{A}_1)$, with its applicants ordered as $\myapp_1(\myroot(\mathcal{T}_{\mathcal{H}^B})), \myapp_2(\myroot(\mathcal{T}_{\mathcal{H}^B})), \dots$.
    Beyond this root, each internal node $u$ has two downward arcs connecting it to its left and right children, denoted as $u_L$ and $u_R$, respectively.
    The left downward arc captures the event where all offers made to the applicants in $\myblock(u)$ are rejected, whereas the right downward arc captures the complementary event, where one of these offers is accepted. 
    As such, when $\mystate(u) = (t, k_t, \mathcal{A}_t)$, we have 
    \begin{align*}
        &\mystate(u_L) ~~=~~ (t + |\myblock(u)|, k_t, \mathcal{A}_t \setminus  \myblock(u) )\\ 
        &\qquad \qquad \text{and} \qquad  \nonumber \mystate(u_R) ~~=~~ (t + |\myblock(u)|, k_t - 1, \mathcal{A}_t \setminus \myblock(u) ) \ . 
    \end{align*}
    This construction recursively proceeds until reaching a leaf node, which is separately discussed below. For ease of notation, we make use of $\mathcal{T}_{\mathcal{H}^B}(u)$ to designate the subtree rooted at node $u$, noting that it can be interpreted as the decision tree of the hiring policy $\mathcal{H}^B$, starting at $\mystate(u)$.

    \item {\em Leaves}: Each leaf $u$ of the decision tree $\mathcal{T}_{\mathcal{H}^B}$ serves as an indication that, upon reaching $\mystate(u)$, the hiring policy ${\mathcal{H}^B}$ cannot extend further offers. In other words, $\mystate(u) = (t, k_t, \mathcal{A}_t)$ happens to be a terminal system state when  $t = T+1$ or $k_t = 0$ or $\mathcal{A}_t = \emptyset$.
    For uniform treatment in subsequent analysis, every leaf will be associated with a virtual block, $B_0=\emptyset$.
\end{itemize}

\paragraph{Expected reward of block-responsive decision trees.}
We conclude this exposition by proposing a convenient way to express the expected reward of a block-responsive decision tree $\mathcal{T}_{\mathcal{H}^B}$.
To this end, for any node $u$ of $\mathcal{T}_{\mathcal{H}^B}$, recall that the order of applicants in $\myblock(u)$ is given by $\myapp_1(u), \dots, \myapp_{|\myblock(u)|}(u)$.
With this notation, it is easy to verify that the expected reward $\mathcal{R}_{\mytree}(\mathcal{T}_{\mathcal{H}^B}(u))$ of the subtree $\mathcal{T}_{\mathcal{H}^B}(u)$ can be written as
\begin{align}
    &\mathcal{R}_{\mytree}(\mathcal{T}_{\mathcal{H}^B}(u)) ~~=~~  \underbrace{\left(\prod_{q=1}^{|\myblock(u)|}(1-p_{\myapp_q(u)})\right)}_{\text{all applicants reject}}  \cdot \mathcal{R}_{\mytree}(\mathcal{T}_{\mathcal{H}^B}(u_L)) \nonumber \\
    & \qquad \mbox{}+\sum_{q=1}^{|\myblock(u)|} \underbrace{ \left(\prod_{\hat{q}=1}^{q-1}(1-p_{\myapp_{\hat{q}}(u)})\right) \cdot  p_{\myapp_q(u)}}_{\text{$q$-th applicant accepts}} \cdot (v_{\myapp_q(u)}+\mathcal{R}_{\mytree}(\mathcal{T}_{\mathcal{H}^B}(u_R))) \ .\label{eq:block_reward_def}
\end{align}
For purposes of analysis, we extend this notation to $\mathcal{R}_{\mytree}(\mathcal{T}_{\mathcal{H}^B}(u), r)$, capturing the expected cumulative reward starting from the $r$-th applicant of $\myblock(u)$.
Similarly to representation~\eqref{eq:block_reward_def}, this measure can be written as
\begin{align}
    &\mathcal{R}_{\mytree}(\mathcal{T}_{\mathcal{H}^B}(u), r) ~~=~~  \underbrace{\left(\prod_{q=r}^{|\myblock(u)|}(1-p_{\myapp_q(u)})\right)}_{\text{all applicants reject}}  \cdot \mathcal{R}_{\mytree}(\mathcal{T}_{\mathcal{H}^B}(u_L)) \nonumber \\
    &\qquad \mbox{} +\sum_{q=r}^{|\myblock(u)|} \underbrace{ \left(\prod_{\hat{q}=r}^{q-1}(1-p_{\myapp_{\hat{q}}(u)})\right) \cdot  p_{\myapp_q(u)}}_{\text{$q$-th applicant accepts}} \cdot (v_{\myapp_q(u)}+\mathcal{R}_{\mytree}(\mathcal{T}_{\mathcal{H}^B}(u_R))) \ .\label{eq:block_reward_def_r}
\end{align}

\subsection{Structural properties of block-responsive decision trees} \label{subsec:block_responsive_structural_properties}
In what follows, we introduce two  structural properties that may or may not be exhibited by an arbitrary block-responsive decision tree, the \lrproperty-block-property and the \vrlproperty-block-property, which will play a crucial role within our analysis.
These properties are analogous to those considered in the context of standard decision trees (see Section~\ref{subsec:structural_properties}).
Along the way, we will describe an efficient procedure for converting any block-responsive decision tree into one satisfying both properties while preserving its expected reward. 
Moving forward, block-responsive decision trees that simultaneously satisfy both properties will be referred to as being ``canonical''. Similarly, block-responsive hiring policies whose decision trees are canonical will be called ``canonical'' as well.

\paragraph{The \lrproperty-block-property.} 
Let $\mathcal{H}^B$ be an arbitrary block-responsive hiring policy, with $\mathcal{T}_{\mathcal{H}^B}$ standing for its decision tree.
We say that $\mathcal{T}_{\mathcal{H}^B}$ satisfies the \lrproperty-block-property when, for every internal node $u$ and every rank $1 \leq r \leq |\myblock(u)|$, the expected reward of $u$'s right subtree $\mathcal{T}_{\mathcal{H}^B}(u_R)$ is dominated by the future expected reward upon rejection by applicant $\myapp_r(u)$. Namely,
\begin{equation} \label{eqn:def_LR_block}
 \mathcal{R}_{\mytree}(\mathcal{T}_{\mathcal{H}^B}(u), r+1) ~~\geq~~ \mathcal{R}_{\mytree}(\mathcal{T}_{\mathcal{H}^B}(u_R)) \text{ ,} \nonumber
\end{equation}
with the convention that $\mathcal{R}_{\mytree}(\mathcal{T}_{\mathcal{H}^B}(u), |\myblock(u)|+1) = \mathcal{R}_{\mytree}(\mathcal{T}_{\mathcal{H}^B}(u_L))$.
In essence, the \lrproperty-block-property asserts that for any internal node, the expected future reward collected upon rejecting an applicant is at least as good as the expected reward of its right subtree, corresponding to an acceptance by some applicant.
Intuitively, this property should be satisfied by ``reasonable'' policies, since upon rejection, open positions are not being filled and we move one step forward.
By contrast, in the right subtree, an open position has already been filled, with the next stage updated by the entire block size.

\paragraph{The \vrlproperty-block-property.} 
The second structural property we wish to highlight states that each offer extended by our hiring policy positively contributes to its overall expected reward. 
Specifically, we say that the decision tree $\mathcal{T}_{\mathcal{H}^B}$ satisfies the \vrlproperty-block-property when, for every internal node $u$ and every rank $1 \leq r \leq |\myblock(u)|$, the expected reward upon rejection by applicant $\myapp_r(u)$ is dominated by the value of this applicant plus the expected reward of its right subtree.
In other words,
\begin{equation} \label{eqn:def_VRL_block}
 \mathcal{R}_{\mytree}(\mathcal{T}_{\mathcal{H}^B}(u), r+1) ~~\leq~~ v_{\myapp_r(u)}+\mathcal{R}_{\mytree}(\mathcal{T}_{\mathcal{H}^B}(u_R)) \ .  \nonumber
\end{equation}
At least intuitively, when $\mathcal{T}_{\mathcal{H}}$ does not exhibit this property, the policy $\mathcal{H}^B$ extends an offer for which rejection is strictly preferable to acceptance, which seems like a suboptimal decision.

\paragraph{Imposing both properties.} 
With these definitions in place, we argue that any block-responsive decision tree can be converted to a canonical one, i.e., simultaneously satisfying the \lrproperty-block-property and the \vrlproperty-block-property, without any loss in its expected reward. The proof of this result is provided in Appendix~\ref{proof:block_lr_and_vrl_modification}. 

\begin{lemma} \label{lemma:block_lr_and_vrl_modification}
Let $\mathcal{T}_{\mathcal{H}^B}$ be a block-responsive decision tree consisting of $N$ nodes. Given $\mathcal{T}_{\mathcal{H}^B}$, we can compute in $O(N T)$ time a canonical block-responsive decision tree $\mathcal{T}_{\tilde{\mathcal{H}}^B}$, with an expected reward of $\mathcal{R}_{\mytree}(\mathcal{T}_{\tilde{\mathcal{H}}^B}) \geq \mathcal{R}_{\mytree}(\mathcal{T}_{\mathcal{H}^B})$.
\end{lemma}

\subsection{The relationship between $\boldsymbol{I}$ and $\boldsymbol{I^{\updownarrow}}$} \label{subsec:inter_summary_block}

Up until now, the preceding discussion considered block-responsive policies that operate on our original instance $I$.
However, as explained in Section~\ref{section:PTAS}, it is substantially more convenient to develop efficient block-responsive policies with respect to the mixed-rounded instance $I^{\updownarrow}$, whose underlying collection of applicants was partitioned into the classes $\mathcal{C}_0, \dots, \mathcal{C}_M$, with their acceptance probabilities and values rounded as described in Section~\ref{subsec:structural_modifications}.
Therefore, the basic question we should address is: Why would an approximation scheme for $I^{\updownarrow}$ lead to an analogous result for $I$?

To create a bridge between these two instances, we dedicate Appendix~\ref{proof:block_canonical_modifying_effects} to establishing the next claim, arguing that the expected reward of any canonical block-responsive policy with respect to $I$ dominates its expected reward with respect to $I^{\updownarrow}$.

\begin{lemma} \label{lemma:block_canonical_modifying_effects}
    Let $\mathcal{H}^B$ be a canonical block-responsive hiring policy with respect to $I^{\updownarrow}$.
    Then, $\mathcal{R}^{I}(\mathcal{H}^B) \geq \mathcal{R}^{I^{\updownarrow}}(\mathcal{H}^B)$.
\end{lemma}

Now, with some wishful thinking, letting ${\mathcal{H}^*}^{\updownarrow}$ be an optimal hiring policy for $I^{\updownarrow}$, suppose that we can compute a block-responsive policy $\mathcal{H}^B$ with respect to this instance such that
\begin{equation}
    \mathcal{R}^{I^{\updownarrow}}(\mathcal{H}^B) ~~\geq~~ (1-\delta) \cdot \mathcal{R}^{I^{\updownarrow}}({\mathcal{H}^*}^{\updownarrow}) \ ,  \label{eq:block_inter}
\end{equation}
where $\delta = \delta(\epsilon, k)$ is a parameter whose value will be specified later on.
The next two observations shed light on the implication of this result in terms of approximating our original instance $I$:
\begin{enumerate}
    \item First, due to the optimality of ${\mathcal{H}^*}^{\updownarrow}$ for $I^{\updownarrow}$, we infer that $\mathcal{R}^{I^{\updownarrow}}({\mathcal{H}^*}^{\updownarrow}) \geq \mathcal{R}^{I^{\updownarrow}}(\mathcal{H}^*)$, where $\mathcal{H}^*$ is a canonical optimal policy for $I$, which is known to exist by Corollary~\ref{cor:rl_vrl_optimal}.
    In addition, following Lemma~\ref{lemma:rounding_up_down}, we have $\mathcal{R}^{I^{\updownarrow}}(\mathcal{H}^*) \geq (1-2\epsilon) \cdot \mathcal{R}^{I}({\mathcal{H}^*})$. \label{block_inter_1}
    \item Next, Lemma~\ref{lemma:block_lr_and_vrl_modification} proves that $\mathcal{H}^B$ can be transformed into a canonical policy  $\tilde{\mathcal{H}}^B$ for $I^{\updownarrow}$, without any loss in its expected reward, i.e., $\mathcal{R}^{I^{\updownarrow}}(\tilde{\mathcal{H}}^B) \geq \mathcal{R}^{I^{\updownarrow}}(\mathcal{H}^B)$. \label{block_inter_2}
\end{enumerate}
Given these findings, we claim that the block-responsive policy $\tilde{\mathcal{H}}^B$ is necessarily near-optimal for our original instance $I$, since
\begin{eqnarray}
    \mathcal{R}^{I}(\tilde{\mathcal{H}}^B) &\geq& \mathcal{R}^{I^{\updownarrow}}(\tilde{\mathcal{H}}^B) \label{eq:block_instances_1}\\
    &\geq& \mathcal{R}^{I^{\updownarrow}}(\mathcal{H}^B) \label{eq:block_instances_2}\\
    &\geq& (1-\delta) \cdot \mathcal{R}^{I^{\updownarrow}}({\mathcal{H}^*}^{\updownarrow}) \label{eq:block_instances_3}\\
    &\geq& (1-2\epsilon) \cdot (1-\delta) \cdot \mathcal{R}^{I}({\mathcal{H}^*}) \ .  \label{eq:block_instances_4}
\end{eqnarray}
Here, inequality~\eqref{eq:block_instances_1} is obtained by instantiating Lemma~\ref{lemma:block_canonical_modifying_effects} with respect to $\tilde{\mathcal{H}}^B$.
Inequalities~\eqref{eq:block_instances_2} and \eqref{eq:block_instances_3} follow from Observation~\ref{block_inter_2} and equation~\eqref{eq:block_inter}.
Finally, inequality~\eqref{eq:block_instances_4} is implied by Observation~\ref{block_inter_1}.

\subsection{Near-optimality of small-sized block-responsive policies?} \label{subsec:thm_block_policy}
With the aim of implementing our wishful thinking, as stated in Section~\ref{subsec:inter_summary_block}, it is worth pointing out that, on the one hand, block-responsive policies allow us to aggregate multiple applicants into single blocks, potentially creating a small-sized decision tree.
On the other hand, multi-applicant blocks may yield inferior expected rewards in comparison to single-applicant selections, as in standard hiring policies, since as soon as one block member accepts an offer, all remaining members are skipped.
This trade-off between decision tree size and expected reward is inherent to block-responsive policies.

Surprisingly, we prove that the expected reward of an optimal policy can nearly be matched by a block-responsive policy whose decision tree is of depth $O(\frac{k}{\epsilon^3}\log\frac{1}{\epsilon} )$, consisting of only $(\frac{1}{\epsilon})^{O(k)}$ nodes.
At the same time, we will argue that this policy satisfies a generalization of the order-by-value rule, defined for standard hiring policies in Section~\ref{subsec:dynamic_programming}.
Specifically, we say that a block-responsive policy $\mathcal{H}^B$ adheres to the block-order-by-value rule when, for any state $(t, k_t, \mathcal{A}_t)$ and any acceptance-probability class $\mathcal{C}_m$, the collection of $\mathcal{C}_m$-applicants in $\mathcal{H}^B(t, k_t, \mathcal{A}_t)$ forms a prefix of the highest-value currently available $\mathcal{C}_m$-applicants.

\begin{theorem}\label{thm:block_policy}
Let ${\mathcal{H}^*}^{\updownarrow}$ be an optimal hiring policy for $I^{\updownarrow}$. 
Then, for any $\eps >0$, there exists a block-responsive policy $\mathcal{H}^B$ such that:
\begin{enumerate}
    \item {\em Expected reward:} $\mathcal{R}^{I^{\updownarrow}}(\mathcal{H}^B) \geq (1 - 4\epsilon^3k) \cdot \mathcal{R}^{I^{\updownarrow}}({\mathcal{H}^*}^{\updownarrow})$.
    
    \item {\em Size:} The decision tree of $\mathcal{H}^B$ consists of $(\frac{1}{\epsilon})^{O(k)}$ nodes and its depth is $O(\frac{k}{\epsilon^3}\log\frac{1}{\epsilon} )$.
    
    \item {\em Block-order-by-value:} $\mathcal{H}^B$ satisfies the block-order-by-value rule.
\end{enumerate}
\end{theorem}
To establish Theorem~\ref{thm:block_policy}, we describe a recursive transformation that converts the optimal policy ${\mathcal{H}^*}^{\updownarrow}$ into being block-responsive, ending up with $\mathcal{H}^B$.
At a high level, as explained in Section~\ref{subsec:policy_transformation}, this transformation employs a carefully-designed method for grouping nodes along the leftmost path of the decision tree $\mathcal{T}_{{\mathcal{H}^*}^{\updownarrow}}$ into a limited number of blocks, each represented by a new node. 
Subsequently, the latter idea is recursively repeated for selected right subtrees descending from this path.
In Section~\ref{subsec:block_responsive_remaining_proof}, we prove that our newly-defined tree $\mathcal{T}_{\mathcal{H}^B}$ satisfies the structural guarantees of Theorem~\ref{thm:block_policy}.
In addition, we argue that the expected reward collected by the block-responsive policy $\mathcal{H}^B$ nearly matches that of our original policy, ${\mathcal{H}^*}^{\updownarrow}$.

\subsection{Proof of Theorem~\ref{thm:block_policy}: Policy transformation} \label{subsec:policy_transformation}
Let ${\mathcal{H}^*}^{\updownarrow}$ be a canonical optimal hiring policy with respect to the rounded instance $I^{\updownarrow}$ that satisfies the order-by-value rule, as shown to exist in Lemmas~\ref{lemma:optimal_Order_by_value} and~\ref{lemma:dynamic_canonical}, with $\mathcal{T}_{{\mathcal{H}^*}^{\updownarrow}}$ being the decision tree of this policy. In what follows, we propose a recursive method for defining a block-responsive policy, $\mathcal{H}^B$, that satisfies the performance guarantees of Theorem~\ref{thm:block_policy}.
To this end, let us consider the so-called leftmost path $P_L(\mathcal{T}_{{\mathcal{H}^*}^{\updownarrow}})$ of $\cal T_{{\mathcal{H}^*}^{\updownarrow}}$, that begins at $\myroot(\cal T_{{\mathcal{H}^*}^{\updownarrow}})$ and proceeds downward to the left child at each node, until reaching a leaf.
Our approach involves identifying a set of $O(\frac{1}{\epsilon^3}\log\frac{1}{\epsilon})$ ``terminal'' nodes along $P_L(\cal T_{{\mathcal{H}^*}^{\updownarrow}})$, utilized to define the block-responsive policy $\mathcal{H}^B$ by suitably splitting the leftmost path into blocks and proceeding by recursion.

\paragraph{Identifying terminal nodes.}
For this purpose, let $u_1, \dots, u_S$ be the sequence of nodes along $P_L(\mathcal{T}_{{\mathcal{H}^*}^{\updownarrow}})$ in root-to-leaf order.
That is, $u_1 = \myroot(\cal T_{{\mathcal{H}^*}^{\updownarrow}})$, $u_2$ is the left child of $u_1$, $u_3$ is the left child of $u_2$, so on and so forth.
In addition, for $s_1 \leq s_2$, we define $\prsub{\mathcal{T}_{{\mathcal{H}^*}^{\updownarrow}}}{u_{s_1} \rightsquigarrow u_{s_2}}$ as the probability of reaching node $u_{s_2}$ from $u_{s_1}$, corresponding to the event where the applicants $\myapp(u_{s_1}), \dots, \myapp(u_{s_2-1})$ reject their offers one after the other.
Out of these nodes, we define our set of terminals $u_{\tau_1}, \dots, u_{\tau_F}$, in root-to-leaf order.
For ease of analysis, we introduce terminals of two potentially-overlapping types, $A$ and $B$, as follows:
\begin{itemize}
    \item {\em Type-$A$ terminals}:
    The first type-$A$ terminal $u_{\tau_1}$ is simply the root of $\mathcal{T}_{{\mathcal{H}^*}^{\updownarrow}}$, i.e., $\tau_1 = 1$.
    Next, we say that an arc $(u_s, u_{s+1})$ on the leftmost path is $A$-crossing when the interval $[\prsub{\mathcal{T}_{{\mathcal{H}^*}^{\updownarrow}}}{u_{\tau_1} \rightsquigarrow u_{s+1}}, \prsub{\mathcal{T}_{{\mathcal{H}^*}^{\updownarrow}}}{u_{\tau_1} \rightsquigarrow u_s}]$ contains an integer power of $1-\epsilon^3$.
    In this case, both $u_s$ and $u_{s+1}$ will be type-$A$ terminals.
    We continue to define type-$A$ terminals via this condition up until reaching the last type-$A$ terminal, denoted as $u_{\tau_F}$, which is the first node with an arrival probability of $\prsub{\mathcal{T}_{{\mathcal{H}^*}^{\updownarrow}}}{u_{\tau_1} \rightsquigarrow u_{\tau_F}} < \epsilon^3$.
    When there is no such node, i.e., $\prsub{\mathcal{T}_{{\mathcal{H}^*}^{\updownarrow}}}{u_{\tau_1} \rightsquigarrow u_{S}} \geq \epsilon^3$, the last type-$A$ terminal will be the leaf node $u_S$.
    \item {\em Type-$B$ terminals}: Next, let us first observe that the sequence of expected rewards $\mathcal{R}^{I^{\updownarrow}}_{\mytree}(\mathcal{T}_{{\mathcal{H}^*}^{\updownarrow}}({u_{1}}_R)), \dots, \mathcal{R}^{I^{\updownarrow}}_{\mytree}(\mathcal{T}_{{\mathcal{H}^*}^{\updownarrow}}({u_{S-1}}_R))$ associated with the right subtrees of $u_1, \ldots, u_{S-1}$ is weakly-decreasing; otherwise, ${\mathcal{H}^*}^{\updownarrow}$ would not have been an optimal policy. In turn, we say that an arc $(u_s, u_{s+1})$ on the leftmost path is $B$-crossing when the interval $[\frac{\mathcal{R}^{I^{\updownarrow}}_{\mytree}(\mathcal{T}_{{\mathcal{H}^*}^{\updownarrow}}({u_{s+1}}_R))}{\mathcal{R}^{I^{\updownarrow}}_{\mytree}(\mathcal{T}_{{\mathcal{H}^*}^{\updownarrow}})}, \frac{\mathcal{R}^{I^{\updownarrow}}_{\mytree}(\mathcal{T}_{{\mathcal{H}^*}^{\updownarrow}}({u_{s}}_R))}{\mathcal{R}^{I^{\updownarrow}}_{\mytree}(\mathcal{T}_{{\mathcal{H}^*}^{\updownarrow}})}]$ contains an integer power of $1-\epsilon^3$.
    That is, for a pair of successive nodes, the ratios of their right subtrees’ expected rewards to the overall tree’s expected reward jumps over an integer power of $1-\epsilon^3$.
    In this case, both $u_s$ and $u_{s+1}$ are type-$B$ terminals.
    We make use of this condition to define type-$B$ terminals up until reaching $\frac{\mathcal{R}^{I^{\updownarrow}}_{\mytree}(\mathcal{T}_{{\mathcal{H}^*}^{\updownarrow}}({u_{s}}_R))}{\mathcal{R}^{I^{\updownarrow}}_{\mytree}(\mathcal{T}_{{\mathcal{H}^*}^{\updownarrow}})} < \epsilon^3$, or until reaching the last type-$A$ terminal $u_{\tau_F}$.
    We denote the index of the last type-$B$ terminal as $\tau_{F_B}$.
\end{itemize}
With these definitions, it is not difficult to verify that the resulting number of terminals $F$ is at most $\frac{13}{\epsilon^3} \ln\frac{1}{\epsilon}$.
To this end, we first show that the number of $A$-crossing arcs is at most $\frac{3}{\epsilon^3} \ln\frac{1}{\epsilon}$.
Indeed, with a strictly greater number of such arcs, it follows that
\begin{equation}
    \prsub{\mathcal{T}_{{\mathcal{H}^*}^{\updownarrow}}}{u_{\tau_1} \rightsquigarrow u_{\tau_{F-1}}} ~~<~~ (1-\epsilon^3)^{\frac{3}{\epsilon^3}\ln\frac{1}{\epsilon}} ~~\leq~~ \epsilon^3  \ ,  \nonumber
\end{equation}
in contradiction to our choice of the last type-$A$ terminal, $u_{\tau_F}$.
Therefore, including the root, there are at most $\frac{6}{\epsilon^3} \ln\frac{1}{\epsilon} +1$ type-$A$ terminals.
A similar argument applies to the number of $B$-crossing arcs, meaning that the number of type-$B$ terminals is at most $\frac{6}{\epsilon^3} \ln\frac{1}{\epsilon}$.

\begin{figure}[htbp!]
    \centering
    \resizebox{\textwidth}{!}{
        \begin{tabular}{c} 
            \begin{tikzpicture}[
                triangle/.style = {fill=gray!30, draw=black, isosceles triangle, shape border rotate=90, font=\bfseries, align=center },
                round/.style = { circle, draw=black, fill=gray!20, minimum size=16mm, font=\bfseries, align=center},
                square/.style = { rectangle, draw=black, fill=gray!20, minimum size=16mm, font=\bfseries, align=center},
                font=\tiny,
                node distance=0.4cm and 0.4cm
            ]
                \node [square](root){$u_1$};
                \node[round, below left=of root](u2){$u_2$};
                \node[triangle, below right=of root, yshift=-1.5cm] (u_R) {$\mathcal{T}_{{\mathcal{H}^*}^{\updownarrow}}(u_{1_R})$};
                \node[round, below left=of u2](u3){$u_3$};
                \node[triangle, below right=of u2, yshift=-1.5cm] (u2_R) {$\mathcal{T}_{{\mathcal{H}^*}^{\updownarrow}}(u_{2_R})$};
                \node[square, below left=of u3](u4){$u_4$};
                \node[triangle, below right=of u3, yshift=-1.5cm] (u3_R) {$\mathcal{T}_{{\mathcal{H}^*}^{\updownarrow}}(u_{3_R})$};
                \node[round, below left=of u4](u5){$u_5$};
                \node[triangle, below right=of u4, yshift=-1.5cm] (u4_R) {$\mathcal{T}_{{\mathcal{H}^*}^{\updownarrow}}(u_{4_R})$};
                \node[triangle, below left=of u5, yshift=-1.5cm] (u5_L) {$\mathcal{T}_{{\mathcal{H}^*}^{\updownarrow}}(u_{5_L})$};
                \node[triangle, below right=of u5, yshift=-1.5cm] (u5_R) {$\mathcal{T}_{{\mathcal{H}^*}^{\updownarrow}}(u_{5_R})$};
                
                \draw[->] (root) --  (u2);
                \draw[->] (root) --  (u_R.north);
                \draw[->] (u2) --  (u3);
                \draw[->] (u2) --  (u2_R.north);
                \draw[->] (u3) --  (u4);
                \draw[->] (u3) --  (u3_R.north);
                \draw[->] (u4) --  (u5);
                \draw[->] (u4) --  (u4_R.north);
                \draw[->] (u5) --  (u5_L.north);
                \draw[->] (u5) --  (u5_R.north);
                
                \node[text width=10cm, font=\large, font=\bfseries, align=center] at (2,-8) {The decision tree $\boldsymbol{\mathcal{T}_{{\mathcal{H}^*}^{\updownarrow}}}$ \\(terminals appear as squares)};
                
                \coordinate (braceTop)    at ([xshift=-2cm, yshift=1cm]root.north west);
                \coordinate (braceBottom) at ([xshift=-2cm]u5_L.south west);
                \draw[decorate, decoration={brace,mirror,amplitude=20pt}, thick]
                    (braceTop) -- (braceBottom)
                    node[midway,left=15pt, yshift=0.5cm, font=\large] {The leftmost path $P_L(\mathcal{T}_{{\mathcal{H}^*}^{\updownarrow}})$};
            \end{tikzpicture} \\
            
            \vspace{0.5cm}
            \begin{tikzpicture}[>=Stealth]
                \draw[->, line width=2pt] (0,0) -- (0,-1.5); 
            \end{tikzpicture} \\
            
            \vspace{0.5cm}
            \begin{tikzpicture}[
                triangle/.style = {fill=gray!30, draw=black, isosceles triangle, shape border rotate=90, font=\bfseries, align=center },
                round/.style = { circle, draw=black, fill=gray!20, minimum size=16mm, font=\bfseries, align=center},
                font=\tiny,
                node distance=0.4cm and 0.4cm
            ]
                \node [round](root){$\myapp(u_3) \rightarrow \myapp(u_1)$ \\ $\rightarrow \myapp(u_2)$};
                \node[round, below left=of root](u2){$\myapp(u_4) \rightarrow \myapp(u_5)$};
                \node[triangle, below right=of root, yshift=-1.5cm] (u_R) {$\mathcal{T}_{\mathcal{H}^B,u_{3_R}}$};
                \node[round, below right=of u2, yshift=-1.5cm] (u2_R) {$B_0$};
                \node[triangle, below left=of u2, yshift=-1.5cm] (u2_L) {$\mathcal{T}_{\mathcal{H}^B, u_{5_L}}$};
                
                \draw[->] (root) --  (u2);
                \draw[->] (root) --  (u_R.north);
                \draw[->] (u2) --  (u2_R.north);
                \draw[->] (u2) --  (u2_L.north);
                
                \node[text width=5cm, font=\large, font=\bfseries, align=center] at (-8,-3) {The decision tree $\boldsymbol{\mathcal{T}_{\mathcal{H}^B}}$};
            \end{tikzpicture}
        \end{tabular}
    }
    \caption{Transforming the optimal decision tree to a block-responsive tree.} 
    \label{fig:block_responsive_construction}
\end{figure}
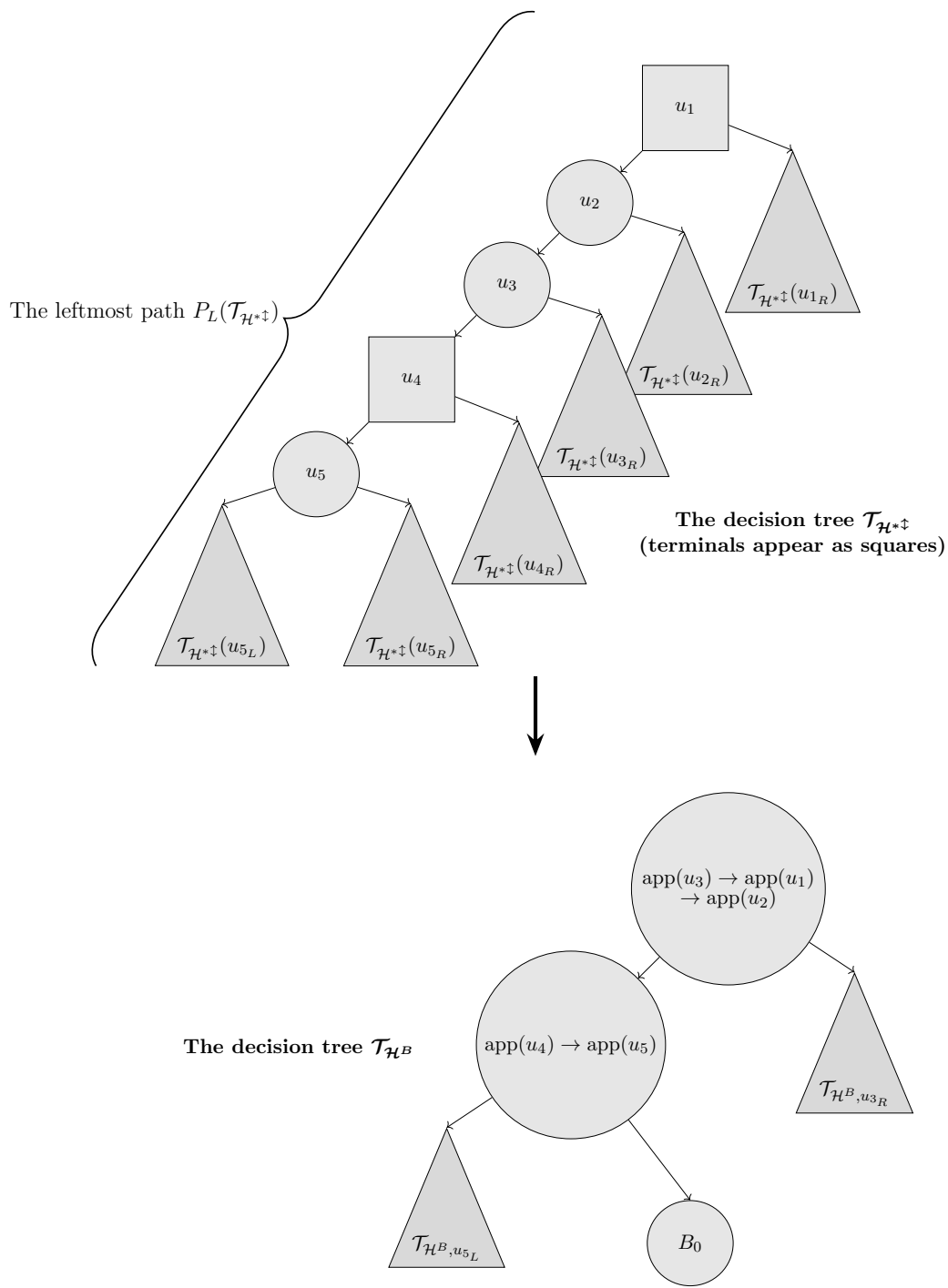

\paragraph{Constructing the decision tree \boldmath{$\mathcal{T}_{\mathcal{H}^B}$}.}
We proceed by explaining how to make use of these terminals to transform the decision tree $\mathcal{T}_{{\mathcal{H}^*}^{\updownarrow}}$ of our original policy ${\mathcal{H}^*}^{\updownarrow}$ into a block-responsive tree, $\mathcal{T}_{\mathcal{H}^B}$.
To better understand the upcoming discussion, we advise the reader to consult the schematic illustration in Figure~\ref{fig:block_responsive_construction}.
As explained in Section~\ref{subsec:block_responsive}, each node of a block-responsive decision tree includes an ordered subset of applicants, referred to as a block. Our approach to construct $\mathcal{T}_{\mathcal{H}^B}$ involves partitioning the leftmost path $P_L(\mathcal{T}_{{\mathcal{H}^*}^{\updownarrow}})$ based on its sequence of terminals $u_{\tau_1}, \dots, u_{\tau_F}$ and grouping the nodes between any two successive terminals into a separate block.
Technically speaking, this procedure operates as follows:
\begin{enumerate}
    \item {\em Root node}: The root of the block-responsive decision tree $\mathcal{T}_{\mathcal{H}^B}$ will be denoted as $u_1^B$.
    Its corresponding block is defined by $\myblock(u_1^B) = \{ \myapp(u_{\tau_1}), \ldots, \myapp(u_{\tau_2 - 1}) \}$.
    In other words, this block consists of the applicants within the nodes $u_{\tau_1}, \dots,u_{\tau_2-1}$ along $P_L(\mathcal{T}_{{\mathcal{H}^*}^{\updownarrow}})$. \label{item:identifying_root_node}
    \item {\em Constructing the leftmost path of $\mathcal{T}_{\mathcal{H}^B}$}: 
    Similarly, the remaining nodes along the leftmost path of $\mathcal{T}_{\mathcal{H}^B}$ will be denoted by $u_2^B, \dots, u_{F}^B$. 
    For every $2\leq f \leq F-1$, we define $\myblock(u_f^B)=\{ \myapp(u_{\tau_f}), \ldots, \myapp(u_{\tau_{f+1} - 1})\}$.
    To complete this path, we include a leaf node $u_F^B$ at its termination, with $\myblock(u_F^B) = \emptyset$.
    In addition, we add left arcs connecting $u_1^B$ to $u_2^B$, $u_2^B$ to $u_3^B$, so on and so forth.
    \label{item:constructing_the_leftmost_path}
    \item {\em Recursive construction}:
    For each $f \leq F-1$, it remains to construct the right subtree $\mathcal{T}_{\mathcal{H}^B}( {u^B_f}_R )$ of $u^B_f$ according to the next two cases, depending on how $f$ is related to the index $F_B$ of the last type-$B$ terminal:
    \begin{itemize}
        \item {\em When $f \leq F_B -1$}: We create $\mathcal{T}_{\mathcal{H}^B}( {u^B_f}_R )$ by recursively converting the right subtree $\mathcal{T}_{{\mathcal{H}^*}^{\updownarrow}}(u_{{\tau_{f+1}-1}_R})$ of $u_{\tau_{f+1}-1}$ into its block-responsive counterpart.
        
        \item {\em When $F_B \leq f \leq F-1$}:
        Here, we define the right subtree $\mathcal{T}_{\mathcal{H}^B}( {u^B_f}_R )$ as a leaf node, containing the empty block, $B_0$.
    \end{itemize}
    This recursive construction terminates upon reaching a leaf node. Its block-responsive equivalent is defined as a single node, again containing the empty block, $B_0$.
    \item {\em Intra-block order}: Finally, we order the applicants within each block in non-increasing order of their value.
    While this convention is not strictly necessary to establish Theorem~\ref{thm:block_policy}, it will be useful in simplifying our proof.  
    \label{item:intra_block_order}
\end{enumerate}

\paragraph{Block rejection probabilities.}
Prior to analyzing this construction, let us momentarily consider any node $u^B_f$ along the leftmost path of $\mathcal{T}_{\mathcal{H}^B}$, and define its rejection probability as $\psi_{u^B_f} = \prod_{q=1}^{|\myblock(u^B_f)|}(1-p^{\updownarrow}_{\myapp_{q}(u^B_f)})$.
In other words, this term is precisely the probability that all applicants in $\myblock(u^B_f)$ reject their offers.
As stated in Observation~\ref{corollary:rejection_prob} below, whenever $\myblock(u^B_f)$ consists of two or more applicants, its rejection probability must be at least $1- \epsilon^3$.
Indeed, as explained in item~2, any such block is comprised of the applicants $ \myapp(u_{\tau_f}), \dots, \myapp(u_{\tau_{f+1}-1})$, where $u_{\tau_f}$ and $u_{\tau_{f+1}}$ are two successive terminals along the leftmost path $P_L(\mathcal{T}_{{\mathcal{H}^*}^{\updownarrow}})$.
However, by definition of type-$A$ terminals, since $u_{\tau_f}$ and $u_{\tau_{f+1}}$ are successive, the path connecting these terminals cannot contain  $A$-crossing arcs, and therefore, 
\[\psi_{u^B_f} ~~=~~ \prod_{s=\tau_f}^{\tau_{f+1}-1}(1-p^{\updownarrow}_{\myapp(u_s)}) ~~=~~ \prsub{\mathcal{T}_{{\mathcal{H}^*}^{\updownarrow}}}{u_{\tau_f} \rightsquigarrow u_{\tau_{f+1}}} ~~\geq~~ 1-\epsilon^3  \ . \]

\begin{observation} \label{corollary:rejection_prob}
$\psi_{u^B_f} \geq 1-\epsilon^3$,
for every $f \leq F-1$ with $|\myblock(u^B_f)| \geq 2$.
\end{observation}

\subsection{Proof of Theorem~\ref{thm:block_policy}: Analysis}\label{subsec:block_responsive_remaining_proof}

\paragraph{Size and depth guarantees.}
Starting with the easy claim, we argue that the number of nodes in the resulting decision tree $\mathcal{T}_{\mathcal{H}^B}$ is only $(\frac{1}{\epsilon})^{O(k)}$.
For this purpose, as explained in Section~\ref{subsec:policy_transformation}, the number of terminal nodes $F$ along the leftmost path $P_L(\mathcal{T}_{{\mathcal{H}^*}^{\updownarrow}})$ is at most $\frac{13}{\epsilon^3} \ln\frac{1}{\epsilon} $.
Since the latter bound is independent of any structural characteristic of $\mathcal{T}_{{\mathcal{H}^*}^{\updownarrow}}$, it applies to the number of terminal nodes obtained when descending by recursion to each of the subtrees $\mathcal{T}_{{\mathcal{H}^*}^{\updownarrow}}(u_{{\tau_2-1}_R}), \dots, \mathcal{T}_{{\mathcal{H}^*}^{\updownarrow}}(u_{{\tau_F-1}_R})$.
To better understand how these bounds are compounding, letting $|\mathcal{T}_{\mathcal{H}^B}(u)|$ be the number of nodes in the subtree of $\mathcal{T}_{\mathcal{H}^B}$ rooted at $u$, it follows that
\begin{equation}
    |\mathcal{T}_{\mathcal{H}^B}| ~~=~~ F + \sum_{f\leq F-1}|\mathcal{T}_{\mathcal{H}^B}({u_f^B}_R)|~~\leq~~ \frac{13}{\epsilon^3}\ln\frac{1}{\epsilon} + \sum_{f\leq F-1}|\mathcal{T}_{\mathcal{H}^B}({u_f^B}_R)|  \ .  \label{eq:block_recursive_size}
\end{equation}
The important observation is that, since the root of each right subtree $\mathcal{T}_{\mathcal{H}^B}({u_f^B}_R)$ corresponds to a state with one fewer position in comparison to that of $\myroot(\mathcal{T}_{\mathcal{H}^B})$, our recursion depth is bounded by the initial number of open positions, $k$.
Consequently, by expanding the recursive relation \eqref{eq:block_recursive_size}, it is easy to verify that $|\mathcal{T}_{\mathcal{H}^B}| = (\frac{1}{\epsilon})^{O(k)}$.

Along the same lines, we observe that $\mydepth(\mathcal{T}_{\mathcal{H}^B}) = O(k  F) = O(\frac{k}{\epsilon^3}\log\frac{1}{\epsilon})$.
Indeed, the leftmost path of every recursively-generated subtree is of length at most $F$, and along any root-to-leaf path, there are at most $k$ right turns, each consuming one available position.
Therefore, any root-to-leaf path in $\mathcal{T}_{\mathcal{H}^B}$ consists of at most $k F$ nodes.

\paragraph{Block-order-by-value.}
We proceed by showing that $\mathcal{T}_{\mathcal{H}^B}$ satisfies the block-order-by-value rule, knowing that $\mathcal{T}_{{\mathcal{H}^*}^{\updownarrow}}$ satisfies the standard order-by-value rule.
To this end, suppose that $i_1$ and $i_2$ are two distinct applicants within the same acceptance-probability class, such that $v^{\updownarrow}_{i_1} > v^{\updownarrow}_{i_2}$.
Let $\myblock(u^B)$ be some block containing applicant $i_2$, and let $\myroot(\mathcal{T}_{\mathcal{H}^B}) =u^B_1 , u^B_2, \dots, u^B_{d_2} = u^B$ be the path connecting $\myroot(\mathcal{T}_{\mathcal{H}^B})$ to this block.
We argue that $i_1 \in \myblock(u^B_{d_1})$ for some $d_1 \leq d_2$.

To verify this claim, according to the construction described in Section~\ref{subsec:policy_transformation}, $\myblock(u^B)$ unifies the applicants belonging to the nodes that lie between two successive terminals on the leftmost path $P_L(\mathcal{T}_{{\mathcal{H}^*}^{\updownarrow}}(v))$ of the subtree rooted at some node $v \in \mathcal{T}_{{\mathcal{H}^*}^{\updownarrow}}$.
Let $u^{i_2} \in P_L(\mathcal{T}_{{\mathcal{H}^*}^{\updownarrow}}(v))$ be the original node containing applicant $i_2$, and let $v = u_1 , u_2, \dots, u_{\hat{d_2}} = u^{i_2}$ be the original path connecting $v$ to $u^{i_2}$.
Then, since $\mathcal{T}_{{\mathcal{H}^*}^{\updownarrow}}$ adheres to the order-by-value rule, and since $v^{\updownarrow}_{i_1} > v^{\updownarrow}_{i_2}$, there exists some $\hat{d}_1 < \hat{d}_2$ for which $\myapp(u_{\hat{d}_1}) =i_1 $.
By construction, the applicants $\myapp(u_1), \dots, \myapp(u_{\hat{d}_2})$ are placed along $\myblock(u^B_1), \dots, \myblock(u^B_{d_2})$, and therefore, there exists some $d_1 \leq d_2$ for which $i_1 \in \myblock(u^B_{d_1})$, as desired.

\paragraph{Approximation guarantee.} Turning to address the more challenging question, Lemma~\ref{lemma:block_responsive_approx_factor} proves that the expected reward attained by our just-constructed block-responsive policy $\mathcal{H}^B$ closely approximates the expected reward of the optimal policy ${\mathcal{H}^*}^{\updownarrow}$.
Leaving the detailed arguments behind this result to be presented in Appendix~\ref{proof:block_responsive_approx_factor}, we will show that each recursive call preserves a fraction of at least $1- 4\epsilon^3$ of our expected reward, thereby ending up with an approximation ratio of at least  $(1-4\epsilon^3)^k \geq (1-4\epsilon^3k)$ by Bernoulli's inequality.

\begin{lemma} \label{lemma:block_responsive_approx_factor}
    $\mathcal{R}^{I^{\updownarrow}}(\mathcal{H}^B) \geq (1 - 4\epsilon^3k) \cdot \mathcal{R}^{I^{\updownarrow}}({\mathcal{H}^*}^{\updownarrow})$.
\end{lemma}

\section{Approximation Scheme in Truly Polynomial Time} \label{section:PTAS}

In this section, block-responsive policies will be exploited to derive our main algorithmic result, an $O(n^{ O(1) } \cdot T^{2^{\tilde{O}(1/\epsilon^2)}})$-time approximation scheme for the sequential hiring problem, as formally stated in Theorem~\ref{thm:ptas}.
To this end, depending on the number of available positions, we introduce two parametric regimes, arguing that the seemingly-simpler one can be handled by the LP-based approach of \cite{EpsteinM22}.
As such, our attention will be dedicated to the second regime, for which a near-optimal block-responsive policy will be efficiently constructed.

\subsection{Parametric regimes} \label{subsec:regimes}
Moving forward, let us begin by examining two parametric regimes within the sequential hiring problem, elucidating the exact circumstances for which an approximation scheme is still missing.
For ease of exposition, these regimes are initially defined in terms of our original instance $I$; along the way, we will explain why one can instead consider its corresponding rounded instance, $I^{\updownarrow}$.

\paragraph{The many-positions regime: $\bs{ k \geq \frac{1}{\epsilon^2}}$.}
We first address scenarios where the number of available positions is relatively large, specifically meaning that $k \geq \frac{1}{\epsilon^2}$. In this case, one can readily exploit the recent work of \cite{EpsteinM22}, who designed a polynomial-time $(1 - \frac{e^{-k}k^k}{k!})$-approximation; we refer the reader back to Section~\ref{subsec:results_questions} for an elaborate discussion on this approach.
To bound the resulting error term, note that
\begin{align*}
\frac{e^{-k}k^k}{k!} ~~\leq~~ \frac{1}{\sqrt{2\pi k}} \cdot e^{-\frac{1}{12k+1}} ~~\leq~~ \frac{1}{\sqrt{2\pi k}} ~~\leq~~ \epsilon \ , 
\end{align*}
where the first inequality follows from an exact Stirling-type bound due to \cite{Robbins55}, stating that $\sqrt{2\pi k}(\frac{k}{e})^k e^{\frac{1}{12k+1}} < k!$, and the last inequality holds since $k \geq \frac{1}{\epsilon^2}$. 
Consequently, in the current regime, this approach suffices to attain a $(1-\epsilon)$-approximation in polynomial time.


\paragraph{The few-positions regime: $\bs{k < \frac{1}{\epsilon^2}}$.}
Consequently, we are left with the technically difficult scenario, where the number of available positions $k$ is relatively small, namely, $k < \frac{1}{\epsilon^2}$.
To deal with such instances, we propose the ``tree enumeration'' algorithm, whose central design idea involves explicitly constructing a polynomially-sized family of block-responsive policies, guaranteed to contain at least one near-optimal policy.
Toward this objective, as explained in Section~\ref{subsec:inter_summary_block}, rather than considering the original instance $I$, our wishful thinking allows us to work instead with the rounded instance $I^{\updownarrow}$. 
In particular, to implement equation~\eqref{eq:block_inter}, we will construct a block-responsive decision tree $\tilde{\mathcal{T}}$ with an expected reward of $\mathcal{R}_{\mytree}^{I^{\updownarrow}}(\tilde{\mathcal{T}}) = (1-O(\epsilon)) \cdot \mathcal{R}_{\mytree}^{I^{\updownarrow}}(\mathcal{T}_{{\mathcal{H}^*}^{\updownarrow}})$. Here, ${\mathcal{H}^*}^{\updownarrow}$ stands for a canonical optimal policy for $I^{\updownarrow}$, whose existence has been established in Corollary~\ref{cor:rl_vrl_optimal}.

For this purpose, by Theorem~\ref{thm:block_policy}, we know that there exists a block-responsive policy $\mathcal{H}^B$ with an expected reward of 
\[ \mathcal{R}^{I^{\updownarrow}}(\mathcal{H}^B) ~~\geq~~ (1 - 4\epsilon^3k) \cdot \mathcal{R}^{I^{\updownarrow}}({\mathcal{H}^*}^{\updownarrow}) ~~\geq~~ (1 - 4\epsilon) \cdot \mathcal{R}^{I^{\updownarrow}}({\mathcal{H}^*}^{\updownarrow}) \ , \]
where the last inequality holds since $k \leq \frac{1}{\epsilon^2}$.
Moreover, the block-responsive decision tree $\mathcal{T}_{\mathcal{H}^B}$ of this policy consists of $(\frac{1}{\eps})^{O(k)} = 2^{\tilde{O}(1/\epsilon^2)}$ nodes, its depth is $O(\frac{k}{\epsilon^3}\log\frac{1}{\epsilon}) = \tilde{O}(\frac{1}{\epsilon^5})$, and it satisfies the block-order-by-value rule.
In what follows, we explain how to approximately reconstruct this tree by means of efficient enumeration.

\subsection{Step 1: Guessing procedure} \label{subsec:step_2}

\paragraph{Length of leftmost path.}
We begin by guessing the number of nodes along $P_L(\mathcal{T}_{\mathcal{H}^B})$, the leftmost path of $\mathcal{T}_{\mathcal{H}^B}$.
As explained in Section~\ref{subsec:policy_transformation}, the number $F$ of such nodes is at most $\frac{13}{\epsilon^3} \log\frac{1}{\epsilon}$, meaning that there are $O(\frac{1}{\epsilon^3} \log\frac{1}{\epsilon})$ guesses to be considered.

\paragraph{Class-to-block contributions.}
Next, suppose that $u^B_1, \dots, u^B_F$ is the sequence of nodes along $P_L(\mathcal{T}_{\mathcal{H}^B})$ in root-to-leaf order, with their corresponding blocks denoted by $B_1 = \myblock(u^B_1), \dots, B_F = \myblock(u^B_F) = \emptyset$. 
Let us recall that, according to how applicants were partitioned in Section~\ref{subsec:partitioning_applicants} into the acceptance-probability classes $\mathcal{C}_0, \dots, \mathcal{C}_M$, those within each class $\mathcal{C}_m$ share the same acceptance probability, which will be referred to as $p^{(m)}$.
Now, for each class $\mathcal{C}_m$ and node $u^B \in \mathcal{T}_{\mathcal{H}^B}$, we define the contribution of class $\mathcal{C}_m$ to $\myblock(u^B)$ as
\[\mathcal{R}^{I^{\updownarrow}}_{\mathcal{T}_{\mathcal{H}^B}}(m, u^B) ~~=~~ p^{(m)} \cdot \sum_{i \in \myblock(u^B) \cap \mathcal{C}_m}  v^{\updownarrow}_{i}  \ .  \]
This expression can be viewed as the expected total reward collected over the $\mathcal{C}_m$-applicants in $\myblock(u^B)$, in the unrealistic scenario where all such applicants are extended offers.
To obtain an upper bound on these quantities in regard to $u^B_1, \dots, u^B_F$, we recall that during the construction of $\mathcal{T}_{\mathcal{H}^B}$ in Section~\ref{subsec:policy_transformation}, the final block $B_F = \emptyset$ corresponds to the last type-$A$ terminal on $P_L(\mathcal{T}_{{\mathcal{H}^*}^{\updownarrow}})$, defined as the first node whose arrival probability drops below $\epsilon^3$.
Therefore, with probability at least $\epsilon^3$, the policy $\mathcal{H}^B$ extends offers to all candidates in blocks $B_1, \dots, B_F$, implying that
\begin{equation}
    \sum_{f\leq F}\sum_{m \leq M}\mathcal{R}^{I^{\updownarrow}}_{\mathcal{T}_{\mathcal{H}^B}}(m, u^B_f) ~~\leq~~ \frac{1}{\epsilon^3} \cdot \mathcal{R}^{I^{\updownarrow}}(\mathcal{H}^B) ~~\leq~~ \frac{1}{\epsilon^3} \cdot \mathcal{R}^{I^{\updownarrow}}({\mathcal{H}^*}^{\updownarrow}) \ .    \label{eq:ptas_contribution_bound}
\end{equation}

\paragraph{Contribution guessing.}
We now employ any constant-factor approximation for the sequential hiring problem, such as the $\frac{1}{2}$-approximation of \cite{PurohitGR19}, to obtain an under-estimate $\widetilde{\opt}^{\updownarrow}$ for the optimal expected reward $\mathcal{R}^{I^{\updownarrow}}({\mathcal{H}^*}^{\updownarrow})$, satisfying $\frac{1}{2} \cdot \mathcal{R}^{I^{\updownarrow}}({\mathcal{H}^*}^{\updownarrow}) \leq \widetilde{\opt}^{\updownarrow} \leq \mathcal{R}^{I^{\updownarrow}}({\mathcal{H}^*}^{\updownarrow})$.
With this parameter, we guess each contribution $\mathcal{R}^{I^{\updownarrow}}_{\mathcal{T}_{\mathcal{H}^B}}(m, u^B_f)$ by enumerating over integer multiples of $\Delta = \frac{\epsilon}{2kM F} \cdot \widetilde{\opt}^{\updownarrow}$, to obtain an under-estimate $\tilde{\mathcal{R}}_{m,u^B_f}$ satisfying
\begin{equation}
   \tilde{\mathcal{R}}_{m,u^B_f} ~~\leq~~ \mathcal{R}^{I^{\updownarrow}}_{\mathcal{T}_{\mathcal{H}^B}}(m, u^B_f) ~~\leq~~ \tilde{\mathcal{R}}_{m,u^B_f} + \Delta \ . \label{eq:contribution_underestimate} 
\end{equation}
For the purpose of bounding the total number of required guesses, note that
\begin{eqnarray}
   \sum_{f \leq F} \sum_{m\leq M} \tilde{\mathcal{R}}_{m, u^B_f} &\leq& \sum_{f \leq F} \sum_{m \leq M}\mathcal{R}^{I^{\updownarrow}}_{\mathcal{T}_{\mathcal{H}^B}}(m, u^B_f) \nonumber\\ 
   &\leq& \frac{1}{\epsilon^3} \cdot \mathcal{R}^{I^{\updownarrow}}({\mathcal{H}^*}^{\updownarrow}) \nonumber \\
   &\leq& \frac{2}{\epsilon^3} \cdot \widetilde{\opt}^{\updownarrow} \nonumber \\
   &=&\frac{4kMF}{\epsilon^4} \cdot \Delta \ , \nonumber
\end{eqnarray}
where the second inequality is precisely~\eqref{eq:ptas_contribution_bound}.
As such, elementary balls-into-bins counting arguments show that the number of guesses for $\{ \tilde{\mathcal{R}}_{m, u^B_f}\}_{f \leq F, m \leq M}$ is at most $\binom{\frac{4kMF}{\epsilon^4}+(M+1)F}{(M+1)F} = O( 2^{O(\frac{kMF}{\epsilon^4})} )$.
Since $k \leq \frac{1}{\epsilon^2}$, $F = O(\frac{1}{\epsilon^3} \log\frac{1}{\epsilon})$, and $M = O(\frac{1}{\eps}\log\frac{T}{\eps})$, we have $O(T^{\tilde{O}(1/\epsilon^{10})})$ options to consider.

\paragraph{Rejection probabilities.}
Following the discussion in Section~\ref{subsec:policy_transformation}, yet another type of $\mathcal{T}_{\mathcal{H}^B}$-related quantity we will guess is the rejection probability $\psi_{u^B} = \prod_{i \in \myblock(u^B)}(1-p^{\updownarrow}_i)$ of each non-leaf node $u^B \in \mathcal{T}_{\mathcal{H}^B}$, capturing the probability that all applicants in $\myblock(u^B)$ reject their offers.
Specifically, letting $\mu = \frac{\epsilon^3}{kF}$, we wish to obtain under-estimates $\tilde{\psi}_{u^B_1}, \dots, \tilde{\psi}_{u^B_{F}}$ for the rejection probabilities of the leftmost-path nodes $u^B_1, \dots, u^B_{F}$, such that $(1-\mu) \cdot \psi_{u^B_f} \leq \tilde{\psi}_{u^B_f} \leq \psi_{u^B_f}$.
To this end, we consider two types of blocks:
\begin{itemize}
    \item {\em Single-applicant blocks}:
    Here, we know that $\psi_{u^B_f} = 1 - p^{(m)}$ for some $m \leq M$, and it suffices to enumerate over the discrete set $\{1- p^{(m)} \}_{m \leq M}$.
    \item {\em Multi-applicants blocks}: Based on Observation~\ref{corollary:rejection_prob}, it follows that the rejection probability $\psi_{u^B_f}$ of such blocks is at least $\psi_{\min} = 1-\epsilon^3$.
    Thus, $\tilde{\psi}_{u^B_f}$ can be restricted to integer powers of $1+\mu$ within the segment $[\psi_{\min}, 1]$
\end{itemize}
Consequently, the number of guesses required for a single estimate is $O(\frac{1}{\mu}\log\frac{1}{\psi_{\min}} +M) = O(\frac{kF}{\epsilon^3} +M) = \tilde{O}(\frac{1}{\epsilon^8}\log T)$ with room to spare, as $\mu = \frac{\epsilon^3}{kF}$, $k \leq \frac{1}{\epsilon^2}$, $F = O(\frac{1}{\epsilon^3}\log\frac{1}{\epsilon})$, and $M=O(\frac{1}{\epsilon}\log\frac{T}{\epsilon})$.
Therefore, along the entire leftmost path, there are $(\log T)^{\tilde{O}(1/\epsilon^3)}$ guesses in total for $\tilde{\psi}_{u^B_1}, \dots, \tilde{\psi}_{u^B_{F}}$.

\subsection{Step 2: Constructing an approximate decision tree} \label{subsec:step_3}
Having guessed the number of nodes $F$ along the leftmost path of $\mathcal{T}_{\mathcal{H}^B}$, the class-to-block contribution estimates $\{ \tilde{\mathcal{R}}_{m, u^B_f}\}_{f \leq F, m \leq M}$, and the rejection probability estimates $\{ \tilde{\psi}_{u^B_f}\}_{f \leq F}$, we are now ready to construct a decision tree $\tilde{\mathcal{T}}$ that approximates $\mathcal{T}_{\mathcal{H}^B}$.
The leftmost path $P_L(\tilde{\mathcal{T}})$ of this tree will consist of the nodes $\tilde{u}^B_1, \dots, \tilde{u}^B_F$ in root-to-leaf order, with corresponding blocks $\tilde{B}_1 = \myblock(\tilde{u}^B_1), \dots, \tilde{B}_F = \myblock(\tilde{u}^B_F) =  \emptyset$.

\paragraph{Assigning applicants on $\bs{P_L(\tilde{\mathcal{T}})}$.}
For simplicity of presentation, we first sort the collection of applicants within each acceptance-probability class $\mathcal{C}_m$ in order of weakly-decreasing values.
For $ q_1 \leq q_2 $, the set of $q_1$-th, $(q_1+1)$-th, $\dots$, $q_2$-th $\mathcal{C}_m$-applicants in this order will be denoted by $\mathcal{C}_m[q_1, q_2]$.
We proceed by sequentially assigning applicants to $\tilde{B}_1, \dots, \tilde{B}_F$, filling each block until its estimated class-to-block contribution is met.
To this end, for any node $\tilde{u}^B$ of our constructed tree $\tilde{\mathcal{T}}$, we make use of $\varphi_{m, \tilde{u}^B}$ to index the first $\mathcal{C}_m$-applicant available upon reaching $\tilde{u}^B$; equivalently, applicants $\mathcal{C}_m[1, \varphi_{m, \tilde{u}^B}-1]$ are precisely those assigned along the path connecting the root of $\tilde{\mathcal{T}}$ to $\tilde{u}^B$.
With this notation, our assignment procedure operates as follows.
\begin{itemize}
    \item {\em Assigning applicants to block $\tilde{B}_1$}:
    For every $0 \leq m \leq M$ with $\tilde{\mathcal{R}}_{m, u^B_1} >0$, let $Q_{m,\tilde{u}^B_1}$ be the minimal index $q \geq 0$ for which $p^{(m)} \cdot \sum_{i \in \mathcal{C}_m[1, q]} v^{\updownarrow}_i \geq \tilde{\mathcal{R}}_{m,u^B_1}$.
    Then, the $\mathcal{C}_m$-applicants that will be assigned to block $\tilde{B}_1$ are the $Q_{m,\tilde{u}^B_1}$-highest value applicants in this class, corresponding to $\mathcal{C}_m[1, Q_{m,\tilde{u}^B_1}]$.
    \item {\em Assigning applicants to block $\tilde{B}_2$}:
    Next, for every $0 \leq m \leq M$ with $\tilde{\mathcal{R}}_{m, u^B_2} >0$, let $Q_{m,\tilde{u}^B_2}$ be the minimal index $q \geq 0$ for which $p^{(m)} \cdot \sum_{i \in \mathcal{C}_m[\varphi_{m, \tilde{u}_2^B}, \varphi_{m, \tilde{u}^B_2} + q]}v^{\updownarrow}_i \geq \tilde{\mathcal{R}}_{m,u^B_2}$, noting that $\varphi_{m, \tilde{u}^B_2} = Q_{m,\tilde{u}^B_1} +1$.
    The $\mathcal{C}_m$-applicants that will be assigned to block $\tilde{B}_2$ are $\mathcal{C}_m[\varphi_{m, \tilde{u}^B_2}, \varphi_{m, \tilde{u}^B_2}+Q_{m,\tilde{u}^B_2}]$.
    \item {\em Assigning applicants to blocks $\tilde{B}_3, \dots, \tilde{B}_F$}:
    In general, by following the same logic, having already handled blocks $\tilde{B}_1, \dots, \tilde{B}_{f-1}$, the $\mathcal{C}_m$-applicants that will be assigned to block $\tilde{B}_f$ are $\mathcal{C}_m[\varphi_{m, \tilde{u}^B_f}, \varphi_{m, \tilde{u}^B_f} + Q_{m,\tilde{u}^B_f}]$.
\end{itemize}
As a side note, one might be concerned that this procedure could be ill-defined, since perhaps $Q_{m,\tilde{u}^B_f}$ does not exist for some $f \leq F$ and $m \leq M$.
This concern will be resolved in Section~\ref{subsec:ptas_feasability}, where we prove that given our guessing procedure, such choices always exist.

\paragraph{Correcting rejection probabilities.}
Additionally, one might question the necessity to guess the rejection probabilities $\psi_{u^B_1}, \dots, \psi_{u^B_F}$ in Section~\ref{subsec:step_2}, since up until now, these guesses have not been utilized to construct $\tilde{\mathcal{T}}$.
For technical reasons that will be clarified in Section~\ref{subsec:ptas_guarantees}, we will deliberately ``correct'' rejection probabilities as follows.

For every $f \leq F$, we introduce a Bernoulli random variable $\Psi_{\tilde{u}^B_f}$, independent of any other source of randomness, with $\Pr[\Psi_{\tilde{u}^B_f}=1] = \tilde{\psi}_{u^B_f}/\psi_{\tilde{u}^B_f}$.
Here, $\tilde{\psi}_{u^B_f}$ is our under-estimate for the rejection probability $\psi_{u^B_f} = \prod_{i \in \myblock(u^B_f)}(1-p^{\updownarrow}_i)$ of node $u^B_f$ with respect to $\mathcal{T}_{\mathcal{H}^B}$, whereas $\psi_{\tilde{u}^B_f} = \prod_{i \in \myblock(\tilde{u}^B_f)}(1-p^{\updownarrow}_i)$ is the rejection probability of node $\tilde{u}^B_f$ with respect to $\tilde{\mathcal{T}}$.
To verify that the ratio $\tilde{\psi}_{u^B_f}/\psi_{\tilde{u}^B_f}$ can indeed serve as a probability, note that $\tilde{\psi}_{u^B} \leq \psi_{u^B}$ by definition of $\tilde{\psi}_{u^B}$; in addition, our analysis will show that $\psi_{u^B} \leq \psi_{\tilde{u}^B}$ (see inequality~\eqref{eq:ptas_gua_1} in Appendix~\ref{proof:rejection_prob_correct}).
For completeness, when $\psi_{\tilde{u}^B_f} = 0$, we define $\Pr[\Psi_{\tilde{u}^B_f}=1] =0$.
Now, upon reaching $\tilde{u}^B_f$, our policy proceeds to take its left downward-arc only when all applicants in $\myblock(\tilde{u}^B_f)$ reject their offers, and in conjunction, $\Psi_{\tilde{u}^B_f} = 1$;
otherwise, the right downward-arc is taken.
As such, we are descending to the left subtree of $\tilde{u}^B_f$ with probability $\psi_{\tilde{u}^B_f} \cdot \Pr[\Psi_{\tilde{u}^B_f}=1] = \tilde{\psi}_{u^B_f}$.

\paragraph{Recursive construction.}
Having completed the construction of our leftmost path $P_L(\tilde{\mathcal{T}})$, we proceed to recursively create the right-subtrees of $\tilde{u}^B_1, \dots, \tilde{u}^B_{F-1}$ via precisely the same guessing and assignment procedures, halting upon reaching a leaf node.
From a running time perspective, since the root of each such subtree corresponds to a state with one fewer available position in comparison to its parent, our recursion depth is upper-bounded by the initial number of open positions, $k$.
Therefore, letting $\eta(\kappa)$ be the number of generated trees with respect to $\kappa$ open positions, we have 
\begin{equation}
   \eta(\kappa) ~~=~~ O( T^{\tilde{O}(1/\epsilon^{10})}) \cdot (\eta(\kappa-1))^{O(\frac{1}{\epsilon^3}\log\frac{1}{\epsilon})} \ . \nonumber
\end{equation}
This relation follows by noting that, at each recursion level, the leftmost path consists of $ O(\frac{1}{\epsilon^3}\log\frac{1}{\epsilon})$ nodes, each leading to a separate recursive call.
In addition, each such call involves $O( T^{\tilde{O}(1/\epsilon^{10})})$ guesses for properties related to the leftmost path in question.
By solving this recursive relation and recalling that $k \leq \frac{1}{\epsilon^2}$, we are generating $O(T^{2^{\tilde{O}(1/\epsilon^2)}})$ decision trees to consider.

\subsection{Analysis: Assignment feasibility} \label{subsec:ptas_feasability}
We begin our analysis by showing that the assignment procedure of Section~\ref{subsec:step_3} can indeed be completed, in spite of utilizing estimates for class-to-block contributions rather than their true values.
To this end, our running assumption will be that $\tilde{\mathcal{T}}$ shares the exact same graphic structure as $\mathcal{T}_{\mathcal{H}^B}$ and that, in analogy to equation~\eqref{eq:contribution_underestimate}, for every node $u^B \in \mathcal{T}_{\mathcal{H}^B}$ and class index $m \leq M$, the class-to-block contribution estimate $\tilde{\mathcal{R}}_{m, u^B}$ satisfies
\begin{equation}
    \tilde{\mathcal{R}}_{m,u^B} ~~\leq~~ \mathcal{R}^{I^{\updownarrow}}_{\mathcal{T}_{\mathcal{H}^B}}(m, u^B) ~~\leq~~ \tilde{\mathcal{R}}_{m,u^B} + \Delta \ . \label{eq:ptas_feasability}
\end{equation}

To show that our assignment procedure is well-defined, we prove that $Q_{m,\tilde{u}^B}$ indeed exists for every node $\tilde{u}^B \in \tilde{\mathcal{T}}$ and acceptance-probability class $\mathcal{C}_m$ for which $\tilde{\mathcal{R}}_{m, u^B} >0$.
To this end, recall that $Q_{m,\tilde{u}^B}$ is the minimal index $q \geq 0$ for which $p^{(m)} \cdot \sum_{i \in \mathcal{C}_m[\varphi_{m, \tilde{u}^B}, \varphi_{m, \tilde{u}^B} + q]}v^{\updownarrow}_i \geq \tilde{\mathcal{R}}_{m,u^B}$. To argue that such an index $q$ exists, it suffices to show that: $(1)$~Both $\mathcal{T}_{\mathcal{H}^B}$ and $\tilde{\mathcal{T}}$ satisfy the block-order-by-value rule; and $(2)$~$\varphi_{m, \tilde{u}^B} \leq \varphi_{m, u^B}$.
With these two conditions, for every acceptance-probability class $\mathcal{C}_m$, the $\mathcal{C}_m$-applicants in both $\myblock(u^B)$ and $\myblock(\tilde{u}^B)$ form a prefix of the highest-value available applicants at $\mystate(u^B)$ and $\mystate(\tilde{u}^B)$, respectively.
As a result, letting $\zeta_{m ,u^B}$ be the number of $\mathcal{C}_m$-applicants in $\myblock(u^B)$, we have
\begin{eqnarray}
     p^{(m)} \cdot \sum_{i \in \mathcal{C}_m[\varphi_{m, \tilde{u}^B}, \varphi_{m, \tilde{u}^B} + \zeta_{m, u^B} -1 ]}v^{\updownarrow}_i &\geq& p^{(m)} \cdot \sum_{i \in \mathcal{C}_m[\varphi_{m, u^B}, \varphi_{m, u^B} +\zeta_{m,u^B} -1]}v^{\updownarrow}_i \nonumber \\
     &=& \mathcal{R}^{I^{\updownarrow}}_{\mathcal{T}_{\mathcal{H}^B}}(m, u^B) \nonumber \\
     &\geq& \tilde{\mathcal{R}}_{m,u^B} \ , \nonumber
\end{eqnarray}
where the first inequality holds since $\varphi_{m, \tilde{u}^B} \leq \varphi_{m, u^B}$, and the second inequality follows from~\eqref{eq:ptas_feasability}.
Therefore, there exists some $q \in [0,\zeta_{m ,u^B}-1]$ for which $p^{(m)} \cdot \sum_{i \in \mathcal{C}_m[\varphi_{m, \tilde{u}^B}, \varphi_{m, \tilde{u}^B} + q ]}v^{\updownarrow}_i \geq \tilde{\mathcal{R}}_{m,u^B}$, meaning that $Q_{m,\tilde{u}^B}$ is well-defined.

\paragraph{Proving condition $\bs{(1)}$.}
First, by Theorem~\ref{thm:block_policy}, we know that $\mathcal{T}_{\mathcal{H}^B}$ satisfies the block-order-by-value rule.
In addition, according to the construction of $\tilde{\mathcal{T}}$ in Section~\ref{subsec:policy_transformation}, upon reaching each node $\tilde{u}^B \in \tilde{\mathcal{T}}$, only a prefix of the highest-value currently available applicants in each class are assigned to $\myblock(\tilde{u}^B)$.
Hence, $\tilde{\mathcal{T}}$ also satisfies the block-order-by-value rule.

\paragraph{Proving condition $\bs{(2)}$.}
In what follows, we prove that $\varphi_{m, \tilde{u}^B} \leq \varphi_{m, u^B}$ by induction on $\mydepth(u^B) = \mydepth(\tilde{u}^B)$ in increasing order.
For the base case, where $u^B = \myroot(\mathcal{T}_{\mathcal{H}^B})$ and $\tilde{u}^B = \myroot(\tilde{\mathcal{T}})$, we clearly have $\varphi_{m, \tilde{u}^B} = \varphi_{m, u^B}=1$, since all $\mathcal{C}_m$-applicants are still available.
In the general case, let $\myroot(\mathcal{T}_{\mathcal{H}^B})=u^B_1, \dots, u^B_d = u^B$ be the root-to-$u^B$ path in $\mathcal{T}_{\mathcal{H}^B}$, and let $\myroot(\tilde{\mathcal{T}})=\tilde{u}^B_1, \dots, \tilde{u}^B_d = \tilde{u}^B$ be its corresponding root-to-$\tilde{u}^B$ path in $\tilde{\mathcal{T}}$.
Since $\mathcal{T}_{\mathcal{H}^B}$ satisfies the block-order-by-value rule, the first available $\mathcal{C}_m$-applicant at $\mystate(u^B_d)$ has an index of $\varphi_{m, u^B_{d}} = \varphi_{m, u^B_{d-1}} + \zeta_{u^B_{d-1},m}$, where as before, $\zeta_{u^B_{d-1},m}$ denotes the number of $\mathcal{C}_m$-applicants in $\myblock(u^B_{d-1})$.
Therefore,
\begin{eqnarray}
     p^{(m)} \cdot \sum_{i \in \mathcal{C}_m[\varphi_{m, \tilde{u}^B_{d-1}}, \varphi_{m, \tilde{u}^B_{d-1}}+\zeta_{m, u^B_{d-1}} -1 ]}v^{\updownarrow}_i &\geq& p^{(m)} \cdot \sum_{i \in \mathcal{C}_m[\varphi_{m, u^B_{d-1}}, \varphi_{m, u^B_{d-1}} + \zeta_{m,u^B_{d-1}}-1]}v^{\updownarrow}_i \nonumber\\
     &=& \mathcal{R}^{I^{\updownarrow}}_{\mathcal{T}_{\mathcal{H}^B}}(m, u^B_{d-1}) \nonumber\\
     &\geq&\tilde{\mathcal{R}}_{m,u^B_{d-1}} \ . \nonumber
\end{eqnarray}
Here, the first inequality holds since both summations are comprised of $\zeta_{m,u^B_{d-1}}$ terms, since the left summation begins no later than the right one, as $\varphi_{m, \tilde{u}^B_{d-1}} \leq \varphi_{m, u^B_{d-1}}$ by the induction hypothesis, and since $\mathcal{C}_m$-applicants are indexed in order of weakly-decreasing values.
Therefore, there exists some $q \in [0,\zeta_{m ,u^B_{d-1}} - 1]$ for which $p^{(m)} \cdot \sum_{i \in \mathcal{C}_m[\varphi_{m, \tilde{u}^B_{d-1}}, \varphi_{m, \tilde{u}^B_{d-1}} + q ]}v^{\updownarrow}_i \geq \tilde{\mathcal{R}}_{m,u^B_{d-1}}$, i.e., $Q_{m,\tilde{u}^B_{d-1}} \in [0, \zeta_{m ,u^B_{d-1}}- 1]$.
In turn, by definition of our assignment procedure,
\[ \varphi_{m, \tilde{u}^B_d} ~~=~~ \varphi_{m, \tilde{u}^B_{d-1}} + Q_{m,\tilde{u}^B_{d-1}} +1 ~~\leq~~ \varphi_{m, u^B_{d-1}} +\zeta_{m ,u^B_{d-1}} ~~=~~ \varphi_{m, u^B_d} \ .\]

\subsection{Analysis: Approximation guarantee} \label{subsec:ptas_guarantees}
In the remainder of this section, we turn our attention to showing that the expected reward of $\tilde{\mathcal{T}}$ closely approximates that of $\mathcal{T}_{{\mathcal{H}^*}^{\updownarrow}}$, as formally stated below.
\begin{lemma} \label{lemma:ptas_approx_factor}
    $\mathcal{R}^{I^{\updownarrow}}_{\mytree}(\tilde{\mathcal{T}}) \geq (1-7\epsilon) \cdot \mathcal{R}_{\mytree}^{I^{\updownarrow}}(\mathcal{T}_{{\mathcal{H}^*}^{\updownarrow}})$.
\end{lemma}

\paragraph{The high-level claim.}
To this end, let $u^B \in \mathcal{T}_{\mathcal{H}^B}$ and $\tilde{u}^B \in \tilde{\mathcal{T}}$ be a pair of corresponding nodes with $\mydepth_{\mathcal{T}_{\mathcal{H}^B}}(u^B) = \mydepth_{\tilde{\mathcal{T}}}(\tilde{u}^B) = d$.
We will prove by downward induction on $d$ that the expected reward of the subtree rooted at $\tilde{u}^B$ can be related to that of the subtree rooted at $u^B$, showing that
\begin{eqnarray}
    \mathcal{R}_{\mytree}^{I^{\updownarrow}}(\tilde{\mathcal{T}} (\tilde{u}^B)) ~~\geq ~~ (1-\epsilon^3) \cdot (1-\mu)^{kF-d} \cdot \left(\mathcal{R}^{I^{\updownarrow}}_{\mytree}(\mathcal{T}_{\mathcal{H}^B}({u^B})) - (M+1)(kF-d)\Delta \right)  \ .  \label{eq:proof_5.1}
\end{eqnarray}
As an immediate conclusion, by instantiating this inequality with the roots of $\tilde{\mathcal{T}}$ and $\mathcal{T}_{\mathcal{H}^B}$, both residing at depth zero, we infer that
    \begin{eqnarray}
        \mathcal{R}_{\mytree}^{I^{\updownarrow}}(\tilde{\mathcal{T}}) 
        &\geq& (1-\epsilon^3) \cdot (1-\mu)^{kF} \cdot  \left(\mathcal{R}_{\mytree}^{I^{\updownarrow}}(\mathcal{T}_{\mathcal{H}^B}) - (M+1) kF \Delta \right) \nonumber\\
        &\geq& (1-\epsilon^3)^2 \cdot \left(\mathcal{R}_{\mytree}^{I^{\updownarrow}}(\mathcal{T}_{\mathcal{H}^B}) - \epsilon \cdot \widetilde{\opt}^{\updownarrow} \right) \nonumber\\
        &\geq& (1-7\epsilon) \cdot \mathcal{R}_{\mytree}^{I^{\updownarrow}}(\mathcal{T}_{{\mathcal{H}^*}^{\updownarrow}}) \ . \nonumber
    \end{eqnarray}
Here, the second inequality holds since $\mu = \frac{\epsilon^3}{kF}$ and $\Delta = \frac{\epsilon}{2kM F} \cdot \widetilde{\opt}^{\updownarrow}$.
The third inequality follows by recalling that $\mathcal{R}_{\mytree}^{I^{\updownarrow}}(\mathcal{T}_{\mathcal{H}^B}) \geq (1-4\epsilon) \cdot \mathcal{R}_{\mytree}^{I^{\updownarrow}}(\mathcal{T}_{{\mathcal{H}^*}^{\updownarrow}})$, as explained in Section~\ref{subsec:regimes}, and that $\widetilde{\opt}^{\updownarrow} \leq \mathcal{R}_{\mytree}^{I^{\updownarrow}}(\mathcal{T}_{{\mathcal{H}^*}^{\updownarrow}})$.

\paragraph{Inductive proof.}
To establish inequality~\eqref{eq:proof_5.1}, we operate in leaf-to-root order.
As explained in Section~\ref{subsec:block_responsive}, every leaf node is treated as a virtual block, $B_0 = \emptyset$, meaning that the desired inequality trivially follows.
In the general case, suppose that $\mydepth_{\mathcal{T}_{\mathcal{H}^B}}(u^B) = \mydepth_{\tilde{\mathcal{T}}}(\tilde{u}^B) = d$.
Then, similarly to how representation~\eqref{eq:block_reward_def} was developed, since we descend from $\tilde{u}^B$ to its left and right subtrees with respective probabilities $\tilde{\psi}_{u^B}$ and $1-\tilde{\psi}_{u^B}$, as guaranteed by our rejection probability correction (see Section~\ref{subsec:step_3}),
\begin{eqnarray}
    \mathcal{R}^{I^{\updownarrow}}_{\mytree}(\tilde{\mathcal{T}}(\tilde{u}^B)) &=& \tilde{\psi}_{u^B}  \cdot \mathcal{R}^{I^{\updownarrow}}_{\mytree}(\tilde{\mathcal{T}}({\tilde{u}^B_L})) + (1 - \tilde{\psi}_{u^B}) \cdot \mathcal{R}^{I^{\updownarrow}}_{\mytree}(\tilde{\mathcal{T}}({\tilde{u}^B_R}))\nonumber \\
    && \mbox{}+\sum_{q=1}^{|\myblock(\tilde{u}^B)|}  \left(\prod_{\hat{q}=1}^{q-1}(1-p^{\updownarrow}_{\myapp_{\hat{q}}(\tilde{u}^B)})\right) \cdot  p^{\updownarrow}_{\myapp_q({\tilde{u}^B})} \cdot v^{\updownarrow}_{\myapp_q({\tilde{u}^B})} \nonumber \\
    &\geq& \underbrace{(1-\mu) \cdot \psi_{u^B} \cdot \mathcal{R}^{I^{\updownarrow}}_{\mytree}(\tilde{\mathcal{T}}({\tilde{u}^B_L})) + (1 - \psi_{u^B}) \cdot \mathcal{R}^{I^{\updownarrow}}_{\mytree}(\tilde{\mathcal{T}}({\tilde{u}^B_R}))}_{\text{(I)}} \nonumber \\
    & & \mbox{}+\underbrace{\sum_{q=1}^{|\myblock(\tilde{u}^B)|}  \left(\prod_{\hat{q}=1}^{q-1}(1-p^{\updownarrow}_{\myapp_{\hat{q}}(\tilde{u}^B)})\right) \cdot  p^{\updownarrow}_{\myapp_q({\tilde{u}^B})} \cdot v^{\updownarrow}_{\myapp_q({\tilde{u}^B})}}_{\text{(II)}}  \ ,  \label{eq:ptas_guarantee_middle}
    \end{eqnarray}
where the inequality above holds since $(1-\mu) \cdot \psi_{u^B} \leq \tilde{\psi}_{u^B} \leq \psi_{u^B}$.

\paragraph{Bounding (I).}
By employing the induction hypothesis with respect to the subtrees rooted at $\tilde{u}^B_L$ and $\tilde{u}^B_R$, both residing at depth $d+1$, expression (I) can be lower-bounded by noticing that
\begin{eqnarray}
    \text{(I)} &\geq& (1-\mu) \cdot\psi_{u^B} \cdot (1-\epsilon^3) \cdot (1-\mu)^{kF-d-1}  \cdot \left(\mathcal{R}^{I^{\updownarrow}}_{\mytree}(\mathcal{T}_{\mathcal{H}^B}({u^B_L})) -(M+1)(kF-d-1)\Delta \right) \nonumber \\
    && \mbox{} +(1-\psi_{u^B})  \cdot(1-\epsilon^3) \cdot (1-\mu)^{kF-d-1} \cdot  \left(\mathcal{R}^{I^{\updownarrow}}_{\mytree}(\mathcal{T}_{\mathcal{H}^B}({u^B_R})) -(M+1)(kF-d-1)\Delta \right) \nonumber \\
    &\geq& (1-\epsilon^3) \cdot (1-\mu)^{kF-d} \cdot \left(   \psi_{u^B}  \cdot \mathcal{R}^{I^{\updownarrow}}_{\mytree}(\mathcal{T}_{\mathcal{H}^B}({u^B_L})) + (1-\psi_{u^B})  \cdot \mathcal{R}^{I^{\updownarrow}}_{\mytree}(\mathcal{T}_{\mathcal{H}^B}({u^B_R}))\right. \nonumber \\
    &\mbox{}& \qquad \qquad \qquad \qquad \qquad \qquad \left.\vphantom{\mathcal{R}^{I^{\updownarrow}}_{\mytree}(\mathcal{T}_{\mathcal{H}^B}({u^B_R}))} -  (M+1)(kF-d-1)\Delta \right) \ . \label{eq:ptas_exp_1}
\end{eqnarray}

\paragraph{Bounding (II).}
To obtain a lower-bound on expression $\text{(II)}$, we first argue that for every $q \leq |\myblock(\tilde{u}^B)|$, 
\begin{equation}
   \prod_{\hat{q}=1}^{q-1}\left(1-p^{\updownarrow}_{\myapp_{\hat{q}}(\tilde{u}^B)}\right) ~~\geq~~ 1-\epsilon^3 \ . \label{eq:ptas_bounding_2}
\end{equation}
For readability purposes, the proof for this inequality is deferred to Appendix~\ref{proof:rejection_prob_correct}.
As a result, the term (II) can be bounded by observing that
\begin{eqnarray}
    \text{(II)} &\geq& (1-\epsilon^3) \cdot \sum_{q=1}^{|\myblock(\tilde{u}^B)|} p^{\updownarrow}_{\myapp_q({\tilde{u}^B})} \cdot v^{\updownarrow}_{\myapp_q({\tilde{u}^B})} \nonumber \\
    &=&(1-\epsilon^3) \cdot \sum_{m \leq M} \mathcal{R}^{I^{\updownarrow}}_{\tilde{\mathcal{T}}}(m, \tilde{u}^B) \nonumber \\
    &\geq& (1-\epsilon^3) \cdot\sum_{m\leq M} \tilde{\mathcal{R}}_{m, u^B} \label{eq:ptas_approx_1} \\
    &\geq& (1-\epsilon^3) \cdot \sum_{m \leq M} \left(\mathcal{R}^{I^{\updownarrow}}_{\mathcal{T}_{\mathcal{H}^B}}(m, u^B) - \Delta \right) \label{eq:ptas_approx_2}\\
    &=& (1-\epsilon^3) \cdot \left( \sum_{q=1}^{|\myblock(u^B)|}p^{\updownarrow}_{\myapp_q(u^B)} \cdot v^{\updownarrow}_{\myapp_q(u^B)} - (M+1) \Delta \right)\ .    \label{eq:ptas_expr_2}
\end{eqnarray}
Here, inequality~\eqref{eq:ptas_approx_1} follows from our assignment rule, which guarantees that each class-to-block contribution satisfies $\mathcal{R}^{I^{\updownarrow}}_{\tilde{\mathcal{T}}}(m, \tilde{u}^B) \geq \tilde{\mathcal{R}}_{m, u^B}$.
Inequality~\eqref{eq:ptas_approx_2} holds since $\mathcal{R}^{I^{\updownarrow}}_{\mathcal{T}_{\mathcal{H}^B}}(m, u^B) \leq \tilde{\mathcal{R}}_{m, u^B} + \Delta$, by inequality~\eqref{eq:ptas_feasability}.

\paragraph{Conclusion.} By plugging inequalities~\eqref{eq:ptas_exp_1} and~\eqref{eq:ptas_expr_2} back into inequality~\eqref{eq:ptas_guarantee_middle}, we have
\begin{eqnarray}    \mathcal{R}^{I^{\updownarrow}}_{\mytree}(\tilde{\mathcal{T}}(\tilde{u}^B)) &\geq& (1-\epsilon^3) \cdot (1-\mu)^{kF-d} \cdot \left(\vphantom{\prod^{q-1}_{q-1}} \psi_{u^B}  \cdot \mathcal{R}^{I^{\updownarrow}}_{\mytree}(\mathcal{T}_{\mathcal{H}^B}({u^B_L})) +  (1-\psi_{u^B})  \cdot \mathcal{R}^{I^{\updownarrow}}_{\mytree}(\mathcal{T}_{\mathcal{H}^B}({u^B_R}))  \right. \nonumber \\
    &\mbox{}& \qquad \mbox{}  -  (M+1)(kF-d-1)\Delta +\sum_{q=1}^{|\myblock(u^B)|}p^{\updownarrow}_{\myapp_q(u^B)} \cdot v^{\updownarrow}_{\myapp_q(u^B)} - (M+1) \Delta \left.\vphantom{\prod^{q-1}_{q-1}}  \right) \nonumber \\
    &\geq& (1-\epsilon^3) \cdot (1-\mu)^{kF-d} \cdot \left(\vphantom{\prod^{q-1}_{q-1}} \psi_{u^B}  \cdot \mathcal{R}^{I^{\updownarrow}}_{\mytree}(\mathcal{T}_{\mathcal{H}^B}({u^B_L})) +  (1-\psi_{u^B})  \cdot \mathcal{R}^{I^{\updownarrow}}_{\mytree}(\mathcal{T}_{\mathcal{H}^B}({u^B_R}))  \right. \nonumber \\
    &\mbox{}& + \sum_{q=1}^{|\myblock(u^B)|} \left(\prod_{\hat{q}=1}^{q-1}(1- p^{\updownarrow}_{\myapp_{\hat{q}}(u^B)}) \right) \cdot  p^{\updownarrow}_{\myapp_q(u^B)} \cdot v^{\updownarrow}_{\myapp_q(u^B)}  -(M+1)(kF-d)\Delta \left.\vphantom{\prod^{q-1}_{q-1}}  \right) \nonumber \\
    &=& (1-\epsilon^3) \cdot (1-\mu)^{kF-d} \cdot \left(\mathcal{R}^{I^{\updownarrow}}_{\mytree}(\mathcal{T}_{\mathcal{H}^B}({u^B})) - (M+1)(kF-d)\Delta \right) \ . \nonumber
\end{eqnarray}

\addcontentsline{toc}{section}{Bibliography}
\bibliographystyle{plainnat}
\bibliography{BIB-Offering}

\changelocaltocdepth{1}
\appendix
\section{Additional Proofs from Section~\ref{sec:background}}

\subsection{Proof of Lemma~\ref{lemma:lr_and_vrl_modification}} \label{proof:lr_and_vrl_modification}
Given a valid decision tree $\mathcal{T}$, we explain how to compute a canonical tree $\tilde{\mathcal{T}}$ with an expected reward of $\mathcal{R}_{\mytree}(\tilde{\mathcal{T}}) \geq \mathcal{R}_{\mytree}(\mathcal{T})$.
It is worth pointing out that, by definition of the \lrproperty-property and the \vrlproperty-property, they can be jointly written as satisfying the next condition for every internal node $u$: 
\begin{equation}
    \mathcal{R}_{\mytree}(\mathcal{T}(u_R)) ~\leq~ \mathcal{R}_{\mytree}(\mathcal{T}(u_L)) ~\leq~ \mathcal{R}_{\mytree}(\mathcal{T}(u_R)) + v_{\myapp(u)} \text{ }. \label{eq:canonical}
\end{equation}
To construct $\tilde{\mathcal{T}}$, we sequentially alter $\mathcal{T}$ by traversing its nodes in depth-first order, ensuring that each internal node $u \in \mathcal{T}$ satisfies condition~\eqref{eq:canonical}.
Specifically, upon reaching a node $u$ for which this condition is not met, we alter the subtree $\mathcal{T}(u)$ according to cases $1$ and $2$ below. It is not difficult to verify that each such modification results in a valid tree, does not decrease its expected reward, and ensures that the nodes we have already traversed still satisfy \eqref{eq:canonical}.

\paragraph{Case 1: $\bs{\mathcal{R}_{\mytree}(\mathcal{T}(u_L)) <  \mathcal{R}_{\mytree}(\mathcal{T}(u_R))}$.} Here, we replace the left subtree $\mathcal{T}(u_L)$ of $u$ by its right subtree, $\mathcal{T}(u_R)$, as explained in Section~\ref{subsec:structural_properties}. 
We remind the reader that, letting $\mystate(u) = (t, k_t, \mathcal{A}_t)$, we have $\mystate(u_L) = (t+1, k_t, \mathcal{A}_t \setminus \{ \myapp(u)\})$ and $\mystate(u_R) = (t+1, k_t-1, \mathcal{A}_t \setminus \{ \myapp(u)\})$. 
As a result, replacing $\mathcal{T}(u_L)$ by $\mathcal{T}(u_R)$ produces a valid tree. 
Moreover, in the modified tree $\tilde{\mathcal{T}}$, the subtrees $\tilde{\mathcal{T}}(u_L)$ and $\tilde{\mathcal{T}}(u_R)$ are now identical, implying that $u$ indeed satisfies condition~\eqref{eq:canonical} as we are operating in depth-first order.
Since $\mathcal{R}_{\mytree}(\tilde{\mathcal{T}}(u_R)) = \mathcal{R}_{\mytree}(\mathcal{T}(u_R))$ and $\mathcal{R}_{\mytree}(\tilde{\mathcal{T}}(u_L)) >  \mathcal{R}_{\mytree}(\mathcal{T}(u_L))$, it follows that $\mathcal{R}_{\mytree}(\tilde{\mathcal{T}}) \geq \mathcal{R}_{\mytree}(\mathcal{T})$.

\paragraph{Case 2: $\bs{\mathcal{R}_{\mytree}(\mathcal{T}(u_L)) > \mathcal{R}_{\mytree}(\mathcal{T}(u_R)) + v_{\myapp(u)}}$.} In this scenario, we replace the entire subtree ${\cal T}( u )$ rooted at $u$ by its left subtree, ${\cal T}( u_L )$. 
Once again, this operation maintains a valid tree since, letting $\mystate(u) = (t, k_t, \mathcal{A}_t)$, we have $\mystate(u_L) = (t+1, k_t, \mathcal{A}_t \setminus \{ \myapp(u)\})$.
Moreover, given that we traverse the overall tree in depth-first order, the node $u_L$ satisfies condition~\eqref{eq:canonical} with respect to $\tilde{\mathcal{T}}$, and since $\mathcal{R}_{\mytree}(\tilde{\mathcal{T}}(u_L)) \geq \mathcal{R}_{\mytree}(\mathcal{T}(u))$, it follows that $\mathcal{R}_{\mytree}(\tilde{\mathcal{T}}) \geq \mathcal{R}_{\mytree}(\mathcal{T})$.

\section{Additional Proofs from Section~\ref{sec:quasi_ptas}}
\subsection{Proof of Lemma~\ref{lemma:rounding_up_down}} \label{app:proof_lemma_rounding_up_down}
Let $\mathcal{H}$ be a canonical policy for our original instance $I$ and let $\mathcal{T}_{\mathcal{H}}$ be its decision tree representation.
We will prove that, for every node $u \in \mathcal{T}_{\mathcal{H}}$, 
\begin{equation}
    \mathcal{R}_{\mytree}^{I^{\updownarrow}}(\mathcal{T}_{\mathcal{H}}(u)) ~~\geq~~ (1-\epsilon) \cdot (1 - \gamma)^{T - \mydepth_{\mathcal{T}_{\mathcal{H}}}(u)} \cdot \mathcal{R}_{\mytree}^{I}(\mathcal{T}_{\mathcal{H}}(u)) \text{ ,} \label{eq:lemma_3.2_1}
\end{equation}
by induction on the depth of $u$ in decreasing order.
Here, $\mydepth_{\mathcal{T}_{\mathcal{H}}}(u)$ designates the number of arcs connecting $\myroot(\mathcal{T}_{\mathcal{H}})$ to $u$.
As an immediate conclusion, we infer that 
\begin{eqnarray}
    \mathcal{R}^{I^{\updownarrow}}(\mathcal{H}) &~=~& \mathcal{R}^{I^{\updownarrow}}_{\mytree}(\myroot(\mathcal{T}_{\mathcal{H}})) \nonumber \\
    &\geq& (1-\epsilon) \cdot (1 - \gamma)^T \cdot \mathcal{R}^I(\mathcal{H}) \label{eq:app_3_1}\\
    &\geq& (1-\epsilon) \cdot (1 - \gamma T) \cdot \mathcal{R}^I(\mathcal{H}) \label{eq:app_3_2}\\
    &\geq& (1 - 2\eps) \cdot \mathcal{R}^I(\mathcal{H}) \ . \label{eq:app_3_3}
\end{eqnarray}
Here, to obtain inequality~\eqref{eq:app_3_1}, we instantiate \eqref{eq:lemma_3.2_1} with $u = \myroot(\mathcal{T}_{\mathcal{H}})$, noting that $\mydepth(\myroot(\mathcal{T}_{\mathcal{H}})) = 0$.
Inequality~\eqref{eq:app_3_2} follows from Bernoulli's inequality.
Finally, inequality~\eqref{eq:app_3_3} is derived by recalling that $\gamma = \frac{\eps}{T}$.

\paragraph{Base case: $\bs{u}$ is a leaf.} 
As explained in Section~\ref{subsec:trees}, every leaf node corresponds to a virtual applicant, whose value and acceptance probability are both 0.
As such, $\mathcal{R}_{\mytree}^{I^{\updownarrow}}(\mathcal{T}_{\mathcal{H}}(u))  = 0 = \mathcal{R}_{\mytree}^{I}(\mathcal{T}_{\mathcal{H}}(u))$.

\paragraph{Induction step: $\bs{u}$ is internal.} Now, suppose that $u$ is an internal node with $\mydepth_{\mathcal{T}_{\mathcal{H}}}(u) = d$, and 
let $u_L$ and $u_R$ be its left and right children, obviously with $\mydepth_{\mathcal{T}_{\mathcal{H}}}(u_L) = \mydepth_{\mathcal{T}_{\mathcal{H}}}(u_R) = d+1$.
We prove  inequality~\eqref{eq:lemma_3.2_1} by considering two cases, depending on how the acceptance probability of $\myapp(u)$ is related to our threshold parameter $\gamma$.
\begin{itemize}
    \item {\em When $p_{\myapp(u)} \leq \gamma$}:
    In this case, $\myapp(u) \in \mathcal{C}_0$. By definition, we know that $p^{\updownarrow}_{\myapp(u)} = \gamma$ and $v^{\updownarrow}_{\myapp(u)} = \frac{r_{\myapp(u)}}{\gamma}$. Therefore, by representation~\eqref{eq:reward_tree_recursive},
    \begin{eqnarray}        \mathcal{R}_{\mytree}^{I^{\updownarrow}}(\mathcal{T}_{\mathcal{H}}(u)) & = & r_{\myapp(u)} + \gamma \cdot \mathcal{R}^{I^{\updownarrow}}_{\mytree}(\mathcal{T}_{\mathcal{H}}(u_R)) + (1 - \gamma) \cdot \mathcal{R}^{I^{\updownarrow}}_{\mytree}(\mathcal{T}_{\mathcal{H}}(u_L)) \nonumber\\
     & \geq & (1-\epsilon) \cdot (1 - \gamma)^{T-(d+1)}  \nonumber \\
     && \qquad \cdot  \left(r_{\myapp(u)} + \gamma \cdot \mathcal{R}_{\mytree}^{I}(\mathcal{T}_{\mathcal{H}}(u_R)) +  (1 - \gamma) \cdot \mathcal{R}_{\mytree}^{I}(\mathcal{T}_{\mathcal{H}}(u_L)) \right) \label{eq:proof_3.2_1}\\
     & \geq &  (1-\epsilon) \cdot (1 - \gamma)^{T-d}  \nonumber\\
     && \qquad \mbox{} \cdot \left( r_{\myapp(u)} + \gamma \cdot \mathcal{R}_{\mytree}^{I}(\mathcal{T}_{\mathcal{H}}(u_R)) + \mathcal{R}_{\mytree}^{I}(\mathcal{T}_{\mathcal{H}}(u_L))\right) \nonumber\\
      & \geq & (1-\epsilon) \cdot (1 - \gamma)^{T-d} \cdot \left(r_{\myapp(u)} + p_{\myapp(u)} \cdot \mathcal{R}_{\mytree}^{I}(\mathcal{T}_{\mathcal{H}}(u_R)) \right. \nonumber \\ 
      && \qquad \qquad \qquad \qquad \qquad + \mbox{} \left. (1- p_{\myapp(u)}) \cdot \mathcal{R}_{\mytree}^{I}(\mathcal{T}_{\mathcal{H}}(u_L))\right) \label{eq:small_p} \\
      & = &(1-\epsilon) \cdot (1 - \gamma)^{T-d} \cdot \mathcal{R}_{\mytree}^{I}(\mathcal{T}_{\mathcal{H}}(u)) \nonumber \text{ },
    \end{eqnarray}
    where inequality~\eqref{eq:proof_3.2_1} follows from the induction hypothesis, and inequality~\eqref{eq:small_p} holds since $p_{\myapp(u)} \leq \gamma$, by the case hypothesis.
    
    \item {\em When $p_{\myapp(u)} > \gamma$}: In this case, we know that $\myapp(u) \in \mathcal{C}_m$ for some class $m \in [M]$, implying that $p^{\updownarrow}_{\myapp(u)} = (1+\eps)^{m-1} \cdot \gamma \leq p_{\myapp(u)} \leq (1+\eps)^{m} \cdot \gamma$ and that $v^{\updownarrow}_{\myapp(u)} = v_{\myapp(u)}$.
    In turn, $r_{\myapp(u)} \leq (1+\epsilon) \cdot r^{\updownarrow}_{\myapp(u)}$.
    Given these observations, by rearranging \eqref{eq:reward_tree_recursive}, we have
    \begin{align}
        \mathcal{R}_{\mytree}^{I}(\mathcal{T}_{\mathcal{H}}(u)) ~~&=~~ r_{\myapp(u)} +  p_{\myapp(u)} \cdot \left( \mathcal{R}_{\mytree}^{I}(\mathcal{T}_{\mathcal{H}}(u_R)) - \mathcal{R}_{\mytree}^{I}(\mathcal{T}_{\mathcal{H}}(u_L)) \right) \nonumber\\
        &\qquad+ \mathcal{R}_{\mytree}^{I}(\mathcal{T}_{\mathcal{H}}(u_L)) \nonumber \\
        &\leq~~ r_{\myapp(u)} +  p^{\updownarrow}_{\myapp(u)} \cdot \left( \mathcal{R}_{\mytree}^{I}(\mathcal{T}_{\mathcal{H}}(u_R)) - \mathcal{R}_{\mytree}^{I}(\mathcal{T}_{\mathcal{H}}(u_L)) \right) \nonumber \\
        &\qquad+ \mathcal{R}_{\mytree}^{I}(\mathcal{T}_{\mathcal{H}}(u_L)) \label{eq:proof_3.2_2} \\
        &\leq~~ (1+\eps) \cdot r^{\updownarrow}_{\myapp(u)} +  p^{\updownarrow}_{\myapp(u)} \cdot \mathcal{R}_{\mytree}^{I}(\mathcal{T}_{\mathcal{H}}(u_R)) \nonumber \\
        & \qquad +(1-p^{\updownarrow}_{\myapp(u)}) \cdot  \mathcal{R}_{\mytree}^{I}(\mathcal{T}_{\mathcal{H}}(u_L)) \label{eq:proof_3.2_3}\\
        &\leq~~ \frac{1}{(1-\epsilon)(1-\gamma)^{T-d}} \cdot \left(r^{\updownarrow}_{\myapp(u)} + p^{\updownarrow}_{\myapp(u)} \cdot \mathcal{R}_{\mytree}^{I^{\updownarrow}}(\mathcal{T}_{\mathcal{H}}(u_R))\right. \nonumber \\
        & \qquad \qquad \qquad \qquad \qquad \qquad \left. \mbox{} + (1 - p^{\updownarrow}_{\myapp(u)}) \cdot \mathcal{R}_{\mytree}^{I^{\updownarrow}}(\mathcal{T}_{\mathcal{H}}(u_L))  \right) \label{eq:proof_3.2_4}\\
        &=~~ \frac{ \mathcal{R}_{\mytree}^{I^{\updownarrow}}(\mathcal{T}_{\mathcal{H}}(u)) }{(1-\epsilon)(1-\gamma)^{T-d}} \ . \nonumber
    \end{align}
    Here, inequality~\eqref{eq:proof_3.2_2} is obtained by recalling that $\mathcal{H}$ is a canonical policy for $I$, implying that $\mathcal{R}_{\mytree}^{I}(\mathcal{T}_{\mathcal{H}}(u_R)) \leq \mathcal{R}_{\mytree}^{I}(\mathcal{T}_{\mathcal{H}}(u_L))$ by the \lrproperty-property.
    In addition, $p_{\myapp(u)} \geq p^{\updownarrow}_{\myapp(u)}$, as observed above.
    Inequality~\eqref{eq:proof_3.2_3} holds since $r_{\myapp(u)} \leq (1+\eps) \cdot r^{\updownarrow}_{\myapp(u)}$. 
    Finally, inequality~\eqref{eq:proof_3.2_4} follows from the induction hypothesis.
\end{itemize}

\subsection{Proof of Lemma~\ref{lemma:idu_smaller_than_i}}\label{proof:idu_smaller_than_i} 
Let $\mathcal{H}$ be a canonical policy with respect to the rounded instance $I^{\updownarrow}$.
For the purpose of showing that $\mathcal{R}^{I}_{\mytree}(\mathcal{T}_{\mathcal{H}}) \geq \mathcal{R}^{I^{\updownarrow}}_{\mytree}(\mathcal{T}_{\mathcal{H}})$, we will prove that $\mathcal{R}^{I}_{\mytree}(\mathcal{T}_{\mathcal{H}}(u)) \geq \mathcal{R}^{I^{\updownarrow}}_{\mytree}(\mathcal{T}_{\mathcal{H}}(u))$ for every node $u \in \mathcal{T}_{\mathcal{H}}$ by induction on $\mydepth_{\mathcal{T}_{\mathcal{H}}}(u)$.

\paragraph{Base case: $\bs{u}$ is a leaf.} 
As explained in Section~\ref{subsec:trees}, every leaf node corresponds to a virtual applicant, whose value and acceptance probability are both 0.
As such, $\mathcal{R}_{\mytree}^{I^{\updownarrow}}(\mathcal{T}_{\mathcal{H}}(u))  = 0 = \mathcal{R}_{\mytree}^{I}(\mathcal{T}_{\mathcal{H}}(u))$.

\paragraph{Induction step: $\bs{u}$ is internal.}
Similarly to the proof of Lemma~\ref{lemma:rounding_up_down}, suppose that $u$ is an internal node with $\mydepth_{\mathcal{T}_{\mathcal{H}}}(u) = d$, and 
let $u_L$ and $u_R$ be its left and right children, respectively, with $\mydepth_{\mathcal{T}_{\mathcal{H}}}(u_L) = \mydepth_{\mathcal{T}_{\mathcal{H}}}(u_R) = d+1$.
By recycling the notation of Appendix~\ref{app:proof_lemma_rounding_up_down}, we consider two cases, one where $p_{\myapp(u)} \leq \gamma$ and the other where $p_{\myapp(u)} > \gamma$.
\begin{itemize}
    \item {\em When $p_{\myapp(u)} \leq \gamma$}: 
    In this case, $\myapp(u) \in \mathcal{C}_0$. By definition, we know that $p^{\updownarrow}_{\myapp(u)} = \gamma$ and $v^{\updownarrow}_{\myapp(u)} = \frac{r_{\myapp(u)}}{\gamma}$. Therefore, according to representation~\eqref{eq:reward_tree_recursive}, 
        \begin{align}
        \mathcal{R}^{I^{\updownarrow}}_{\mytree}(\mathcal{T}_{\mathcal{H}}(u)) ~~&=~~ p^{\updownarrow}_{\myapp(u)} \cdot \left(v^{\updownarrow}_{\myapp(u)} + \mathcal{R}^{I^{\updownarrow}}_{\mytree}(\mathcal{T}_{\mathcal{H}}(u_R))\right) + (1 - p^{\updownarrow}_{\myapp(u)}) \cdot \mathcal{R}^{I^{\updownarrow}}_{\mytree}(\mathcal{T}_{\mathcal{H}}(u_L)) \nonumber \\
        &=~~ r_{\myapp(u)} + \gamma \cdot \left(\mathcal{R}^{I^{\updownarrow}}_{\mytree}(\mathcal{T}_{\mathcal{H}}(u_R)) - \mathcal{R}^{I^{\updownarrow}}_{\mytree}(\mathcal{T}_{\mathcal{H}}(u_L))\right) + \mathcal{R}^{I^{\updownarrow}}_{\mytree}(\mathcal{T}_{\mathcal{H}}(u_L))  \nonumber\\
        &\leq~~ r_{\myapp(u)} + p_{\myapp(u)} \cdot \left(\mathcal{R}^{I^{\updownarrow}}_{\mytree}(\mathcal{T}_{\mathcal{H}}(u_R)) - \mathcal{R}^{I^{\updownarrow}}_{\mytree}(\mathcal{T}_{\mathcal{H}}(u_L))\right) \nonumber \\
        & \qquad + \mathcal{R}^{I^{\updownarrow}}_{\mytree}(\mathcal{T}_{\mathcal{H}}(u_L)) \label{eq:quasi_3}\\
        &\leq~~ r_{\myapp(u)} + p_{\myapp(u)} \cdot \mathcal{R}^{I}_{\mytree}(\mathcal{T}_{\mathcal{H}}(u_R)) + (1 - p_{\myapp(u)}) \cdot \mathcal{R}^{I}_{\mytree}(\mathcal{T}_{\mathcal{H}}(u_L)) \label{eq:quasi_5}\\
        &=~~ \mathcal{R}^{I}_{\mytree}(\mathcal{T}_{\mathcal{H}}(u)) \text{ }. \nonumber
    \end{align}
    Here, inequality~\eqref{eq:quasi_3} is obtained by recalling that $\mathcal{H}$ is canonical for $I^{\updownarrow}$, implying that $\mathcal{R}_{\mytree}^{I^{\updownarrow}}(\mathcal{T}_{\mathcal{H}}(u_R)) \leq \mathcal{R}_{\mytree}^{I^{\updownarrow}}(\mathcal{T}_{\mathcal{H}}(u_L))$ by the \lrproperty-property.
    In addition, $p_{\myapp(u)} \leq \gamma$, by the case hypothesis. Inequality~\eqref{eq:quasi_5} follows from the induction hypothesis.

    \item {\em When $p_{\myapp(u)} > \gamma$}: In this case, we know that $\myapp(u) \in \mathcal{C}_m$ for some class $m \in [M]$, implying that $p^{\updownarrow}_{\myapp(u)} = (1+\eps)^{m-1} \cdot \gamma \leq p_{\myapp(u)} \leq (1+\eps)^{m} \cdot \gamma$, and that $v^{\updownarrow}_{\myapp(u)} = v_{\myapp(u)}$.
    Given these observations, by representation~\eqref{eq:reward_tree_recursive}, 
    \begin{align}
    \mathcal{R}^{I^{\updownarrow}}_{\mytree}(\mathcal{T}_{\mathcal{H}}(u)) ~~&=~~ p^{\updownarrow}_{\myapp(u)} \cdot \left(v^{\updownarrow}_{\myapp(u)} + \mathcal{R}^{I^{\updownarrow}}_{\mytree}(\mathcal{T}_{\mathcal{H}}(u_R)) - \mathcal{R}^{I^{\updownarrow}}_{\mytree}(\mathcal{T}_{\mathcal{H}}(u_L))\right) + \mathcal{R}^{I^{\updownarrow}}_{\mytree}(\mathcal{T}_{\mathcal{H}}(u_L)) \nonumber\\
        &\leq~~ p_{\myapp(u)} \cdot \left(v_{\myapp(u)} + \mathcal{R}^{I^{\updownarrow}}_{\mytree}(\mathcal{T}_{\mathcal{H}}(u_R)) - \mathcal{R}^{I^{\updownarrow}}_{\mytree}(\mathcal{T}_{\mathcal{H}}(u_L))\right) \nonumber \\
        & \qquad + \mathcal{R}^{I^{\updownarrow}}_{\mytree}(\mathcal{T}_{\mathcal{H}}(u_L)) \label{eq:quasi_7}\\
        &=~~ r_{\myapp(u)} + p_{\myapp(u)} \cdot \mathcal{R}^{I^{\updownarrow}}_{\mytree}(\mathcal{T}_{\mathcal{H}}(u_R)) + (1 - p_{\myapp(u)}) \cdot \mathcal{R}^{I^{\updownarrow}}_{\mytree}(\mathcal{T}_{\mathcal{H}}(u_L)) \nonumber\\
        &\leq~~ r_{\myapp(u)} + p_{\myapp(u)} \cdot\mathcal{R}^{I}_{\mytree}(\mathcal{T}_{\mathcal{H}}(u_R)) + (1 - p_{\myapp(u)}) \cdot \mathcal{R}^{I}_{\mytree}(\mathcal{T}_{\mathcal{H}}(u_L)) \label{eq:quasi_9}\\
        &=~~ \mathcal{R}^{I}_{\mytree}(\mathcal{T}_{\mathcal{H}}(u)) \text{ }. \nonumber
    \end{align}
    Here, inequality~\eqref{eq:quasi_7} is obtained by recalling that $\mathcal{H}$ is canonical for $I^{\updownarrow}$, implying that $v^{\updownarrow}_{\myapp(u)} + \mathcal{R}_{\mytree}^{I^{\updownarrow}}(\mathcal{T}_{\mathcal{H}}(u_R)) \geq \mathcal{R}_{\mytree}^{I^{\updownarrow}}(\mathcal{T}_{\mathcal{H}}(u_L))$ by the \vrlproperty-property.
    In addition, $p^{\updownarrow}_{\myapp(u)} \leq p_{\myapp(u)}$ and $v^{\updownarrow}_{\myapp(u)} = v_{\myapp(u)}$, as observed above.    Inequality~\eqref{eq:quasi_9} follows from the induction hypothesis.
\end{itemize}

\subsection{Proof of Lemma~\ref{lemma:optimal_Order_by_value}}\label{proof:optimal_order_by_value}
For any policy $\mathcal{H}$, we say that a node $u$ of its decision tree $\mathcal{T}_{\mathcal{H}}$ violates the order-by-value rule when the applicant chosen at $u$ fails to satisfy this rule.
Formally, assuming that $\mystate(u) = (t, k_t, \mathcal{A}_t)$ and $\myapp(u) \in \mathcal{C}_m$, then $u$ is a violator when either $v^{\updownarrow}_{\myapp(u)} < \max_{i \in \mathcal{A}_t \cap  \mathcal{C}_m} v^{\updownarrow}_i$ or, among the applicants in $\mathcal{A}_t \cap  \mathcal{C}_m$ attaining the latter maximum, $\myapp(u)$ is not the lexicographically first.

With this terminology, across all optimal policies, let $\mathcal{H}^*$ be one where the first violator in a BFS-scan (starting at the root) appears as late as possible.
When $\mathcal{T}_{\mathcal{H}^*}$ does not contain any violator, $\mathcal{H}^*$ is already an optimal policy that satisfies the order-by-value rule.
In the opposite case, we will construct a different optimal policy, $\tilde{\mathcal{H}}$, whose first BFS-scan violator occurs strictly later than that of $\mathcal{H}^*$, thus contradicting the choice of $\mathcal{H}^*$.

To this end, let $u$ be the above-mentioned violator. 
For convenience, suppose that $\myapp(u) = i^- \in \mathcal{C}_m$, and let $i^+ \in \mathcal{C}_m$ be the applicant that should have been chosen at node $u$ under the order-by-value rule.
By our definition of a violator, $v^{\updownarrow}_{i^+} \geq v^{\updownarrow}_{i^-}$, and since both $i^+$ and $i^-$ belong to class $\mathcal{C}_m$, they share the same acceptance probability, say $p$.
We now define a modified policy $\tilde{\mathcal{H}}$ by altering the decision tree $\mathcal{T}_{\mathcal{H}^*}$, touching only the subtree $\mathcal{T}_{\mathcal{H}^*}(u)$ rooted at the violator $u$. Specifically, at node $u$, the selected applicant $i^-$ is replaced by $i^+$. Then, on every downward path connecting $u$ to a leaf, whenever $\mathcal{H}^*$ selects $i^+$, our modified policy $\tilde{\mathcal{H}}$ selects $i^-$.

By construction, it is easy to verify that $u$ is no longer a violator of $\mathcal{T}_{\tilde{\mathcal{H}}}$, and moreover, no earlier node in BFS-order turned into a violator.
Consequently, it remains to show that $\tilde{\mathcal{H}}$ is an optimal policy.
Noting that the decision trees $\mathcal{T}_{\mathcal{H}^*}$ and $\mathcal{T}_{\tilde{\mathcal{H}}}$ only differ in their subtrees $\mathcal{T}_{\mathcal{H}^*}(u)$ and $\mathcal{T}_{\tilde{\mathcal{H}}}(u)$, it suffices to show that $\mathcal{R}^{I^{\updownarrow}}_{\mytree}(\mathcal{T}_{\tilde{\mathcal{H}}}(u)) \geq \mathcal{R}^{I^{\updownarrow}}_{\mytree}(\mathcal{T}_{\mathcal{H}^*}(u))$.
To this end, let $p_R$ and $p_L$ be the probabilities that applicant $i^+$ is extended an offer with respect to $\mathcal{T}_{\mathcal{H}^*}(u_R)$ and $\mathcal{T}_{\mathcal{H}^*}(u_L)$.
Since $i^+$ and $i^-$ are replaced between $\mathcal{T}_{\mathcal{H}^*}(u)$ and $\mathcal{T}_{\tilde{\mathcal{H}}}(u)$, it follows that $p_R$ and $p_L$ are also the probabilities that $i^-$ is extended an offer in the corresponding subtrees of $\mathcal{T}_{\tilde{\mathcal{H}}}$.
Using this notation, we observe that
\begin{eqnarray*}    \mathcal{R}^{I^{\updownarrow}}_{\mytree}(\mathcal{T}_{\tilde{\mathcal{H}}}(u)) - \mathcal{R}^{I^{\updownarrow}}_{\mytree}(\mathcal{T}_{\mathcal{H}^*}(u))  & = & \left( p \cdot (v^{\updownarrow}_{i^+} +  p_R \cdot p \cdot v^{\updownarrow}_{i^-}) + (1-p) \cdot p_L \cdot p \cdot v^{\updownarrow}_{i^-} \right)  \\
&& \mbox{} - \left( p \cdot (v^{\updownarrow}_{i^-} + p_R \cdot p \cdot v^{\updownarrow}_{i^+}) + (1-p) \cdot p_L \cdot p \cdot v^{\updownarrow}_{i^+} \right)  \\
& = & (v^{\updownarrow}_{i^+} - v^{\updownarrow}_{i^-}) \cdot p \cdot (1 - (p \cdot p_R + (1-p) \cdot p_L)) \\
& \geq & 0 \ . 
\end{eqnarray*}
To obtain the last inequality, note that the term $p \cdot p_R + (1-p) \cdot p_L$ is a convex combination of $p_R$ and $p_L$ and, thus lying in $[0,1]$. Since $v^{\updownarrow}_{i^+} \geq v^{\updownarrow}_{i^-}$, the entire expression is non-negative.

\subsection{Proof of Lemma~\ref{lemma:dynamic_canonical}} \label{proof:dynamic_canonical}
To argue that $\mathcal{H}^{\updownarrow}$ is canonical for $I^{\updownarrow}$, consider any node $u$ of the decision tree $\mathcal{T}_{\mathcal{H}^{\updownarrow}}$, and let $(t, k_t, \mathcal{A}_t)$ be its state.
Recalling that the policy $\mathcal{H}^{\updownarrow}$ was generated by our dynamic program, we know that $\mathcal{A}_t$ cannot be some arbitrary subset of applicants. 
Rather, out of each class $\mathcal{C}_m$, this set must be comprised of the applicants $\mathcal{C}_m[\ell_{t, m}+1], \mathcal{C}_m[\ell_{t, m}+2], \dots$, for some $0 \leq \ell_{t, m} \leq |\mathcal{C}_m|$.
This structural property ensures that $\mathcal{A}_t$ corresponds to a unique vector $L_t = (\ell_{t, 0}, \dots, \ell_{t, M})$, where each $\ell_{t, m}$ represents the number of $\mathcal{C}_m$-applicants selected in earlier stages.
With this correspondence, we have $\mathcal{R}^{I^{\updownarrow}}_{\mytree}(\mathcal{T}_{\mathcal{H}^{\updownarrow}}(u)) = F(t, k_t, L_t)$, implying that $\mathcal{T}_{\mathcal{H}^{\updownarrow}}(u)$ is an optimal subtree at $\mystate(u)$, since $F$ attains the maximum-possible expected reward at each recursion depth.
The same observation applies to the left and right subtrees, i.e., $\mathcal{R}^{I^{\updownarrow}}_{\mytree}(\mathcal{T}_{\mathcal{H}^{\updownarrow}}(u_L)) = F(t+1, k_t, L_t+\mathrm{e}_m)$ and $\mathcal{R}^{I^{\updownarrow}}_{\mytree}(\mathcal{T}_{\mathcal{H}^{\updownarrow}}(u_R)) = F(t+1, k_t-1, L_t+\mathrm{e}_m)$, for some $0 \leq m \leq M$.

We proceed by proving that $\mathcal{H}^{\updownarrow}$ is indeed canonical, separately considering the two required properties.
\begin{itemize}
    \item {\em The \lrproperty-property}:
    Recall that $F(t+1, k_t, L_t+\mathrm{e}_m)$ represents the maximum expected reward attainable by any order-by-value policy at state $(t+1, k_t, L_t+\mathrm{e}_m)$.
    Since every policy valid at $(t+1, k_t-1, L_t+\mathrm{e}_m)$ is also valid at $(t+1, k_t, L_t+\mathrm{e}_m)$, we have 
    $F(t+1, k_t, L_t+\mathrm{e}_m) \geq F(t+1, k_t-1, L_t+\mathrm{e}_m)$, and therefore, $\mathcal{R}^{I^{\updownarrow}}_{\mytree}(\mathcal{T}_{\mathcal{H}^{\updownarrow}}(u_L)) \geq \mathcal{R}^{I^{\updownarrow}}_{\mytree}(\mathcal{T}_{\mathcal{H}^{\updownarrow}}(u_R))$.

    \item {\em The \vrlproperty-property}:
    Let $\mathcal{C}_m$ be the class to which $\myapp(u)$ belongs, implying that 
    \begin{eqnarray*}
    F(t, k_t, L_t) & = &   p^{\updownarrow}_{\mathcal{C}_{m}[{\ell}_{t, m}+1]} \cdot (v^{\updownarrow}_{\mathcal{C}_{m}[{\ell}_{t, m}+1]} + F(t+1, k_t - 1, L_t  +\mathrm{e}_m)) \\
    && \mbox{} + (1 - p^{\updownarrow}_{\mathcal{C}_{m}[{\ell}_{t, m}+1]}) \cdot F(t+1, k_t, L_t +\mathrm{e}_m) \ .     
    \end{eqnarray*}
    
    Since every policy valid at state $(t+1, k_t, L_t+\mathrm{e}_m)$ is also valid at $(t, k_t, L_t+\mathrm{e}_m)$, we have $F(t, k_t, L_t) \geq F(t+1, k_t, L_t+\mathrm{e}_m)$.
    Consequently,
    \begin{eqnarray}
    \mathcal{R}^{I^{\updownarrow}}_{\mytree}(\mathcal{T}_{\mathcal{H}^{\updownarrow}}(u_L)) &=& F(t+1, k_t, L_t+\mathrm{e}_m) \nonumber \\
      &\leq& F(t, k_t, L_t) \nonumber \\
      &=& p^{\updownarrow}_{\mathcal{C}_{m}[{\ell}_{t, m}+1]} \cdot (v^{\updownarrow}_{\mathcal{C}_{m}[{\ell}_{t, m}+1]} + F(t+1, k_t - 1, L_t +\mathrm{e}_m)) \nonumber \\
      && \mbox{} + (1 - p^{\updownarrow}_{\mathcal{C}_{m}[{\ell}_{t, m}+1]}) \cdot F(t+1, k_t, L_t +\mathrm{e}_m) \nonumber \\
      &=& p^{\updownarrow}_{\mathcal{C}_{m}[{\ell}_{t, m}+1]} \cdot (v^{\updownarrow}_{\mathcal{C}_{m}[{\ell}_{t, m}+1]} + \mathcal{R}^{I^{\updownarrow}}_{\mytree}(\mathcal{T}_{\mathcal{H}^{\updownarrow}}(u_R))) \nonumber \\
      && \mbox{} + (1 - p^{\updownarrow}_{\mathcal{C}_{m}[{\ell}_{t, m}+1]}) \cdot \mathcal{R}^{I^{\updownarrow}}_{\mytree}(\mathcal{T}_{\mathcal{H}^{\updownarrow}}(u_L)) \ , \nonumber      
    \end{eqnarray}
and by rearranging this inequality,
\[ v^{\updownarrow}_{\mathcal{C}_{m}[{\ell}_{t, m}+1]} + \mathcal{R}^{I^{\updownarrow}}_{\mytree}(\mathcal{T}_{\mathcal{H}^{\updownarrow}}(u_R)) ~~\geq~~ \mathcal{R}^{I^{\updownarrow}}_{\mytree}(\mathcal{T}_{\mathcal{H}^{\updownarrow}}(u_L)) \ . \]
\end{itemize}

\section{Additional Proofs from Section~\ref{sec:block_responsive}}

\subsection{Proof of Lemma~\ref{lemma:block_lr_and_vrl_modification}} \label{proof:block_lr_and_vrl_modification}

Given a block-responsive decision tree $\mathcal{T}_{\mathcal{H}^B}$ consisting of $N$ nodes, we explain how to compute in $O(NT)$ time a canonical tree $\mathcal{T}_{\mathcal{\tilde{H}}^B}$ with an expected reward of $\mathcal{R}_{\mytree}(\mathcal{T}_{\mathcal{\tilde{H}}^B}) \geq \mathcal{R}_{\mytree}(\mathcal{T}_{\mathcal{H}^B})$.
It is worth pointing out that, by definition of the \lrproperty-block-property and the \vrlproperty-block-property, they can be jointly written as satisfying the next condition for every internal node $u$ and rank $1 \leq r \leq |\myblock(u)|$: 
\begin{equation}
    \mathcal{R}_{\mytree}(\mathcal{T}_{\mathcal{H}^B}(u_R)) ~~\leq~~ \mathcal{R}_{\mytree}(\mathcal{T}_{\mathcal{H}^B}(u), r+1) ~~\leq~~ v_{\myapp_r(u)}+\mathcal{R}_{\mytree}(\mathcal{T}_{\mathcal{H}^B}(u_R)) \text{ }. \label{eq:canonical_block}
\end{equation}
To construct $\mathcal{T}_{\mathcal{\tilde{H}}^B}$, we sequentially alter $\mathcal{T}_{\mathcal{H}^B}$ by traversing its nodes in depth-first order, ensuring that each internal node $u$ and its associated applicants satisfy condition~\eqref{eq:canonical_block}.
Within each such node, we traverse applicants in reverse order, starting with $\myapp_{|\myblock(u)|}(u)$ and proceeding toward $\myapp_{1}(u)$.
When this condition is violated, we alter the subtree $\mathcal{T}_{\mathcal{H}^B}(u)$ according to cases $1$ and $2$ below.
It is not difficult to verify that each such modification results in a valid tree, does not decrease its expected reward, and ensures that the nodes we have already traversed still satisfy \eqref{eq:canonical_block}.
As far as running time is concerned, there are at most $T$ applicants in each block, and therefore, the entire tree can be traversed in $O(N T)$ iterations. As we proceed to explain, each modification can be implemented in $O(1)$ time, meaning that the overall running time of our algorithm is $O(N T)$.

\paragraph{Case 1: $\bs{\mathcal{R}_{\mytree}(\mathcal{T}_{\mathcal{H}^B}(u), r+1) <  \mathcal{R}_{\mytree}(\mathcal{T}_{\mathcal{H}^B}(u_R))}$.}
When $r$ coincides with the highest rank in $\myblock(u)$, i.e., $r = |\myblock(u)|$, we have $\mathcal{R}_{\mytree}(\mathcal{T}_{\mathcal{H}^B}(u), r+1) = \mathcal{R}_{\mytree}(\mathcal{T}_{\mathcal{H}^B}(u_L))$ by definition.
Here, we replace the left subtree $\mathcal{T}_{\mathcal{H}^B}(u_L)$ of $u$ by its right subtree, $\mathcal{T}_{\mathcal{H}^B}(u_R)$, exactly as in case $1$ of Appendix~\ref{proof:lr_and_vrl_modification} for standard decision trees. 
In the complementary case, where $r <|\myblock(u)|$, we observe that due to traversing the applicants of $\myblock(u)$ in reverse order, we have already ensured that $\mathcal{R}_{\mytree}(\mathcal{T}_{\mathcal{\tilde{H}}^B}(u), r+2) \geq  \mathcal{R}_{\mytree}(\mathcal{T}_{\mathcal{\tilde{H}}^B}(u_R))$.
As such, without further modifications, we actually have a stronger inequality, since
\begin{eqnarray}
    \mathcal{R}_{\mytree}(\mathcal{T}_{\mathcal{\tilde{H}}^B}(u), r+1) &=& p_{\myapp_{r+1}(u)} \cdot (v_{\myapp_{r+1}(u)} + \mathcal{R}_{\mytree}(\mathcal{T}_{\mathcal{\tilde{H}}^B}(u_R))) \nonumber \\
    && \mbox{}+ (1 - p_{\myapp_{r+1}(u)}) \cdot \mathcal{R}_{\mytree}(\mathcal{T}_{\mathcal{\tilde{H}}^B}(u), r+2) \nonumber \\
    &\geq&\mathcal{R}_{\mytree}(\mathcal{T}_{\mathcal{\tilde{H}}^B}(u_R)) + p_{\myapp_{r+1}(u)} \cdot v_{\myapp_{r+1}(u)} \ .  \nonumber
\end{eqnarray}

\paragraph{Case 2: $\bs{\mathcal{R}_{\mytree}(\mathcal{T}_{\mathcal{H}^B}(u), r+1) > v_{\myapp_r(u)}+\mathcal{R}_{\mytree}(\mathcal{T}_{\mathcal{H}^B}(u_R))}$.}
In this scenario, we simply eliminate $\myapp_r(u)$ from $\myblock(u)$.
In the event where $\myblock(u)$ becomes empty, to maintain structural integrity, we connect $u$'s left child, $u_L$, directly to $u$'s parent node and delete its right subtree $\mathcal{T}_{\mathcal{H}^B}(u_R)$.
Following this modification, $\mathcal{R}_{\mytree}(\mathcal{T}_{\mathcal{\tilde{H}}^B}(u), r) = \mathcal{R}_{\mytree}(\mathcal{T}_{\mathcal{H}^B}(u), r+1)$, due to eliminating $\myapp_r(u)$ from $\myblock(u)$.
Concurrently, in terms of expected reward,
\begin{eqnarray}
    \mathcal{R}_{\mytree}(\mathcal{T}_{\mathcal{\tilde{H}}^B}(u), r) &=& \mathcal{R}_{\mytree}(\mathcal{T}_{\mathcal{H}^B}(u), r+1) \nonumber\\
    &=& p_{\myapp_r(u)} \cdot \mathcal{R}_{\mytree}(\mathcal{T}_{\mathcal{H}^B}(u), r+1) + (1-p_{\myapp_r(u)}) \cdot \mathcal{R}_{\mytree}(\mathcal{T}_{\mathcal{H}^B}(u), r+1) \nonumber \\
    &>&p_{\myapp_r(u)} \cdot (v_{\myapp_r(u)}+\mathcal{R}_{\mytree}(\mathcal{T}_{\mathcal{H}^B}(u_R)))\nonumber\\
    && \mbox{}+ (1-p_{\myapp_r(u)}) \cdot \mathcal{R}_{\mytree}(\mathcal{T}_{\mathcal{H}^B}(u), r+1) \nonumber \\
    &=&\mathcal{R}_{\mytree}(\mathcal{T}_{\mathcal{H}^B}(u), r) \ , \nonumber 
\end{eqnarray}
where the sole inequality above follows from our case hypothesis.

\subsection{Proof of Lemma~\ref{lemma:block_canonical_modifying_effects}} \label{proof:block_canonical_modifying_effects}

Let $\mathcal{H}^B$ be a canonical block-responsive policy with respect to the rounded instance $I^{\updownarrow}$.
For every node $u \in \mathcal{T}_{\mathcal{H}^B}$ and rank $1 \leq r \leq |\myblock(u)|$, we define the ``depth'' of applicant $\myapp_r(u)$ as the maximum number of applicants who could have been offered positions up to and including $\myapp_r(u)$.
In other words, when $\mystate(u) = (t_u, k_u, \mathcal{A}_u)$, we have $\mydepth_{_{\mathcal{T}_{\mathcal{H}^B}}}(\myapp_r(u)) = t_u +r$.
For the purpose of proving that $\mathcal{R}^{I}_{\mytree}(\mathcal{T}_{\mathcal{H}^B}) \geq \mathcal{R}^{I^{\updownarrow}}_{\mytree}(\mathcal{T}_{\mathcal{H}^B})$, we will show that $\mathcal{R}^{I}_{\mytree}(\mathcal{T}_{\mathcal{H}^B}(u), r) \geq \mathcal{R}^{I^{\updownarrow}}_{\mytree}(\mathcal{T}_{\mathcal{H}^B}(u), r)$ for every node $u$ and rank $1 \leq r \leq |\myblock(u)|$.
Our proof works by induction on $\mydepth_{\mathcal{T}_{\mathcal{H}^B}}(\myapp_r(u))$ in decreasing order.

\paragraph{Base case: $\bs{u}$ is a leaf.}
As explained in Section~\ref{subsec:block_responsive}, every leaf node is treated as a virtual block, $B_0 = \emptyset$.
As such, the desired inequality trivially follows.

\paragraph{Induction step: $\bs{u}$ is internal.}
For an internal node $u$ and rank $1 \leq r \leq |\myblock(u)|$, suppose that $\mydepth_{_{\mathcal{T}_{\mathcal{H}^B}}}(\myapp_r(u)) = d$.
Clearly, $\mydepth_{_{\mathcal{T}_{\mathcal{H}^B}}}(\myapp_{r+1}(u)) >d$ and $\mydepth_{_{\mathcal{T}_{\mathcal{H}^B}}}(u_R) >d$, meaning that by the induction hypothesis, $\mathcal{R}^{I}_{\mytree}(\mathcal{T}_{\mathcal{H}^B}(u), r+1) \geq \mathcal{R}^{I^{\updownarrow}}_{\mytree}(\mathcal{T}_{\mathcal{H}^B}(u), r+1)$ and $\mathcal{R}^{I}_{\mytree}(\mathcal{T}_{\mathcal{H}^B}(u_R)) \geq \mathcal{R}^{I^{\updownarrow}}_{\mytree}(\mathcal{T}_{\mathcal{H}^B}(u_R))$.
We consider two cases, one where $p_{\myapp_r(u)} \leq \gamma$ and the other where $p_{\myapp_r(u)} > \gamma$.
\begin{itemize}
    \item {\em When $p_{\myapp_r(u)} \leq \gamma$}:
    In this case, $\myapp_r(u) \in \mathcal{C}_0$. By definition, we know that $p^{\updownarrow}_{\myapp_r(u)} = \gamma$ and $v^{\updownarrow}_{\myapp_r(u)} = \frac{r_{\myapp_r(u)}}{\gamma}$. Therefore,
    \begin{align}
        \mathcal{R}^{I^{\updownarrow}}_{\mytree}(\mathcal{T}_{\mathcal{H}^B}(u), r) ~~&=~~ p^{\updownarrow}_{\myapp_r(u)} \cdot \left(v^{\updownarrow}_{\myapp_r(u)} + \mathcal{R}^{I^{\updownarrow}}_{\mytree}(\mathcal{T}_{\mathcal{H}^B}(u_R))\right)\nonumber\\
        & \qquad +(1 - p^{\updownarrow}_{\myapp_r(u)}) \cdot \mathcal{R}^{I^{\updownarrow}}_{\mytree}(\mathcal{T}_{\mathcal{H}^B}(u),r+1) \nonumber \\
        &=~~ r_{\myapp_r(u)} + \gamma \cdot \left(\mathcal{R}^{I^{\updownarrow}}_{\mytree}(\mathcal{T}_{\mathcal{H}^B}(u_R)) - \mathcal{R}^{I^{\updownarrow}}_{\mytree}(\mathcal{T}_{\mathcal{H}^B}(u),r+1)\right) \nonumber\\
        & \qquad + \mathcal{R}^{I^{\updownarrow}}_{\mytree}(\mathcal{T}_{\mathcal{H}^B}(u), r+1)  \nonumber\\
        &\leq~~ r_{\myapp_r(u)} + p_{\myapp_r(u)} \cdot \left(\mathcal{R}^{I^{\updownarrow}}_{\mytree}(\mathcal{T}_{\mathcal{H}^B}(u_R)) - \mathcal{R}^{I^{\updownarrow}}_{\mytree}(\mathcal{T}_{\mathcal{H}^B}(u), r+1)\right) \nonumber \\
        & \qquad + \mathcal{R}^{I^{\updownarrow}}_{\mytree}(\mathcal{T}_{\mathcal{H}^B}(u), r+1) \label{eq:block_1}\\
        &\leq~~ r_{\myapp_r(u)} + p_{\myapp_r(u)} \cdot \mathcal{R}^{I}_{\mytree}(\mathcal{T}_{\mathcal{H}^B}(u_R)) \nonumber\\
        & \qquad + (1 - p_{\myapp_r(u)}) \cdot \mathcal{R}^{I}_{\mytree}(\mathcal{T}_{\mathcal{H}^B}(u), r+1) \label{eq:block_2}\\
        &=~~ \mathcal{R}^{I}_{\mytree}(\mathcal{T}_{\mathcal{H}^B}(u),r) \text{ }. \nonumber
    \end{align}
    Here, inequality~\eqref{eq:block_1} is obtained by recalling that $\mathcal{H}^B$ is canonical for $I^{\updownarrow}$, implying that $\mathcal{R}_{\mytree}^{I^{\updownarrow}}(\mathcal{T}_{\mathcal{H}^B}(u_R)) \leq \mathcal{R}_{\mytree}^{I^{\updownarrow}}(\mathcal{T}_{\mathcal{H}}(u), r+1)$ by the \lrproperty-block-property.
    In addition, $p_{\myapp_r(u)} \leq \gamma$, by our case hypothesis. Inequality~\eqref{eq:block_2} follows from the induction hypothesis.

    \item {\em When $p_{\myapp_r(u)} > \gamma$}:
    In this case, we know that $\myapp_r(u) \in \mathcal{C}_m$ for some class $m \in [M]$, implying that $p^{\updownarrow}_{\myapp_r(u)} = (1+\eps)^{m-1} \cdot \gamma \leq p_{\myapp_r(u)} \leq (1+\eps)^{m} \cdot \gamma$, and that $v^{\updownarrow}_{\myapp_r(u)} = v_{\myapp_r(u)}$.
    Given these observations,
    \begin{align}
        \mathcal{R}^{I^{\updownarrow}}_{\mytree}(\mathcal{T}_{\mathcal{H}^B}(u), r) ~~&=~~ p^{\updownarrow}_{\myapp_r(u)} \cdot \left(v^{\updownarrow}_{\myapp_r(u)} + \mathcal{R}^{I^{\updownarrow}}_{\mytree}(\mathcal{T}_{\mathcal{H}^B}(u_R)) - \mathcal{R}^{I^{\updownarrow}}_{\mytree}(\mathcal{T}_{\mathcal{H}^B}(u), r+1)\right) \nonumber\\
        & \qquad + \mathcal{R}^{I^{\updownarrow}}_{\mytree}(\mathcal{T}_{\mathcal{H}^B}(u), r+1) \nonumber\\
        &\leq~~ p_{\myapp_r(u)} \cdot \left(v_{\myapp_r(u)} + \mathcal{R}^{I^{\updownarrow}}_{\mytree}(\mathcal{T}_{\mathcal{H}^B}(u_R)) - \mathcal{R}^{I^{\updownarrow}}_{\mytree}(\mathcal{T}_{\mathcal{H}^B}(u), r+1)\right) \nonumber \\
        & \qquad + \mathcal{R}^{I^{\updownarrow}}_{\mytree}(\mathcal{T}_{\mathcal{H}^B}(u), r+1) \label{eq:block_3}\\
        &=~~ r_{\myapp_r(u)} + p_{\myapp_r(u)} \cdot \mathcal{R}^{I^{\updownarrow}}_{\mytree}(\mathcal{T}_{\mathcal{H}^B}(u_R)) \nonumber\\
        & \qquad + (1 - p_{\myapp_r(u)}) \cdot \mathcal{R}^{I^{\updownarrow}}_{\mytree}(\mathcal{T}_{\mathcal{H}^B}(u), r+1) \nonumber\\
        &\leq~~ r_{\myapp_r(u)} + p_{\myapp_r(u)} \cdot\mathcal{R}^{I}_{\mytree}(\mathcal{T}_{\mathcal{H}^B}(u_R)) \nonumber\\
        & \qquad + (1 - p_{\myapp_r(u)}) \cdot \mathcal{R}^{I}_{\mytree}(\mathcal{T}_{\mathcal{H}^B}(u), r+1) \label{eq:block_4}\\
        &=~~ \mathcal{R}^{I}_{\mytree}(\mathcal{T}_{\mathcal{H}^B}(u), r) \text{ }. \nonumber
    \end{align}
    Here, inequality~\eqref{eq:block_3} is obtained by recalling that $\mathcal{H}^B$ is canonical for $I^{\updownarrow}$, implying that $v^{\updownarrow}_{\myapp_r(u)} + \mathcal{R}_{\mytree}^{I^{\updownarrow}}(\mathcal{T}_{\mathcal{H}^B}(u_R)) \geq \mathcal{R}_{\mytree}^{I^{\updownarrow}}(\mathcal{T}_{\mathcal{H}^B}(u), r+1)$ by the \vrlproperty-block-property.
    In addition, $p^{\updownarrow}_{\myapp_r(u)} \leq p_{\myapp_r(u)}$ and $v^{\updownarrow}_{\myapp_r(u)} = v_{\myapp_r(u)}$, as observed above.
    Inequality~\eqref{eq:block_4} follows from the induction hypothesis.
\end{itemize}

\subsection{Proof of Lemma~\ref{lemma:block_responsive_approx_factor}} \label{proof:block_responsive_approx_factor}
In what follows, for every node $u$ of the decision tree $\mathcal{T}_{{\mathcal{H}^*}^{\updownarrow}}$, with $\mystate(u) = (t_u, k_u, \mathcal{A}_u)$, we will prove by induction on $k_u$ in increasing order that
\begin{equation}
    \mathcal{R}_{\mytree}^{I^{\updownarrow}}(\mathcal{T}_{\mathcal{H}^B, u}) ~~\geq~~ (1-4\epsilon^3)^{k_u} \cdot \mathcal{R}_{\mytree}^{I^{\updownarrow}}(\mathcal{T}_{{\mathcal{H}^*}^{\updownarrow}}(u))  \ ,  \label{eq:block_proof_1}
\end{equation}
where $\mathcal{T}_{\mathcal{H}^B, u}$ is the block-responsive tree obtained by applying our recursive construction to $\mathcal{T}_{{\mathcal{H}^*}^{\updownarrow}}(u)$.
By instantiating  this inequality with $u = \myroot(\mathcal{T}_{{\mathcal{H}^*}^{\updownarrow}})$, we conclude that
\begin{equation*}
    \mathcal{R}^{I^{\updownarrow}}_{\mytree}(\mathcal{T}_{\mathcal{H}^B}) ~~\geq~~ (1-4\epsilon^3)^k \cdot \mathcal{R}^{I^{\updownarrow}}_{\mytree}(\mathcal{T}_{{\mathcal{H}^*}^{\updownarrow}}) ~~\geq~~ (1-4\epsilon^3k) \cdot \mathcal{R}^{I^{\updownarrow}}_{\mytree}(\mathcal{T}_{{\mathcal{H}^*}^{\updownarrow}}) \ ,
\end{equation*}
where the last transition follows from Bernoulli's inequality.

\paragraph{Base case: $\bs{k_u = 0}$.} In this case, $u$ must be a leaf. As explained in Sections~\ref{subsec:trees} and~\ref{subsec:block_responsive_structural_properties}, every leaf node corresponds to a virtual applicant, whose value and acceptance probability are both 0.
As such, $\mathcal{R}_{\mytree}^{I^{\updownarrow}}(\mathcal{T}_{\mathcal{H}^B, u}) =0=\mathcal{R}_{\mytree}^{I^{\updownarrow}}(\mathcal{T}_{{\mathcal{H}^*}^{\updownarrow}}(u))$.

\paragraph{Induction step: $\bs{k_u \geq 1}$.} When $u$ is a leaf, the argument is identical to that of $k_u = 0$, and we therefore consider the case where $u$ is an internal node, with $\mystate(u) = (t_u, k_u, \mathcal{A}_u)$.
By adopting the notation of Section~\ref{subsec:policy_transformation}, let $u=u_1, \dots, u_{S}$ be the nodes along $P_L(\mathcal{T}_{{\mathcal{H}^*}^{\updownarrow}}(u))$ in root-to-leaf order, and let $u_{\tau_1}, \dots, u_{\tau_F}$ be the sequence of terminals along this path.
We remind the reader that $u^B_1, \dots, u^B_{F}$ are the nodes comprising $P_L(\mathcal{T}_{\mathcal{H}^B, u})$, meaning in particular that $\mathcal{T}_{\mathcal{H}^B, u} = \mathcal{T}_{\mathcal{H}^B}(u^B_1)$.
In addition, as explained in Section~\ref{subsec:policy_transformation}, for every $f \leq F_B-1$, the right subtree of each node $u^B_f$ in this tree is constructed by applying our recursive procedure to $\mathcal{T}_{{\mathcal{H}^*}^{\updownarrow}}( {u_{\tau_{f+1}-1}}_R)$, i.e., $\mathcal{T}_{\mathcal{H}^B}({u^B_f}_R) = \mathcal{T}_{\mathcal{H}^B, {u_{\tau_{f+1}-1}}_R}$.
For $F_B \leq f \leq F$, the right subtree of $u^B_f$ is a leaf node corresponding to the empty block, $B_0$.
In this regard, since $k_{ {u_s}_R } \leq k_u - 1$ for every $s \leq S$, our induction hypothesis implies that
\begin{equation}
    \mathcal{R}_{\mytree}^{I^{\updownarrow}}(\mathcal{T}_{\mathcal{H}^B, {u_s}_R}) ~~\geq~~ (1-4\epsilon^3)^{k_u-1} \cdot \mathcal{R}_{\mytree}^{I^{\updownarrow}}(\mathcal{T}_{{\mathcal{H}^*}^{\updownarrow}}({u_s}_R))  \ .  \label{eq:proof_4.4_hypo}
\end{equation}

\paragraph{Bounding $\boldsymbol{\mathcal{R}_{\mytree}^{I^{\updownarrow}}(\mathcal{T}_{{\mathcal{H}^*}^{\updownarrow}}(u))}$ and $\boldsymbol{\mathcal{R}_{\mytree}^{I^{\updownarrow}}(\mathcal{T}_{\mathcal{H}^B, u})}$.}
In Lemma~\ref{lemma:regular_bound}, we establish an upper bound on $\mathcal{R}_{\mytree}^{I^{\updownarrow}}(\mathcal{T}_{{\mathcal{H}^*}^{\updownarrow}}(u))$
by expressing this expected reward conditional on the identity of the first accepting applicant.
To this end, our analysis decomposes the leftmost path of $\mathcal{T}_{{\mathcal{H}^*}^{\updownarrow}}(u)$ into three subpaths, using terminals as breakpoints, and derives an upper bound on the contribution of each subpath toward $\mathcal{R}_{\mytree}^{I^{\updownarrow}}(\mathcal{T}_{{\mathcal{H}^*}^{\updownarrow}}(u))$.
The finer details of this proof are provided in Appendix~\ref{proof:regular_bound}, where to avoid cumbersome notation, we define
\[\Delta_f ~~=~~ \sum_{s=\tau_f}^{\tau_{f+1}-1} p^{\updownarrow}_{\myapp(u_s)}  \cdot \left(v^{\updownarrow}_{\myapp(u_s)}+\mathcal{R}^{I^{\updownarrow}}_{\mytree}(\mathcal{T}_{{\mathcal{H}^*}^{\updownarrow}}({u_{\tau_{f+1}-1}}_R)) \right)  \ . \]

\begin{lemma} \label{lemma:regular_bound}
The expected reward of $\mathcal{T}_{{\mathcal{H}^*}^{\updownarrow}}(u)$ can be upper-bounded as follows:
\begin{eqnarray}
\mathcal{R}_{\mytree}^{I^{\updownarrow}}(\mathcal{T}_{{\mathcal{H}^*}^{\updownarrow}}(u)) &\leq& \frac{1}{1-3\epsilon^3} \cdot \left(\sum_{f=1}^{F_B-1}\prsub{\mathcal{T}_{{\mathcal{H}^*}^{\updownarrow}}}{u_{1} \rightsquigarrow u_{\tau_f}} \cdot \Delta_f \right. \nonumber \\
&& \mbox{} \left. + \sum_{f=F_B}^{F-1} \prsub{\mathcal{T}_{{\mathcal{H}^*}^{\updownarrow}}}{u_{1} \rightsquigarrow u_{\tau_f}} \cdot \sum_{s=\tau_f}^{\tau_{f+1}-1} p^{\updownarrow}_{\myapp(u_s)}  \cdot v^{\updownarrow}_{\myapp(u_s)}\right) \ . \nonumber
\end{eqnarray}
\end{lemma}

Along these lines, Lemma~\ref{lemma:block_bound} provides a lower bound on $\mathcal{R}_{\mytree}^{I^{\updownarrow}}(\mathcal{T}_{\mathcal{H}^B, u})$, by employing similar arguments in the opposite direction.
The full proof of this result is provided in Appendix~\ref{proof:block_bound}, where we make use of the  shorthand notation
\[\Delta_f^B ~~=~~ \sum_{s=1}^{|\myblock(u^B_f)|}  p^{\updownarrow}_{\myapp_s(u^B_f)} \cdot \left(v^{\updownarrow}_{\myapp_s(u^B_f)}+\mathcal{R}^{I^{\updownarrow}}_{\mytree}(\mathcal{T}_{\mathcal{H}^B}({u^B_f}_R))\right)  \ . \]

\begin{lemma} \label{lemma:block_bound}
The expected reward of $\mathcal{T}_{\mathcal{H}^B, u}$ can be lower-bounded as follows:
\begin{eqnarray}
    \mathcal{R}_{\mytree}^{I^{\updownarrow}}(\mathcal{T}_{\mathcal{H}^B, u}) &\geq& (1-\epsilon^3) \cdot \left( \sum_{f=1}^{F_B-1} \prsub{\mathcal{T}_{{\mathcal{H}^*}^{\updownarrow}}}{u_{1} \rightsquigarrow u_{\tau_f}} \cdot \Delta_f^B \right. \nonumber \\
    && \mbox{} + \left.  \sum_{f=F_B}^{F-1} \prsub{\mathcal{T}_{{\mathcal{H}^*}^{\updownarrow}}}{u_{1} \rightsquigarrow u_{\tau_f}} \cdot \sum_{s=\tau_f}^{\tau_{f+1}-1}  p^{\updownarrow}_{\myapp(u_s)} \cdot v^{\updownarrow}_{\myapp(u_s)}\right) \ .  \nonumber
\end{eqnarray}
\end{lemma}

\paragraph{Deriving inequality~\eqref{eq:block_proof_1}.} Consequently, let us notice that, by our induction hypothesis \eqref{eq:proof_4.4_hypo}, we have $\Delta_f^B \geq (1-4\epsilon^3)^{k_u - 1} \cdot \Delta_f$, for every $f \leq F_B-1$. Plugging this relation into Lemma~\ref{lemma:block_bound}, we end up with
\begin{eqnarray}
    \mathcal{R}_{\mytree}^{I^{\updownarrow}}(\mathcal{T}_{\mathcal{H}^B, u}) &\geq& (1-\epsilon^3) \cdot (1-4\epsilon^3)^{k_u - 1} \cdot \left( \sum_{f=1}^{F_B-1} \prsub{\mathcal{T}_{{\mathcal{H}^*}^{\updownarrow}}}{u_{1} \rightsquigarrow u_{\tau_f}} \cdot \Delta_f \right. \nonumber \\
    && \qquad \mbox{} + \left.  \sum_{f=F_B}^{F-1} \prsub{\mathcal{T}_{{\mathcal{H}^*}^{\updownarrow}}}{u_{1} \rightsquigarrow u_{\tau_f}} \cdot \sum_{s=\tau_f}^{\tau_{f+1}-1}  p^{\updownarrow}_{\myapp(u_s)} \cdot v^{\updownarrow}_{\myapp(u_s)}\right) \nonumber \\
    &\geq& (1-4\epsilon^3)^{k_u } \cdot \mathcal{R}_{\mytree}^{I^{\updownarrow}}(\mathcal{T}_{{\mathcal{H}^*}^{\updownarrow}}(u))  \ , \nonumber
\end{eqnarray}
where the last inequality follows from Lemma~\ref{lemma:regular_bound}.

\subsection{Proof of Lemma~\ref{lemma:regular_bound}} \label{proof:regular_bound}
We begin by expanding representation~\eqref{eq:reward_tree_recursive} of the expected reward function to express $\mathcal{R}_{\mytree}^{I^{\updownarrow}}(\mathcal{T}_{{\mathcal{H}^*}^{\updownarrow}}(u))$ conditional on the identity of the first accepting applicant, namely, 
\begin{eqnarray}
    \mathcal{R}_{\mytree}^{I^{\updownarrow}}(\mathcal{T}_{{\mathcal{H}^*}^{\updownarrow}}(u)) &=& \sum_{s=1}^{S-1} \prsub{\mathcal{T}_{{\mathcal{H}^*}^{\updownarrow}}}{u_{1} \rightsquigarrow u_{s}}  \cdot p^{\updownarrow}_{\myapp(u_s)}  \cdot \left(v^{\updownarrow}_{\myapp(u_s)}+\mathcal{R}^{I^{\updownarrow}}_{\mytree}(\mathcal{T}_{{\mathcal{H}^*}^{\updownarrow}}({u_{s}}_R))\right)  \ .  \nonumber 
\end{eqnarray}
Next, we partition the latter summation into three ranges: (I) $1 \leq s \leq \tau_{F_B}-1$; (II) $\tau_{F_B} \leq s \leq \tau_F-1$; and (III) $\tau_F \leq s \leq S-1$, providing upper bounds for each of these parts.

\paragraph{Bounding $\bs{(\mathrm{I})}$.}
To upper-bound (I), note that
\begin{eqnarray}
    \text{(I)} &=&\sum_{f=1}^{F_B-1}\prsub{\mathcal{T}_{{\mathcal{H}^*}^{\updownarrow}}}{u_{1} \rightsquigarrow u_{\tau_f}} \cdot \sum_{s=\tau_f}^{\tau_{f+1}-1}\prsub{\mathcal{T}_{{\mathcal{H}^*}^{\updownarrow}}}{u_{\tau_f} \rightsquigarrow u_{s}} \cdot p^{\updownarrow}_{\myapp(u_s)}  \nonumber \\
    &\mbox{}& \qquad\qquad\qquad\qquad\qquad\qquad\qquad \cdot\left(v^{\updownarrow}_{\myapp(u_s)}+\mathcal{R}^{I^{\updownarrow}}_{\mytree}(\mathcal{T}_{{\mathcal{H}^*}^{\updownarrow}}({u_{s}}_R)) \right) \nonumber \\
    &\leq&\sum_{f=1}^{F_B-1}\prsub{\mathcal{T}_{{\mathcal{H}^*}^{\updownarrow}}}{u_{1} \rightsquigarrow u_{\tau_f}} \cdot \sum_{s=\tau_f}^{\tau_{f+1}-1} p^{\updownarrow}_{\myapp(u_s)}  \cdot \left(v^{\updownarrow}_{\myapp(u_s)}+\mathcal{R}^{I^{\updownarrow}}_{\mytree}(\mathcal{T}_{{\mathcal{H}^*}^{\updownarrow}}({u_{s}}_R)) \right) \nonumber \\
    &\leq&\sum_{f=1}^{F_B-1}\prsub{\mathcal{T}_{{\mathcal{H}^*}^{\updownarrow}}}{u_{1} \rightsquigarrow u_{\tau_f}} \cdot \sum_{s=\tau_f}^{\tau_{f+1}-1} p^{\updownarrow}_{\myapp(u_s)} \nonumber \\
    &\mbox{}& \qquad\qquad\qquad\qquad\qquad\qquad \cdot \left(v^{\updownarrow}_{\myapp(u_s)}+\frac{1}{1-\epsilon^3} \cdot\mathcal{R}^{I^{\updownarrow}}_{\mytree}(\mathcal{T}_{{\mathcal{H}^*}^{\updownarrow}}({u_{\tau_{f+1}-1}}_R)) \right) \label{eq:proof_c_1_2} \\
    &\leq&\frac{1}{1-\epsilon^3} \cdot\sum_{f=1}^{F_B-1}\prsub{\mathcal{T}_{{\mathcal{H}^*}^{\updownarrow}}}{u_{1} \rightsquigarrow u_{\tau_f}} \cdot \Delta_f  \ . \nonumber 
\end{eqnarray}
Here, to obtain inequality~\eqref{eq:proof_c_1_2},  we notice that for every $\tau_f \leq s \leq \tau_{f+1}-1$, there are no $B$-crossing arcs on the path connecting $u_f$ to $u_{\tau_{f+1}-1}$ on the leftmost path of $\mathcal{T}_{{\mathcal{H}^*}^{\updownarrow}}(u)$, implying that $\mathcal{R}^{I^{\updownarrow}}_{\mytree}(\mathcal{T}_{{\mathcal{H}^*}^{\updownarrow}}({u_{s}}_R)) \leq \frac{1}{1-\epsilon^3} \cdot\mathcal{R}^{I^{\updownarrow}}_{\mytree}(\mathcal{T}_{{\mathcal{H}^*}^{\updownarrow}}({u_{\tau_{f+1}-1}}_R))$.

\paragraph{Bounding $\bs{(\mathrm{II})}$.}
To bound (II), we recall that $u_{F_B}$ is the first node along $P_L(\mathcal{T}_{{\mathcal{H}^*}^{\updownarrow}}(u))$ whose right subtree has an expected reward of at most $\epsilon^3 \cdot \mathcal{R}^{I^{\updownarrow}}_{\mytree}(\mathcal{T}_{{\mathcal{H}^*}^{\updownarrow}}(u))$.
Since $\mathcal{T}_{{\mathcal{H}^*}^{\updownarrow}}$ is an optimal decision tree, the nodes $u_{\tau_{F_B+1}}, \dots, u_s$ must be satisfying this property as well. Therefore,
\begin{eqnarray}
  \text{(II)} &\leq&\sum_{s=\tau_{F_B}}^{\tau_F-1} \prsub{\mathcal{T}_{{\mathcal{H}^*}^{\updownarrow}}}{u_{1} \rightsquigarrow u_{s}}  \cdot p^{\updownarrow}_{\myapp(u_s)}  \cdot \left(v^{\updownarrow}_{\myapp(u_s)}+\epsilon^3 \cdot\mathcal{R}^{I^{\updownarrow}}_{\mytree}(\mathcal{T}_{{\mathcal{H}^*}^{\updownarrow}}(u)) \right) \nonumber\\
    &\leq& \sum_{s=\tau_{F_B}}^{\tau_F-1} \prsub{\mathcal{T}_{{\mathcal{H}^*}^{\updownarrow}}}{u_{1} \rightsquigarrow u_{s}}  \cdot p^{\updownarrow}_{\myapp(u_s)}  \cdot v^{\updownarrow}_{\myapp(u_s)} + \epsilon^3 \cdot\mathcal{R}^{I^{\updownarrow}}_{\mytree}(\mathcal{T}_{{\mathcal{H}^*}^{\updownarrow}}(u))\nonumber \\
    &\leq& \sum_{f=F_B}^{F-1} \prsub{\mathcal{T}_{{\mathcal{H}^*}^{\updownarrow}}}{u_{1} \rightsquigarrow u_{\tau_f}} \cdot \sum_{s=\tau_f}^{\tau_{f+1}-1} p^{\updownarrow}_{\myapp(u_s)}  \cdot v^{\updownarrow}_{\myapp(u_s)} + \epsilon^3 \cdot\mathcal{R}^{I^{\updownarrow}}_{\mytree}(\mathcal{T}_{{\mathcal{H}^*}^{\updownarrow}}(u)) \ . \nonumber  
\end{eqnarray}

\paragraph{Bounding $\bs{(\mathrm{III})}$.}
Finally, the expected reward over  (III) can be equivalently written as 
\begin{eqnarray}
    \text{(III)} &=& \prsub{\mathcal{T}_{{\mathcal{H}^*}^{\updownarrow}}}{u_{1} \rightsquigarrow u_{\tau_F}} \cdot \mathcal{R}^{I^{\updownarrow}}_{\mytree}(\mathcal{T}_{{\mathcal{H}^*}^{\updownarrow}}(u_{\tau_F})) \nonumber \\
    &\leq& \epsilon^3 \cdot \mathcal{R}^{I^{\updownarrow}}_{\mytree}(\mathcal{T}_{{\mathcal{H}^*}^{\updownarrow}}(u_{\tau_F})) \label{eq:proof_c1_3} \\
    &\leq& \epsilon^3 \cdot \mathcal{R}^{I^{\updownarrow}}_{\mytree}(\mathcal{T}_{{\mathcal{H}^*}^{\updownarrow}}(u)) \ . \label{eq:proof_c1_4}
\end{eqnarray}
Here, inequalities~\eqref{eq:proof_c1_3} and~\eqref{eq:proof_c1_4}  respectively hold since $\prsub{\mathcal{T}_{{\mathcal{H}^*}^{\updownarrow}}}{u_{1} \rightsquigarrow u_{\tau_F}} < \epsilon^3$, by definition of $\tau_F$, and since  $\mathcal{T}_{{\mathcal{H}^*}^{\updownarrow}}$ is an optimal decision tree.

\paragraph{Putting it all together.}
By combining our upper bounds on (I), (II), and (III), we have
\begin{eqnarray}
    \mathcal{R}^{I^{\updownarrow}}_{\mytree}(\mathcal{T}_{{\mathcal{H}^*}^{\updownarrow}}(u)) &\leq& 2\epsilon^3 \cdot \mathcal{R}^{I^{\updownarrow}}_{\mytree}(\mathcal{T}_{{\mathcal{H}^*}^{\updownarrow}}(u)) +  \sum_{f=F_B}^{F-1} \prsub{\mathcal{T}_{{\mathcal{H}^*}^{\updownarrow}}}{u_{1} \rightsquigarrow u_{\tau_f}} \cdot \sum_{s=\tau_f}^{\tau_{f+1}-1} p^{\updownarrow}_{\myapp(u_s)}  \cdot v^{\updownarrow}_{\myapp(u_s)} \nonumber \\
    && \mbox{}+\frac{1}{1-\epsilon^3} \cdot\sum_{f=1}^{F_B-1}\prsub{\mathcal{T}_{{\mathcal{H}^*}^{\updownarrow}}}{u_{1} \rightsquigarrow u_{\tau_f}} \cdot \Delta_f  \ , \nonumber
\end{eqnarray}
and we obtain the desired claim by rearranging this inequality.

\subsection{Proof of Lemma~\ref{lemma:block_bound}} \label{proof:block_bound}
By expanding representation~\eqref{eq:block_reward_def} for the expected reward of a block-responsive policy and conditioning on the identity of the first accepting applicant, we have 
\begin{eqnarray}
\mathcal{R}_{\mytree}^{I^{\updownarrow}}(\mathcal{T}_{\mathcal{H}^B, u}) &=& \sum_{f=1}^{F-1} \prsub{\mathcal{T}_{{\mathcal{H}^*}^{\updownarrow}}}{u_{1} \rightsquigarrow u_{\tau_f}} \cdot\left(\sum_{q=1}^{|\myblock(u^B_f)|} \left( \prod_{\hat{q}=1}^{q-1}(1-p^{\updownarrow}_{\myapp_{\hat{q}}(u^B_f)}) \right) \cdot p^{\updownarrow}_{\myapp_q(u^B_f)} \right. \nonumber \\
&& \qquad \qquad \qquad \qquad \qquad \qquad \qquad \left. \vphantom{\sum_{q=1}^{|\myblock(u^B_f)|} \left( \prod_{\hat{q}=1}^{q-1}(1-p^{\updownarrow}_{\myapp_{\hat{q}}(u^B_f)}) \right)} \cdot (v^{\updownarrow}_{\myapp_q(u^B_f)}+\mathcal{R}^{I^{\updownarrow}}_{\mytree}(\mathcal{T}_{\mathcal{H}^B, {u_{\tau_{f+1}-1}}_R}))\right) \nonumber\\
    &\geq& (1-\epsilon^3) \cdot \sum_{f=1}^{F-1} \prsub{\mathcal{T}_{{\mathcal{H}^*}^{\updownarrow}}}{u_{1} \rightsquigarrow u_{\tau_f}} \nonumber\\
    &\mbox{}& \qquad\qquad\qquad \cdot \sum_{q=1}^{|\myblock(u^B_f)|}  p^{\updownarrow}_{\myapp_q(u^B_f)} \cdot (v^{\updownarrow}_{\myapp_q(u^B_f)}+\mathcal{R}^{I^{\updownarrow}}_{\mytree}(\mathcal{T}_{\mathcal{H}^B, {u_{\tau_{f+1}-1}}_R})) \ . \label{eq:proof_c2_1}    
\end{eqnarray}
To better understand this inequality, we remind the reader that $\myblock(u^B_f)$ consists of the applicants $\myapp(u_{\tau_f}), \dots, \myapp(u_{\tau_{f+1}-1})$.
As such, $\prod_{\hat{q}=1}^{q-1}(1-p^{\updownarrow}_{\myapp_{\hat{q}}(u^B_f)}) \geq \prod_{s=\tau_f}^{\tau_{f+1}-1}(1-p^{\updownarrow}_{\myapp(u_s)}) \geq 1-\epsilon^3$, since there are no $A$-crossing arcs in the path connecting $u_{\tau_f}$ to $u_{\tau_{f+1}}$.
Next, we partition the outer summation in \eqref{eq:proof_c2_1} into two ranges: (I) $1 \leq f \leq F_B-1$; and (II) $F_B \leq f \leq F-1$, providing lower bounds for each of these parts.

\paragraph{Bounding $\bs{(\mathrm{I})}$.}
To lower bound (I), we notice that $\mathcal{T}_{\mathcal{H}^B}({u^B_f}_R) = \mathcal{T}_{\mathcal{H}^B, {u_{\tau_{f+1}-1}}_R}$ for every $f \leq F_B-1$. As such,
\begin{eqnarray}
    \text{(I)} &=& \sum_{f=1}^{F_B-1} \prsub{\mathcal{T}_{{\mathcal{H}^*}^{\updownarrow}}}{u_{1} \rightsquigarrow u_{\tau_f}} \cdot\left(\sum_{q=1}^{|\myblock(u^B_f)|}  p^{\updownarrow}_{\myapp_q(u^B_f)} \cdot (v^{\updownarrow}_{\myapp_q(u^B_f)}+\mathcal{R}^{I^{\updownarrow}}_{\mytree}(\mathcal{T}_{\mathcal{H}^B}({u^B_f}_R)))\right)  \nonumber \\
    &=& \sum_{f=1}^{F_B-1} \prsub{\mathcal{T}_{{\mathcal{H}^*}^{\updownarrow}}}{u_{1} \rightsquigarrow u_{\tau_f}} \cdot \Delta_f^B  \ . \nonumber
\end{eqnarray}

\paragraph{Bounding $\bs{(\mathrm{II})}$.}
Here, to lower bound (II), we recall that $\myblock(\mathcal{T}_{\mathcal{H}^B}({u^B_f}_R)) = B_0 = \emptyset$ for every $F_B \leq f \leq F-1$.
Therefore,
\begin{eqnarray}
    \text{(II)} &=&\sum_{f=F_B}^{F-1} \prsub{\mathcal{T}_{{\mathcal{H}^*}^{\updownarrow}}}{u_{1} \rightsquigarrow u_{\tau_f}} \cdot \sum_{q=1}^{|\myblock(u^B_f)|}  p^{\updownarrow}_{\myapp_q(u^B_f)} \cdot v^{\updownarrow}_{\myapp_q(u^B_f)} \nonumber \\
    &=&\sum_{f=F_B}^{F-1} \prsub{\mathcal{T}_{{\mathcal{H}^*}^{\updownarrow}}}{u_{1} \rightsquigarrow u_{\tau_f}} \cdot \sum_{q=\tau_f}^{\tau_{f+1}-1}  p^{\updownarrow}_{\myapp(u_q)} \cdot v^{\updownarrow}_{\myapp(u_q)} \ . \nonumber
\end{eqnarray}

\paragraph{Putting it all together.}
By combining our lower bounds on (I) and (II), we have
\begin{eqnarray}
    \mathcal{R}_{\mytree}^{I^{\updownarrow}}(\mathcal{T}_{\mathcal{H}^B, u}) &\geq& (1-\epsilon^3) \cdot \left( \sum_{f=1}^{F_B-1} \prsub{\mathcal{T}_{{\mathcal{H}^*}^{\updownarrow}}}{u_{1} \rightsquigarrow u_{\tau_f}} \cdot \Delta_f^B \right. \nonumber \\
    &\mbox{}&~~ \left. + \sum_{f=F_B}^{F-1} \prsub{\mathcal{T}_{{\mathcal{H}^*}^{\updownarrow}}}{u_{1} \rightsquigarrow u_{\tau_f}} \cdot \sum_{s=\tau_f}^{\tau_{f+1}-1}  p^{\updownarrow}_{\myapp(u_s)} \cdot v^{\updownarrow}_{\myapp(u_s)}\right) \ . \nonumber
\end{eqnarray}

\section{Additional Proofs from Section~\ref{section:PTAS}}

\subsection{Proof of inequality~\eqref{eq:ptas_bounding_2}} \label{proof:rejection_prob_correct}

When $q=1$, the desired claim is trivial. When $2 \leq q \leq |\myblock(\tilde{u}^B)|$, we observe that
\begin{eqnarray}
    \prod_{\hat{q}=1}^{q-1}\left(1-p^{\updownarrow}_{\myapp_{\hat{q}}(\tilde{u}^B)}\right) &\geq& \prod_{\hat{q}=1}^{|\myblock(\tilde{u}^B)|}\left(1-p^{\updownarrow}_{\myapp_{\hat{q}}(\tilde{u}^B)}\right) \nonumber \\
    &=& \psi_{\tilde{u}^B} \nonumber\\
    &\geq&  \psi_{u^B} \label{eq:ptas_gua_1}\\
    &\geq& 1-\epsilon^3 \ . \label{eq:ptas_gua_2}
\end{eqnarray}
In the remainder of this proof, we establish inequality~\eqref{eq:ptas_gua_1}, showing along the way that $|\myblock(u^B)| \geq |\myblock(\tilde{u}^B)|$.
The latter claim shows in particular that $|\myblock(u^B)| \geq 2$, meaning that inequality~\eqref{eq:ptas_gua_2} directly follows from Observation~\ref{corollary:rejection_prob}.
To relate the rejection probabilities $\psi_{\tilde{u}^B}$ and $\psi_{u^B}$, we note that
\begin{eqnarray}
    \psi_{\tilde{u}^B} &=&\prod_{\hat{q}=1}^{|\myblock(\tilde{u}^B)|}\left(1-p^{\updownarrow}_{\myapp_{\hat{q}}(\tilde{u}^B)}\right) \nonumber \\
    &=& \prod_{m \leq M} \left( 1-p^{(m)}\right)^{|\myblock(\tilde{u}^B)\cap \mathcal{C}_m|} \label{eq:rej_correct_1} \\
    &\geq&  \prod_{m \leq M} \left( 1-p^{(m)}\right)^{|\myblock(u^B)\cap \mathcal{C}_m|} \label{eq:rej_correct_2}\\
    &=& \prod_{\hat{q}=1}^{|\myblock(u^B)|}\left(1-p^{\updownarrow}_{\myapp_{\hat{q}}(u^B)}\right) \nonumber\\
    &=& \psi_{u^B} \ .  \nonumber
\end{eqnarray}
Here, inequality~\eqref{eq:rej_correct_1} is obtained by recalling that all $\mathcal{C}_m$-applicants share the same acceptance probability, $p^{(m)}$.
To arrive at inequality~\eqref{eq:rej_correct_2}, the important observation is that our feasibility analysis in Section~\ref{subsec:ptas_feasability} shows en route that $|\myblock(\tilde{u}^B) \cap \mathcal{C}_m| \leq |\myblock(u^B) \cap \mathcal{C}_m|$, i.e., the number of $\mathcal{C}_m$-applicants in $\myblock(\tilde{u}^B)$ is upper-bounded by that of $\myblock(u^B)$, for every $m \leq M$.

\end{document}